%% file: CAN_template.tex
\documentclass[twoside, a4paper, 12pt]{article}

\usepackage{amsfonts}
\usepackage{amsmath}
\usepackage{amssymb}
\usepackage{array}

\usepackage[english]{babel}
\usepackage{bm}

\usepackage{caption}
\usepackage{color}

\usepackage[hang, flushmargin]{footmisc} 

\usepackage[left = 2.9cm, right = 2.9cm, top = 3cm, bottom = 4.5cm]{geometry}
\usepackage{graphicx}
\usepackage{graphics}

\usepackage{hyperref}

\usepackage{makecell}
\usepackage{multirow}

\usepackage{nicefrac}

\usepackage{orcidlink}

\usepackage{pifont}
\usepackage{placeins}

\usepackage{rotating}

\usepackage{setspace}
\usepackage{siunitx}
\usepackage[small, bf, hang, raggedright]{subfigure}

\usepackage{textcomp}

\usepackage{verbatim}

\usepackage{wrapfig}

\usepackage{xspace}

\newcommand{\ahcal}{{A{\textsc{hcal}}}\xspace}
\newcommand{\atlas}{{A{\textsc{tlas}}}\xspace}
\newcommand{\calice}{{C{\textsc{alice}}}\xspace}
\newcommand{\cern}{{C{\textsc{ern}}}\xspace}
\newcommand{\Geant}{{G{\textsc{eant4}}}\xspace}
\newcommand{\cms}{{C{\textsc{ms}}}\xspace}
\newcommand{\hgcal}{{H{\textsc{gcal}}}\xspace}

\DeclareSIUnit\mip{MIP}

\begin{document}
    \selectlanguage{english}
    \begin{titlepage}
        \begin{flushright}
            Prepared for submission to JINST\\
            2026\\
        \end{flushright}
        \begin{center}
            \vspace*{\fill}
            \begin{LARGE}
                \textbf{An investigation of fast simulation techniques for pion showers using kernel density estimators with the CALICE AHCAL Technological Prototype}
            \end{LARGE}\\[10ex]
            \begin{large}
                \textbf{Abstract}\\
            \end{large}
        \end{center}
        In this article, the development and investigation of fast hadron shower simulation methods is presented. A test beam dataset has been recorded in 2018 at \cern with the \ahcal Technological Prototype of the \calice Collaboration, where the calorimeter prototype was exposed to electron, muon, and negatively charged pion beams of various initial energies. The pion shower dataset, covering energies between $\SI{10}{\giga\electronvolt}$ and $\SI{200}{\giga\electronvolt}$, has been used to develop a data-driven fast simulation algorithm of the \ahcal response to pion showers. The resulting shower model demonstrates excellent agreement with measured shower observables. In addition, a method for simulating pion showers at arbitrary energies is introduced, based upon interpolation between simulated showers at neighbouring beam energies.
        \vspace*{1cm}
        \begin{center}
            \begin{large}
                \textbf{Keywords}
            \end{large}
        \end{center}
        Calorimeters; Detector alignment and calibration methods; Detector design and construction technologies and materials; Detector modelling and simulations I (interaction of radiation with matter, interaction of photons with matter, interaction of hadrons with matter, etc.)
    \end{titlepage}
    \begin{center}
        \begin{large}
            \textbf{The CALICE Collaboration}
        \end{large}
    \end{center}
    A. Wilhahn\orcidlink{0009-0006-3588-9403}\footnote[1]{Corresponding author: \texttt{andre.wilhahn@uni-goettingen.de}}\setcounter{footnote}{0}\renewcommand{\thefootnote}{\alph{footnote}}\footnotemark, J. Utehs\footnotemark[\value{footnote}], Z. Ghafoor\footnotemark[\value{footnote}], G. Eigen\footnotemark[\value{footnote}], S. Lai\footnotemark[\value{footnote}]\footnotetext{II. Physikalisches Institut, Georg-August-Universit\"at G\"ottingen, Friedrich-Hund-Platz 1, D-37077 G\"ottingen, Germany},
    O. Bach\footnotemark, E. Brianne\footnotemark[\value{footnote}], K. Gadow\footnotemark[\value{footnote}], D. Heuchel\footnotemark[\value{footnote}], K. Kr\"{u}ger\footnotemark[\value{footnote}], J. Kvasnicka\footnotemark[\value{footnote}]\footnotetext[2]{DESY, Notkestrasse 85, D-22603 Hamburg, Germany}\footnote{Also at Institute of Physics, The Czech Academy of Sciences}, A.  Laudrain\footnotemark[2], O. Pinto\footnotemark[2], M. Reinecke\footnotemark[2], F. Sefkow\footnotemark[2], M. De Silva\footnotemark[2],
    E. Garutti\footnotemark, G. Kasieczka\footnotemark[\value{footnote}], S. Martens\footnotemark[\value{footnote}], J. Rolph\footnotemark[\value{footnote}]\footnotetext{Univ. Hamburg,  Physics Department,  Institut f\"ur Experimentalphysik,  Luruper Chaussee 149,  22761 Hamburg, Germany},
    F. Hummer\footnotemark, F. Simon\footnotemark[\value{footnote}]\footnotetext{Karlsruhe Institute of Technology, Institute for Data Processing and Electronics, Kaiserstr. 12, D-76131 Karlsruhe, Germany},
    A. Brogna\footnotemark, V. B\"uscher\footnotemark[\value{footnote}], L. Masetti\footnotemark[\value{footnote}], A. Rosmanitz\footnotemark[\value{footnote}], C. Schmitt\footnotemark[\value{footnote}], Q. Weitzel\footnotemark[\value{footnote}]\footnotetext{Institut f\"ur Physik und PRISMA$^{++}$ Cluster of Excellence, Universit\"at Mainz, Staudinger Weg 7, D-55099 Mainz, Germany},
    W. Ootani\footnotemark, T. Suehara\footnotemark[\value{footnote}]\footnotetext{ICEPP, The University of Tokyo, 7-3-1 Hongo, Bunkyo-ku, Tokyo 113-0033, Japan},
    A. Irles\footnote{Instituto de F\'isica Corpuscular (IFIC), CSIC-Universitat de Val\`encia, Parque Cient\'ifico, Catedr\'atico Jos\'e Beltr\'an, 2 | E-46980 Paterna, Espa\~na}
    \setcounter{footnote}{0}
    \renewcommand{\thefootnote}{\arabic{footnote}}
    \vspace*{1cm}

    \clearpage
    \tableofcontents
    \newpage
    
    \input{chapters/introduction}
    \clearpage
    \input{chapters/ahcal}
    \clearpage
    \input{chapters/kde}
    \clearpage
    \input{chapters/hitlevel}
    \clearpage
    \input{chapters/interpolation}
    \clearpage
    \input{chapters/conclusion}
    \clearpage
    
    \addcontentsline{toc}{section}{Bibliography}
    \bibliographystyle{CAN_template}
    \bibliography{references}
    \clearpage
\end{document}

%% file: chapters/introduction.tex
\section{Introduction}
\label{sec: introduction}

Sophisticated particle detectors are an essential tool of modern high-energy particle physics. In order to detect and study the properties of particles, particle detectors comprise multiple components, each fulfilling a different purpose. These detector components are carefully developed, built, and tested to ensure reliable functionality, high performance, as well as fine resolution. However, modern high-energy physics experiments not only require well-functioning particle detectors for data collection, but also the extensive simulation of particle interactions with the detector material beforehand. This serves two main purposes: firstly, detector simulations allow for design enhancements by evaluating their performance under realistic conditions. Secondly, simulations facilitate a sensible interpretation of experimental data. Detailed simulations are therefore crucial in order to validate detector performance and to provide accurate predictions to compare with measurements.
\par
Currently, \Geant (“\textbf{Ge}ometry \textbf{an}d \textbf{t}racking”) \cite{geant4_1, geant4_2, geant4_3} is the standard framework for the simulation of particle interactions with matter in high-energy physics experiments. Based on Monte-Carlo (MC) methods, \Geant is able to generate event kinematics, track single particles through complex detector geometries, and simulate interactions between particles and matter while taking into account a comprehensive set of physics processes. Although MC simulations yield highly accurate predictions, they are usually computationally expensive and can suffer from residual mismodelling. To mitigate these problems, fast simulations can be employed, for they are a useful tool to capture and provide the most important information about particle interactions without relying on large amounts of resources and computing time. By now, they have become indispensable for large-scale high-energy physics experiments, and particle physics collaborations such as \atlas or \cms are extensively employing them to model a wide range of physics objects \cite{atl_fast_3, cms_fast_sim}. Among many approaches, these fast simulations can be implemented as data-driven simulations through which the need to explicitly compute all underlying physical processes can be easily avoided.
\par
In this article, a data-driven fast simulation is developed and tested for a highly granular calorimeter prototype. This fast simulation algorithm is based on test beam data, and with the help of kernel density estimators (KDEs) \cite{kde_paper_1, kde_paper_2}, it is applied to simulate pion showers at various energies. The calorimeter prototype is the \ahcal technological prototype \cite{ahcal_paper}, developed by the \calice Collaboration to test modern calorimeter reconstruction methods, such as particle flow \cite{pfa_cms, pfa_atlas} and software compensation. The test beam data was recorded at \cern in June 2018. It is shown that this fast simulation algorithm can perform as well or better than full simulation \cite{olin_pinto} in describing many pion shower variables. Moreover, a method to simulate pion showers at energies not available in the test beam is also presented. For the comparison of full simulation with data and fast simulation, pion shower samples have been simulated with CaliceSoft version 4.15.
\par
Other approaches to use fast simulation of calorimeter responses to high energy particles are detailed in Refs. \cite{generative_models_1, generative_models_2, generative_models_3}, where neural network based algorithms \cite{wasserstein_GAN, autoencoder} were trained on simulated test beam data. In this article however, the dataset upon which the simulation is based is true test beam data, complete with realistic experimental conditions. The usage of KDEs avoids the complex training procedure of neural network based algorithms, along with optimising many arbitrary architectures.
\par
Data-driven simulations offer a powerful way of modelling physical processes and detector responses, since they reflect real-world behaviour more accurately than conventional simulation methods due to the empirical evidence they are based upon. Implementing simulations in a data-driven manner allows to make predictions with higher confidence and to probe physics at any scale. Furthermore, they also allow to identify physics beyond the established theoretical framework, since unknown processes are already, by default, incorporated into the underlying dataset. However, data-driven simulations also come with certain drawbacks. In particular, detector imperfections and flawed experimental setups are inevitably included in the recorded data and, therefore, become incorporated into the simulation framework too, which results in intrinsic mismodelling of physical processes. If such imperfections are unique only to the experimental setups in the specific dataset upon which the simulation algorithm is based, and not present in general data-taking conditions, then this will result in mismodelling effects. In addition, data-driven simulations are confined to the energies that are available in the underlying dataset, and sophisticated algorithms need to be developed in order to inter- and extrapolate between the given energies. This problem, in particular, will be addressed in this article.
\par
The main method of generating pion shower probability density functions (PDFs) used throughout this article are KDEs. They have been used to estimate and model high-dimensional distributions of particle showers based upon an input subset of the aforementioned test beam dataset, and to subsequently simulate pion showers by sampling from the resulting PDFs. From these simulated event samples, kinematic variables were then computed in order to assess the performance of the fast simulation algorithm.
\par
The structure of this article is as follows. Section \ref{sec: the ahcal prototype} begins with a brief overview of the \ahcal technological prototype. In Section \ref{sec: kernel density estimators for pion showers}, kernel density estimators are introduced, and Section \ref{sec: simulation of hit energy distributions using kernel density estimators} then follows with the application of KDEs to the aforementioned dataset in order to estimate PDFs and to simulate pion showers at various initial energies. After this, Section \ref{sec: interpolation studies of hit energy distributions using kernel density estimators} presents an interpolation algorithm for hit energy distributions at various energies. Finally, a conclusion and an outlook are given in Section \ref{sec: conclusion}.

%% file: chapters/ahcal.tex
\section{The AHCAL Prototype}
\label{sec: the ahcal prototype}

The \ahcal technological prototype (henceforth simply referred to as ``the \ahcal'') is a sampling calorimeter that uses non-magnetic stainless steel as absorber material with a total of $38$ active scintillator layers integrated into the absorber structure. Each absorber layer has a thickness of $\SI{17.2}{\milli\meter}$, corresponding to approximately one radiation length or $\SI{0.1}{}$ nuclear absorption lengths, while each active layer is only $\SI{5.4}{\milli\meter}$ thick. This amounts to a total depth of about $\SI{4.4}{}$ interaction lengths for the fully assembled \ahcal.
\par
Each active layer of the \ahcal consists of four HCAL Base Units (HBUs), each covering an area of $36\times\SI{36}{\centi\meter\squared}$. Together, they are arranged quadratically, such that one active layer spans $72\times\SI{72}{\centi\meter\squared}$. A single HBU is equipped with $144$ ($12\times 12$) active scintillator tiles of $3\times\SI{3}{\centi\meter\squared}$, which means that one active layer encompasses a grid of $576$ ($24\times 24$) scintillator tiles. Consequently, the full detector prototype comprises $\SI{21888}{}$ channels that are individually read out via silicon photomultipliers (SiPMs). The SiPM model selected for the \ahcal is the Hamamatsu MPPC of type S13360-1325PE. In addition, every tile is individually wrapped in reflector foil in order to minimise optical crosstalk between the tiles. Images of the \ahcal can be found, for example, in Ref. \cite{jack_rolph}.
\par
The dataset that has been used for this investigation was recorded in $2018$ at the \cern Super Proton Synchrotron beam test facility. The test beam campaign was conducted in three separate periods during May, June, and October. During these runs, the \ahcal was exposed to muon beams (for calibration) as well as to electron or negatively charged pion beams of various energies, which include $\{10,\ 20,\ 30,\ 40,\ 60,\ 80,\ 100\}\, \SI{}{\giga\electronvolt}$ for electrons and $\{10,\ 15,\ 20,\ 30,\ 40,\ 50,\ 60,\ 80,\\ 100,\ 120,\ 160,\ 200,\ 350\}\, \SI{}{\giga\electronvolt}$ for pions. The beams were directed perpendicular to the detector's $xy$-plane, and in order to move the detector up and down or left and right, it was placed on a movable platform. This allowed particles to be detected in, and to probe, different regions of the detector volume.
\par
Data recorded in May was found to be of lower quality, and during the October campaign, the \ahcal was positioned downstream of the \cms\ \hgcal \cite{hgcal1, hgcal2}, such that only shower tails reached the first few \ahcal layers. Therefore, only the pion dataset acquired in June has been used for this study. This dataset includes events at nine initial pion energies: $E_{\text{initial}}=\{10,\ 20,\ 30,\ 40,\ 60,\ 80,\ 120,\ 160,\ 200\}\, \SI{}{\giga\electronvolt}$. During the campaign in June, the $38$th active layer was replaced by a module equipped with larger $6\times\SI{6}{\centi\meter\squared}$ scintillator tiles. The replaced layer was instead reinstalled in the $41$st absorber gap. In addition, a single HBU was installed in front of the detector, and a tail catcher was setup in the rear. Throughout the campaign, the detector was operated in power-pulsing mode.
\par
Several detailed studies have already been carried out using \ahcal data. These include investigations of the calorimeter response and the ratio $h/e$ from longitudinal shower profiles as well as parameterisations of their profiles \cite{hadronic_shower_profiles}, the time development of hadronic showers \cite{eldwan_brianne}, precision time and energy measurements \cite{christian_graf}, particle flow reconstruction \cite{daniel_heuchel}, shower shape modelling \cite{olin_pinto}, and shower separation in five dimensions using machine learning techniques \cite{jack_rolph}. This work, in contrast, focuses on the use of test beam data to develop multidimensional kernel density estimators, to sample simulated events from them, and to validate this data-driven approach against experimental data.

%% file: chapters/kde.tex
\section{Kernel Density Estimators for Pion Shower PDFs}
\label{sec: kernel density estimators for pion showers}

As is often the case in experimental physics, the underlying density function of a given dataset might not always be known because the established theoretical framework is not able to provide an analytical prediction. In such cases, one has to approximate the distribution directly from data. One non-parametric approach to achieve this is by using KDEs. In this investigation, KDEs are used to model the energy distributions of individual calorimeter tiles. This section provides a brief overview of their mathematical foundation.
\par
Consider a set of $n$ data points ($x_{1}$, $x_{2}$, ..., $x_{n}$) whose underlying PDF is unknown, for example repeated measurements of a certain physical variable, $x$. In order to estimate the underlying (unknown) distribution, one can estimate it in the following way:
\begin{equation}\label{eq: kde definition}
	f(x)=\frac{1}{nh}\sum\limits_{i\, =\, 1}^{n}K\left(\frac{x-x_{i}}{h}\right)\, .
\end{equation}
Here, the left-hand side represents the approximated distribution of the dataset. The right-hand side includes a parameter $h>0$ called the bandwidth, which controls the degree of smoothing of the final PDF, and a sum of kernels, $K$, running over all data points $x_{i}$. The kernel function can be any non-negative normalised density function that describes the contribution of a single data point adequately enough. For this investigation, a Gaussian normal distribution has been used:
\begin{equation}\label{eq: 1D Gaussian distribution}
	K(x)=\frac{1}{\sqrt{2\pi}}\exp\left(-\frac{1}{2}x^{2}\right)\, .
\end{equation}
Substituting Equation (\ref{eq: 1D Gaussian distribution}) into Equation (\ref{eq: kde definition}), the estimated PDF becomes a sum of Gaussian distributions centred around each data point and normalised to the prefactor $nh$.
\par
\begin{figure}[ht!]
	\centering
	\includegraphics[width = 1\textwidth]{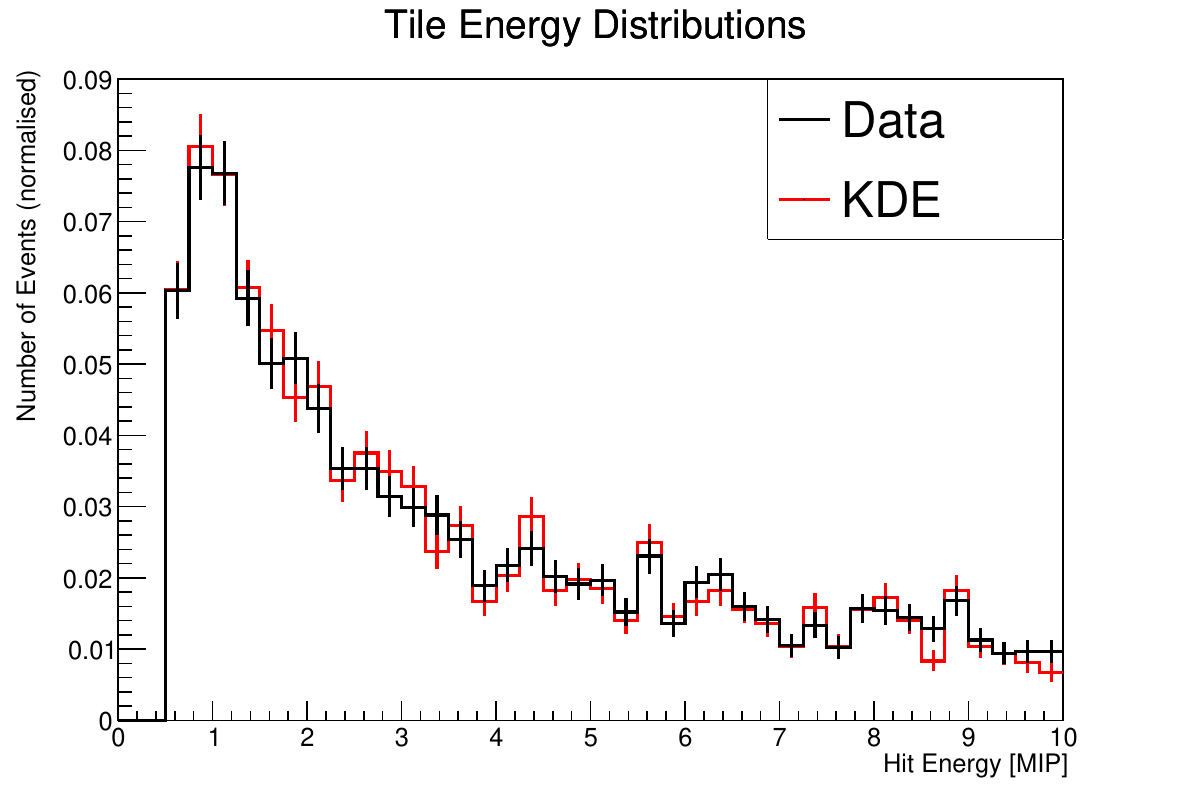}
	\caption{An example of a PDF estimation using Gaussian kernels. The hit energy distribution obtained from data is shown in black. It was used as input for Equation (\ref{eq: kde definition}) from which simulated events have been sampled. These are represented by the red distribution. For this example, the pion shower dataset includes $\SI{10000}{}$ events, and a bandwidth of $h=\SI{0.01}{\mip}$ was used for the estimation. In addition, zero-energy hits have been excluded for better visibility.}
	\label{fig: tile energy distributions}
\end{figure}

An example is displayed in Figure \ref{fig: tile energy distributions} which shows results of applying KDEs to real \ahcal data. In this example, the hit energy distributions of a specific \ahcal tile are shown, comparing data with simulated events obtained via Equation (\ref{eq: kde definition}). A dataset of $n=\SI{10000}{}$ events was used for the estimation, and all histograms are normalised to unity. For this particular case, a bandwidth of $\SI{0.01}{\mip}$ was chosen, and zero-energy hits have been excluded from the figure in order to avoid sharp peaks at zero that would otherwise obscure much of the structure of the remaining distributions. The sampling is done by randomly selecting a single event from the test beam dataset, and using its corresponding Gaussian kernel to generate a single simulated event. The event generation, in turn, from the Gaussian kernel is done via a standard Gaussian random number generator, as implemented in Python.
\par
For the application to test beam data, one has to generalise Equation (\ref{eq: kde definition}) to $d$ dimensions. This is necessary because a pion shower, recorded by the \ahcal, does not represent the measurement of a single value but instead the measurement of multiple correlated hit energies. Therefore, contrary to $n$ single values, one is now dealing with a set of $n$ $d$-dimensional data vectors ($\mathbf{x}_{1}$, $\mathbf{x}_{2}$, ..., $\mathbf{x}_{n}$). Each of these $\mathbf{x}_{i}$ now corresponds to one pion shower, and the entries of the vectors are the individual hit energies. The corresponding KDE is then defined as
\begin{equation}\label{eq: generalised kde}
	f(\mathbf{x})=\frac{1}{n}\sum\limits_{i\, =\, 1}^{n}|\mathbf{H}|^{-1/2}K\left(\mathbf{H}^{-1/2}(\mathbf{x}-\mathbf{x}_{i})\right)\, ,
\end{equation}
where $\mathbf{H}$ is the symmetric, positive definite $d\times d$ bandwidth matrix and $|\mathbf{H}|$ is its corresponding determinant. Furthermore, $\mathbf{x}$ is a vector of the hit energies of the calorimeter tiles for a simulated pion shower. In this way, a PDF for the hit energies can be constructed, with the probability of obtaining a specific set of hit energies within $[\mathbf{x},\ \mathbf{x}+\text{d}\mathbf{x}]$ being given by $f(\mathbf{x})\text{d}\mathbf{x}$. For this investigation, $n$ was set to $\SI{10000}{}$, meaning that a real pion shower dataset of $\SI{10000}{}$ events was used as basis for the estimation, and a multivariate Gaussian distribution has been used as kernel function in this generalised case:
\begin{equation}\label{eq: multivariate kernel}
	K(\mathbf{x})=\frac{1}{(2\pi)^{d/2}}\exp\left(-\frac{1}{2}\mathbf{x}^{\text{T}}\mathbf{H}^{-1}\mathbf{x}\right)\, .
\end{equation}
Here, $\mathbf{x}^{\text{T}}$ is the transposed vector of $\mathbf{x}$. In principle, $\mathbf{H}$ can be chosen to be any symmetric, positive definite matrix. However, the amount of parameters that needs to be chosen grows quadratically in $d$. That is why it is often convenient to choose a more simplified form for $\mathbf{H}$. For this study, the bandwidth matrix
\begin{equation}\label{eq: chosen bandwidth matrix}
	\mathbf{H}=h^{2}\mathbf{I}_{d\times d}
\end{equation}
has been used, where $h$ is the chosen positive bandwidth and $\mathbf{I}_{d\times d}$ is the unit matrix in $d$ dimensions. This choice is not only convenient but also natural, since Equation (\ref{eq: chosen bandwidth matrix}) ensures that the energy distributions of all calorimeter tiles are uniformly smoothed. Substituting Equations (\ref{eq: multivariate kernel}) and (\ref{eq: chosen bandwidth matrix}) into Equation (\ref{eq: generalised kde}) then yields the KDE definition that has been applied to the \ahcal test beam dataset:
\begin{equation}\label{eq: kde definition with covariance bandwidth}
	f(\mathbf{x})=\frac{1}{nh^{d}\left(2\pi\right)^{d/2}}\sum\limits_{i\, =\, 1}^{n}\exp\left(-\frac{1}{2h^{2}}(\mathbf{x}-\mathbf{x}_{i})^{\text{T}}(\mathbf{x}-\mathbf{x}_{i})\right)\, .
\end{equation}
Here, too, the bandwidth value is an important parameter because it has a significant impact on the smoothness of the estimated PDF. A value of $h=\SI{0.01}{\mip}$ as the bandwidth was shown to provide well-behaved PDFs that captured the structure of the underlying distributions, without being overly sensitive to statistical fluctuations in the sample. If $h$ were to be too small (e.g. $h=\SI{0.001}{\mip}$), the influence of each data point on the final PDF would be too large, causing the PDFs to peak at individual data points. On the other hand, if $h$ is too large (for example, $h=\SI{1}{\mip}$), then much of the underlying structure of the PDF will be lost.

%% file: chapters/hitlevel.tex
\section{Simulation of Hit Energy Distributions using Kernel Density Estimators}
\label{sec: simulation of hit energy distributions using kernel density estimators}

This section presents the simulation of hit energies and shower shape variables using KDEs, which are employed to estimate PDFs obtained from the dataset as sums of Gaussian distributions. The freely selectable bandwidth $h$ was optimised to the value $\SI{0.01}{\mip}$, giving each data point an adequate contribution to the overall PDF without over-modelling the data.
\par
A detector coordinate system was chosen for this investigation in which the $x$- and $y$-axis are defined as the bottom and left edge of the \ahcal, respectively, such that the origin lies in the bottom left corner. In addition, the $z$-axis runs perpendicularly through the bottom left corner of each detector plane (i.e. through the origin). In this way, the absolute (that is, physical) position of a hit is always given by three natural numbers in units of tiles for the $x$- and $y$- and in units of layers for the $z$-axis. However, in order to decouple the simulation from any fixed shower start position or global shower axis, hit positions are additionally described relative to the centre of gravity (CoG) in the $xy$-plane and the shower start layer (and thus, in integer multiples of tiles and layers instead of natural numbers). Details on how the shower start layer is determined are given in Refs. \cite{daniel_heuchel, olin_pinto}. This choice allows one to independently place a randomly generated shower start and CoG$_{xy}$ within the calorimeter and it ensures that the simulation of the overall kinematic behaviour of the shower remains unaffected by its position within the \ahcal. Only when the shower is placed close to the boundaries of the detector volume, leakage becomes relevant.
\par
Section \ref{subsec: cuts on the centre of gravity} introduces an event selection and the coordinate system that describes hit positions relative to the shower start layer as well as to the lateral CoG. These coordinates have been used consistently throughout the course of this investigation. Furthermore, for practical reasons, cuts have been applied to the lateral CoG, which will also be explained in the following section. Afterwards, results of simulating individual hit energy distributions are shown in Section \ref{subsec: simulated distributions of kinematic shower variables}, whereas Section \ref{subsec: simulated correlations between kinematic shower variables} compares simulated and measured correlation factors between kinematic shower variables. An evaluation of the computational performance of the fast simulation algorithm is given in Section \ref{subsec: computational requirements KDEs}.

\subsection{Selection of Data Events}
\label{subsec: cuts on the centre of gravity}

All events had to pass a particle identification, followed by the removal from the first physical \ahcal layer from the whole dataset. This removal minimises uncertainties of the shower start finding algorithm. The third step involved a cut on the shower start of all events, which was confined to the first ten physical layers. Every event where the shower started later was removed in order to avoid leakage effects. Finally, a low-energy threshold was applied to the remaining events in order to exclude those where the pion decays into a muon and a neutrino, since such events usually do not pass the event selection criteria of real data analyses anyway.
\par
\begin{figure}[ht]
    \hspace*{-10mm}
    \centering
    \includegraphics[width = 1.25\linewidth]{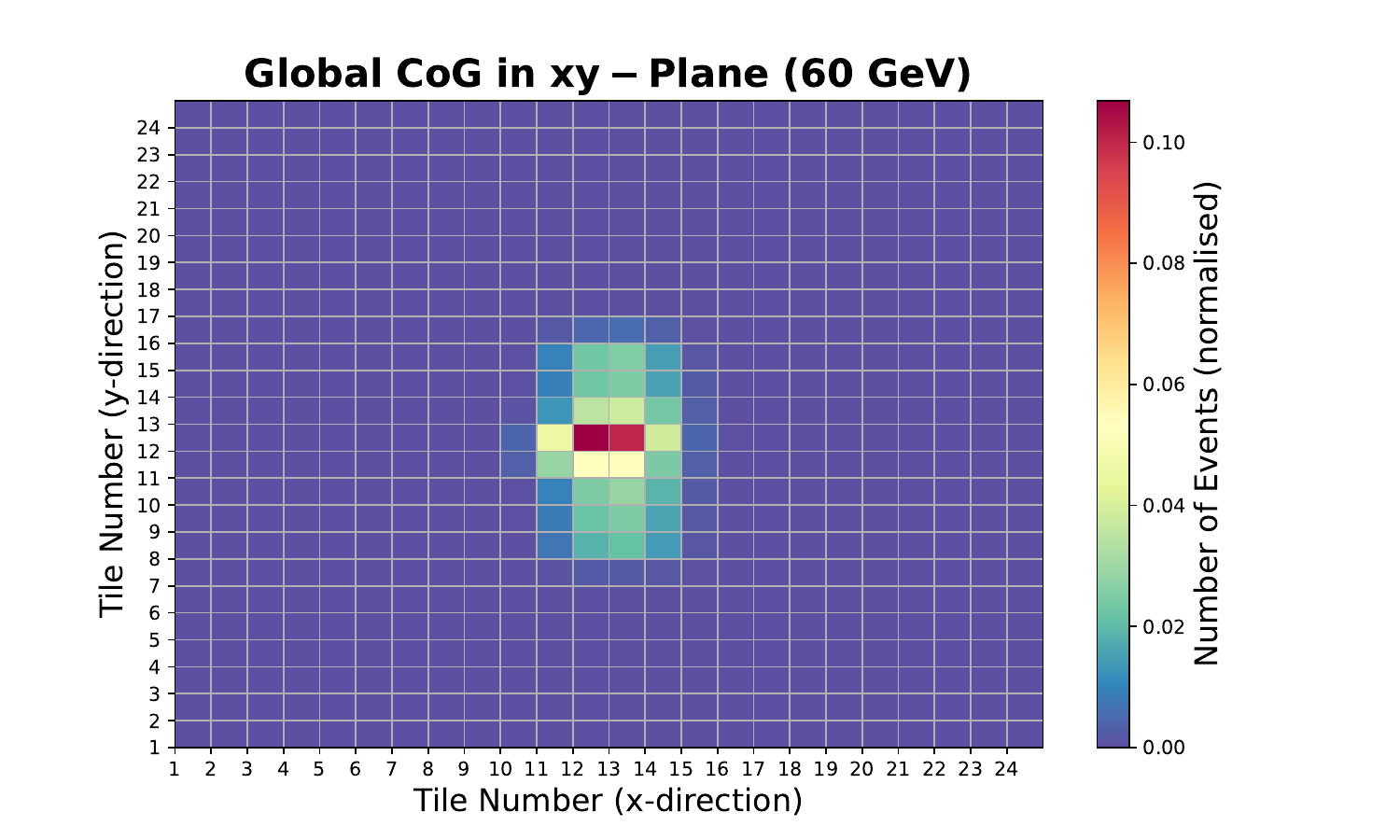}
    \caption{Distribution of global CoGs in the $xy$-plane in $2018$ \ahcal test beam data for $\SI{60}{\giga\electronvolt}$ pions. The $x$- and $y$-axis represent the position of the CoG and the colourbar depicts the normalised number of events. The majority of CoGs falls into a small region around the detector centre.}
    \label{fig: CoGs in the xy plane}
\end{figure}

For a fast simulation on single-tile level, it is more practical to describe the position of each hit in relative rather than absolute coordinates -- that is, with respect to the shower start layer and the CoG in the $xy$-plane. The \ahcal, however, already comprises a total of $24\times24\times38=\SI{21888}{}$ readout channels, and describing hit position in relative coordinates only increases this number even further, since the CoG can vary from shower to shower. In order to limit this complexity, it is therefore reasonable to constrain the lateral CoG to a narrow range of tiles, which is supported by Figure \ref{fig: CoGs in the xy plane} which is based on $2018$ test beam data for $\SI{60}{\giga\electronvolt}$ pions. Here, the $x$- and $y$-axis depict the absolute tile coordinates of the \ahcal, and the colour bar represents the normalised event count. The majority of events is concentrated close to the detector centre, while minor asymmetries can be attributed to the fact that the beam line was not aligned exactly between the four central tiles, but instead slightly shifted towards the top left in Figure \ref{fig: CoGs in the xy plane}.
\par
Based on these observations, only events whose lateral CoGs fall within the range $x,\, y\in[9,\, 16]$ have been used for the following density estimation. This selection was applied consistently across all beam energies. Under this constraint, the maximum possible distance between a hit and a CoG of the same event is $15$ tiles. Since a hit may either occur on the extreme left- or extreme right-hand side of the detector plane, the number of allowed relative positions per axis doubles. Additionally, it may happen that a hit is detected in the same tile into which the CoG falls, introducing another possible configuration. Consequently, there are $31$ possible $x$- and  $y$-values that a hit position can assume. Combining this with the $38$ layers defined relative to the shower start layer, and one finds that there are a total of $31\times31\times38=\SI{36518}{}$ possible hit positions per event. Therefore, in the following, each of the data vectors (i.e., each event) used as input for the KDE is going to contain $\SI{36518}{}$ hit energies, which must be simulated event by event and also simultaneously in order to preserve correlations between individual tiles. A simulated event will thus also contain $\SI{36518}{}$ simulated hit energies; their relative positions in the new coordinate system will be labelled by $x_{\text{hit}}$, $y_{\text{hit}}$, and $z_{\text{hit}}$. The former two are given with respect to the global, lateral CoG and the latter with respect to the shower start layer. The results of simulating these hit energies are shown in the following.

\subsection{Simulated Distributions of Kinematic Shower Variables}
\label{subsec: simulated distributions of kinematic shower variables}

KDEs have been exploited in order to estimate hit energy distributions and to sample simulated events from them. For this investigation, $\SI{100000}{}$ simulated events, each with $\SI{36518}{}$ simultaneously generated hit energies, have been compared with the whole dataset, which includes events at the same order of magnitude, as well as with simulated events obtained from full simulation. In order to validate the performance and accuracy of the fast simulation algorithm, various kinematic shower variables have been computed based on the simulated hit energies, and their distributions have been compared among the three datasets. These shower variables include: the number of hits per event, hit energy distributions, and the total energy,
\begin{equation}
    E_{\text{tot}}=\sum\limits_{\text{hits}}E_{\text{hit}}\, .
\end{equation}
Based upon the total energy, the CoG along the $z$-axis,
\begin{equation}
    \text{CoG}_{z}=\frac{1}{E_{\text{tot}}}\sum\limits_{\text{hits}}E_{\text{hit}}z_{\text{hit}}\, ,
\end{equation}
and the mean shower radius,
\begin{equation}
    r_{\text{mean}}=\frac{1}{E_{\text{tot}}}\sum\limits_{\text{hits}}E_{\text{hit}}r_{\text{hit}}
\end{equation}
with
\begin{equation}\label{eq: hit radius definition}
    r_{\text{hit}}=\sqrt{(x_{\text{hit}}-\text{CoG}_{x})^{2}+(y_{\text{hit}}-\text{CoG}_{y})^{2}}\, ,
\end{equation}
have also been computed. Furthermore, two energy fractions have been adapted from Ref. \cite{olin_pinto}, which has already demonstrated their validity as effective selection variables, namely the energy fraction deposited within the first $22$ layers from the shower start layer,
\begin{equation}
    f_{22}=\frac{1}{E_{\text{tot}}}\sum\limits_{\text{hits}}E_{\text{hit}}\text{ if }z_{\text{hit}}<22\, ,
\end{equation}
and the energy fraction deposited within a radius of $\SI{30}{\milli\meter}$ from the shower axis,
\begin{equation}
    f_{\text{central}}=\frac{1}{E_{\text{tot}}}\sum\limits_{\text{hits}}E_{\text{hit}}\text{ if }r_{\text{hit}}\leq\SI{30}{\milli\meter}\, .
\end{equation}
Here, $r_{\text{hit}}$ is the same as defined by Equation (\ref{eq: hit radius definition}) and the shower axis is defined as the straight line that passes perpendicularly through all detector layers with the intersection points located at the position of the global, lateral CoG. Lastly, three different shower moments have been investigated. These shower moments were calculated for all three spatial dimensions and include the variance,
\begin{equation}
    \text{Var}_{i}=\frac{1}{E_{\text{tot}}}\sum\limits_{\text{hits}}E_{\text{hit}}(i_{\text{hit}}-\text{CoG}_{i})^{2}\, ,
\end{equation}
the skewness,
\begin{equation}
    \text{Skew}_{i}=\frac{1}{E_{\text{tot}}}\sum\limits_{\text{hits}}E_{\text{hit}}\left(\frac{i_{\text{hit}}-\text{CoG}_{i}}{\sigma_{i}}\right)^{3}\, ,
\end{equation}
and the kurtosis,
\begin{equation}
    \text{Kurt}_{i}=\frac{1}{E_{\text{tot}}}\sum\limits_{\text{hits}}E_{\text{hit}}\left(\frac{i_{\text{hit}}-\text{CoG}_{i}}{\sigma_{i}}\right)^{4}\, ,
\end{equation}
where $\sigma_{i}=\sqrt{\text{Var}_{i}}$ is the respective standard deviation and $i$ is either $x$, $y$, or $z$. Table \ref{tab: summary kinematic variables} summarises all kinematic variables and gives short descriptions of their physical interpretations.
\begin{table}[ht]
    \centering
    \caption{A summary of all kinematic variables that have been computed for this investigation as well as short descriptions of their physical interpretations. The variance, the skewness, and the kurtosis have been computed for all three spatial dimensions.}
    \begin{tabular}{|c|c|}
        \hline
        Variable & Interpretation\\
        \hline
        \hline
        Number of Hits & Number of all non-zero hits per event.\\
        \hline
        Hit Energy & Distribution of all non-zero hit energies.\\
        \hline
        Total Energy & Sum of all non-zero hit energies per event.\\
        \hline
        Longitudinal CoG & \makecell{Energy-weighted, event-wise sum of all longitudinal\\distances of all hits to the shower start layer.}\\
        \hline
        Mean Shower Radius & \makecell{Energy-weighted, event-wise sum of radial\\distances of all hits to the global CoG$_{xy}$.}\\
        \hline
        Fraction-$22$ & \makecell{Fraction of energy per event deposited within\\the first $22$ layers from the shower start layer.}\\
        \hline
        Central Fraction & \makecell{Fraction of energy per event deposited within \\a radius of $\SI{30}{\milli\meter}$ from the shower axis.}\\
        \hline
        Variance & Mean squared distance of hits to the CoG.\\
        \hline
        Skewness & Measure of the asymmetry of a shower.\\
        \hline
        Kurtosis & \makecell{Measure of how sharply the energy\\distribution of the shower peaks.}\\
        \hline
    \end{tabular}
    \label{tab: summary kinematic variables}
\end{table}
\par
\begin{figure}[hp]
    \centering
    \subfigure[]{\includegraphics[width = 0.49\textwidth]{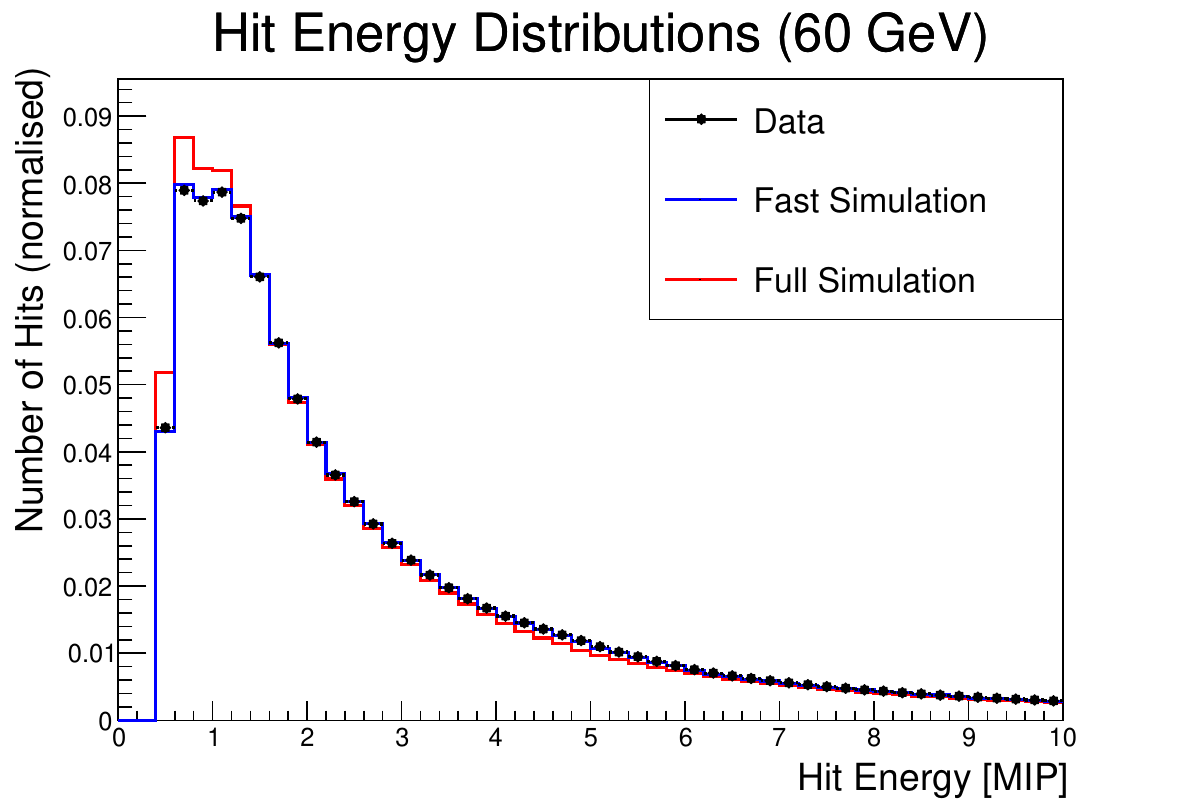}}
    \subfigure[]{\includegraphics[width = 0.49\textwidth]{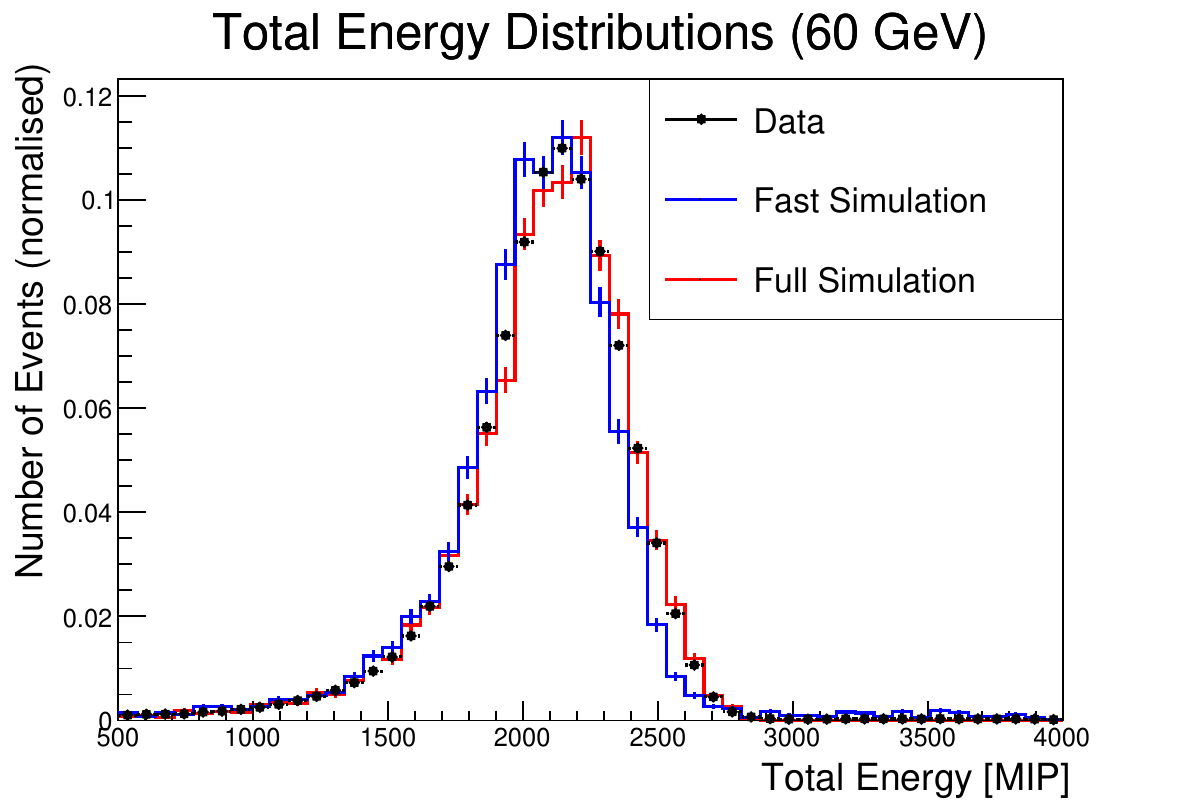}}
    \subfigure[]{\includegraphics[width = 0.49\textwidth]{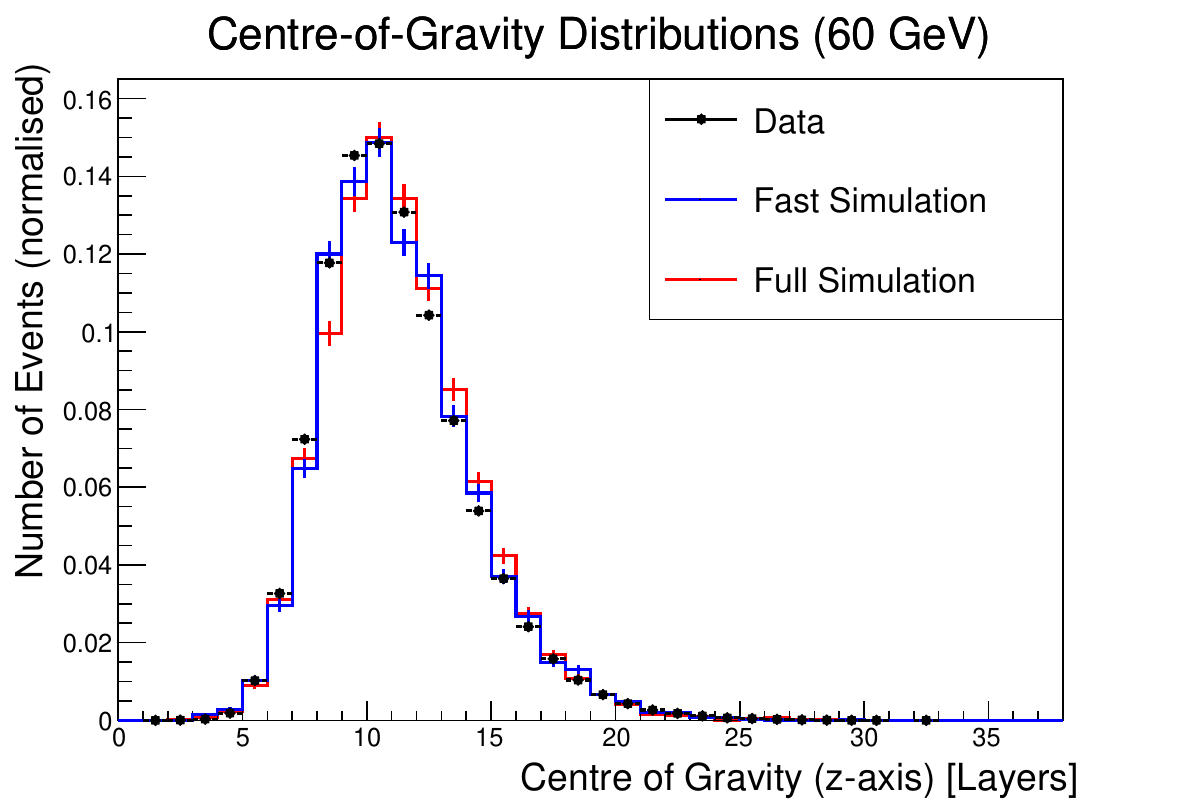}}
    \subfigure[]{\includegraphics[width = 0.49\textwidth]{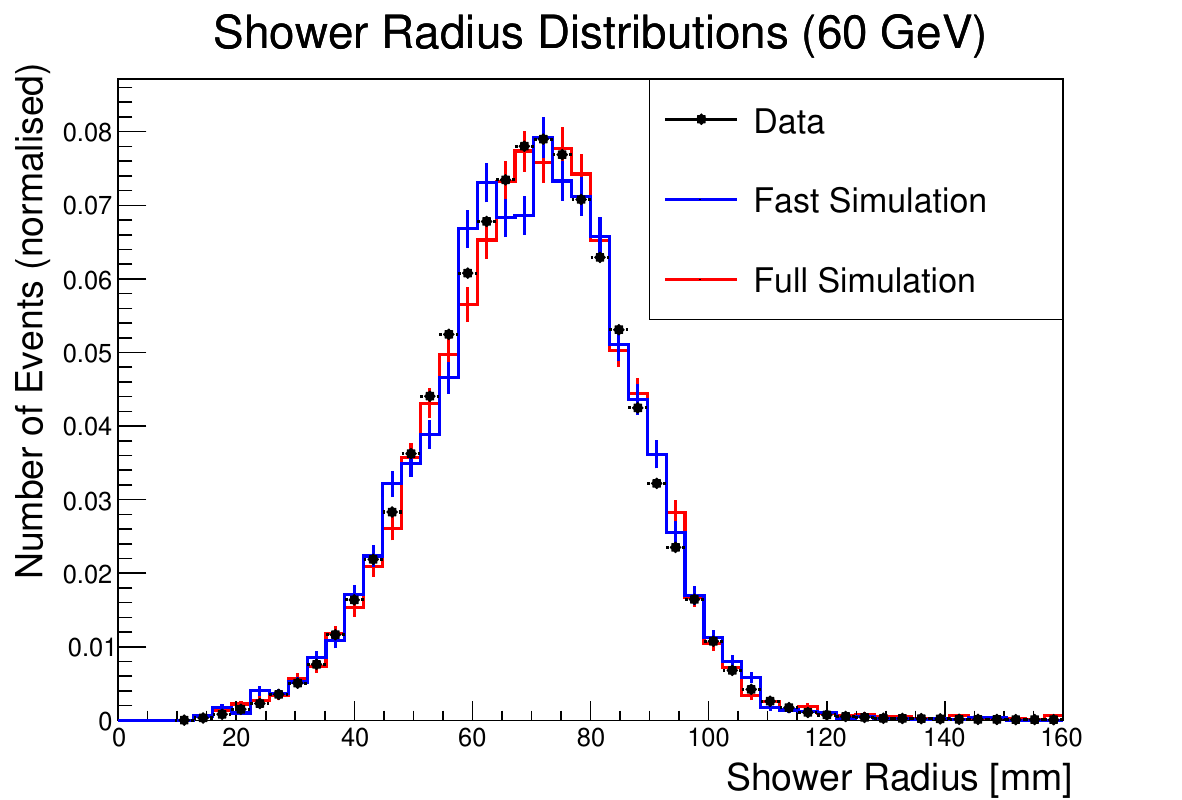}}
    \subfigure[]{\includegraphics[width = 0.49\textwidth]{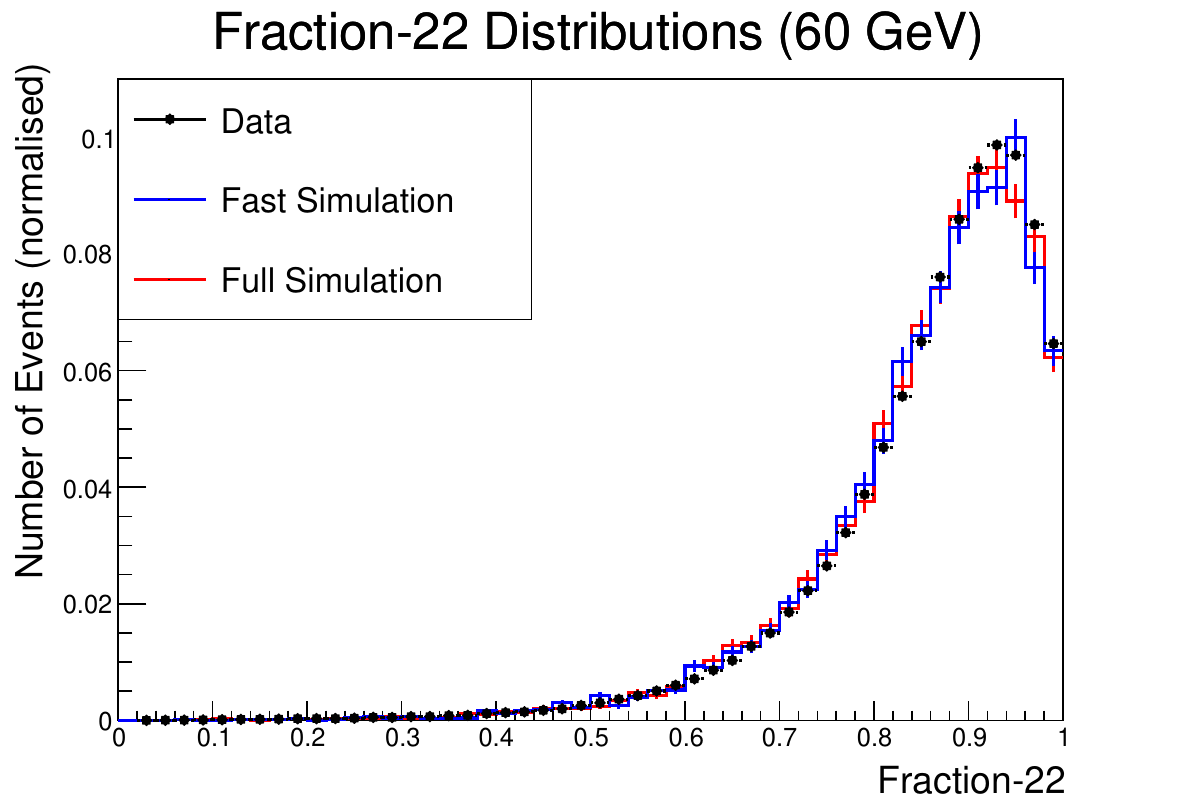}}
    \subfigure[]{\includegraphics[width = 0.49\textwidth]{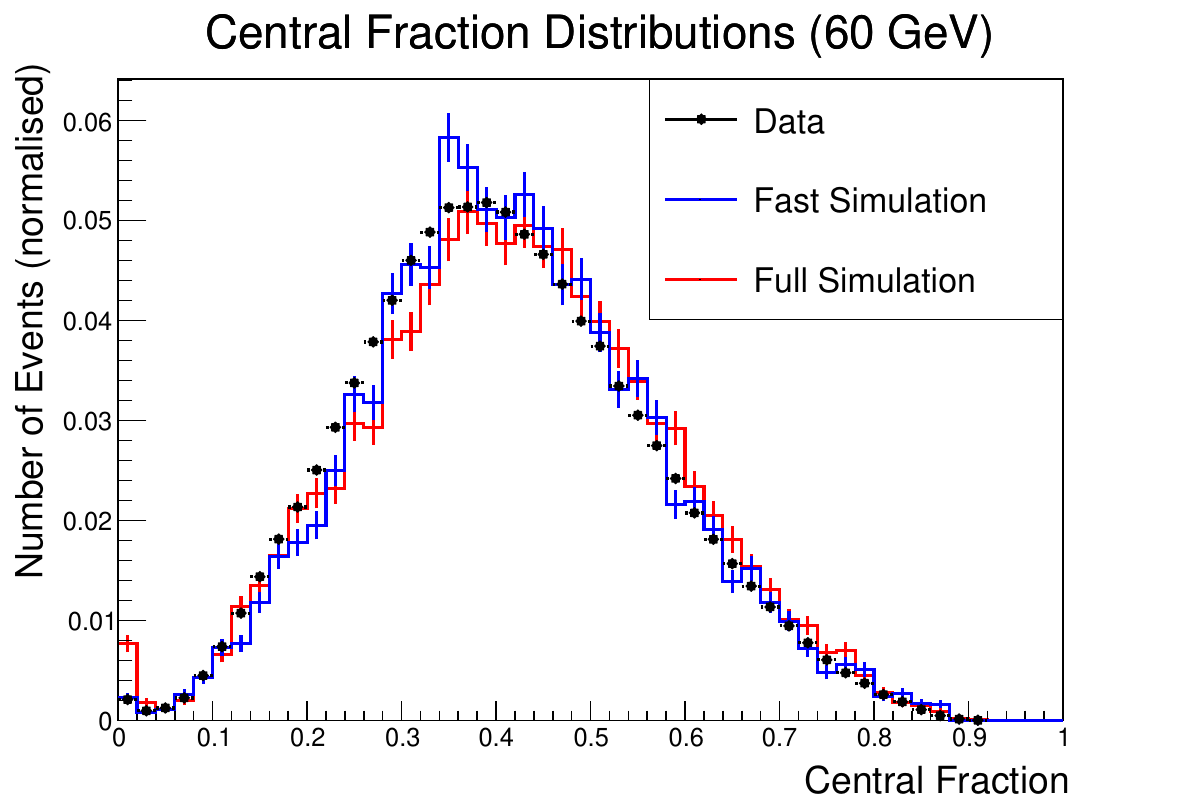}}
    \caption{Distributions of different kinematic shower variables for $\SI{60}{\giga\electronvolt}$ pions. Histograms are shown for (a) the hit energies, (b) the total energy, (c) the CoG along the $z$-axis, (d) the mean shower radius, (e) the energy fraction within the first $22$ layers, and (f) the energy fraction within a cylinder of radius $\SI{30}{\milli\meter}$. Black points represent the complete dataset, dark blue curves the fast simulation, and red curves depict full simulation. The fast simulation approach agrees very well with data and even outperforms the full simulation in parts.}
    \label{fig: kinematic shower variables for simulation with KDEs 60 GeV}
\end{figure}
\begin{figure}[hp]
    \centering
    \subfigure[]{\includegraphics[width = 0.49\textwidth]{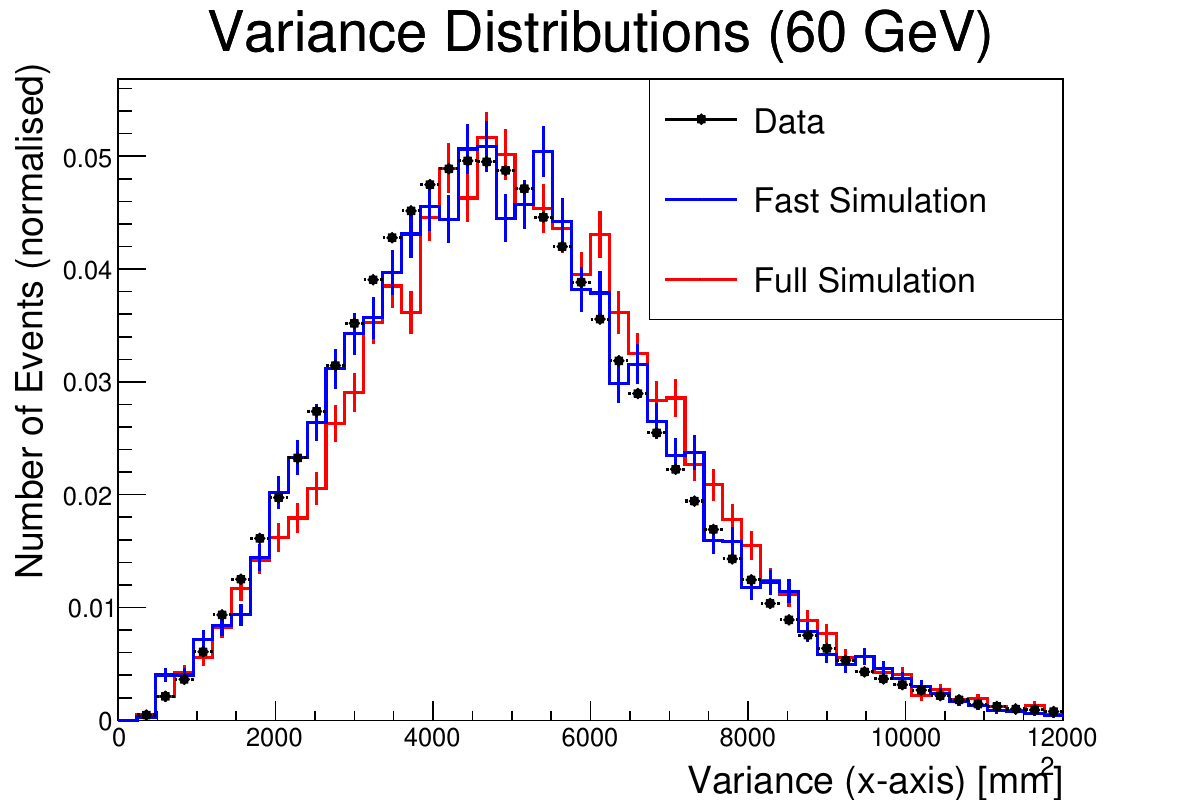}}
    \subfigure[]{\includegraphics[width = 0.49\textwidth]{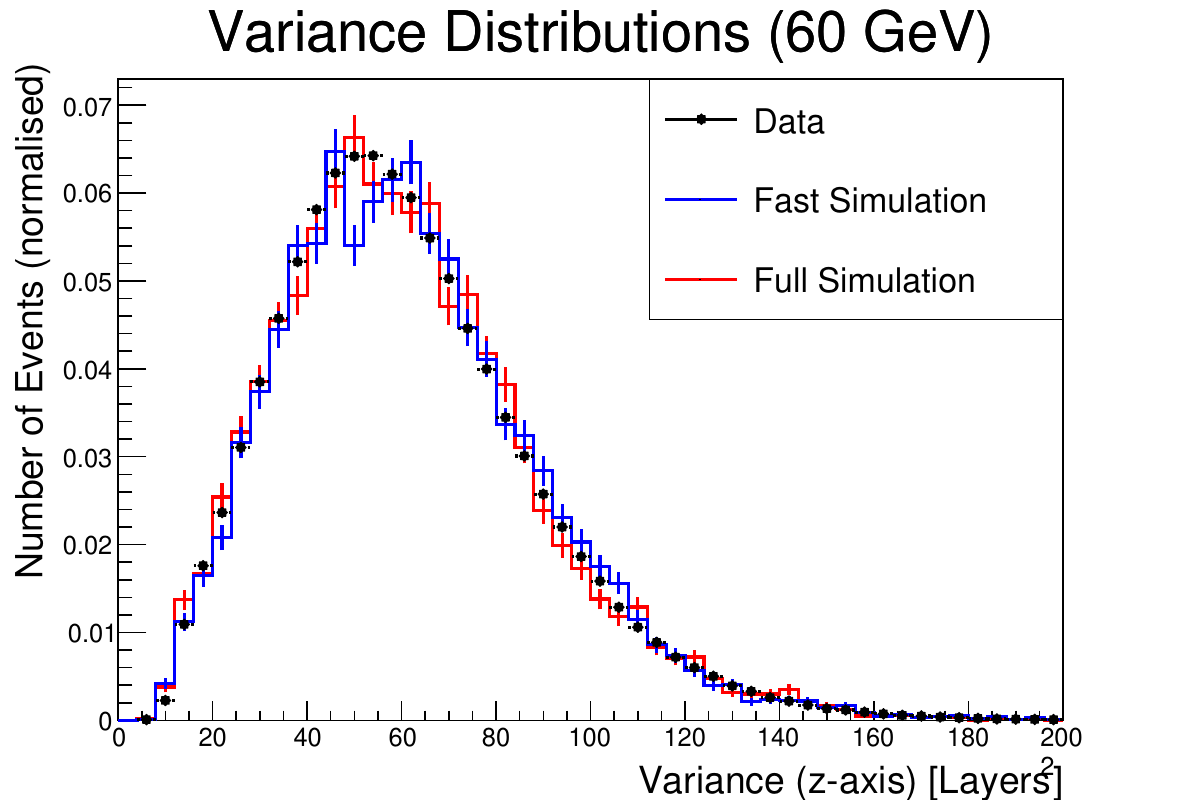}}
    \subfigure[]{\includegraphics[width = 0.49\textwidth]{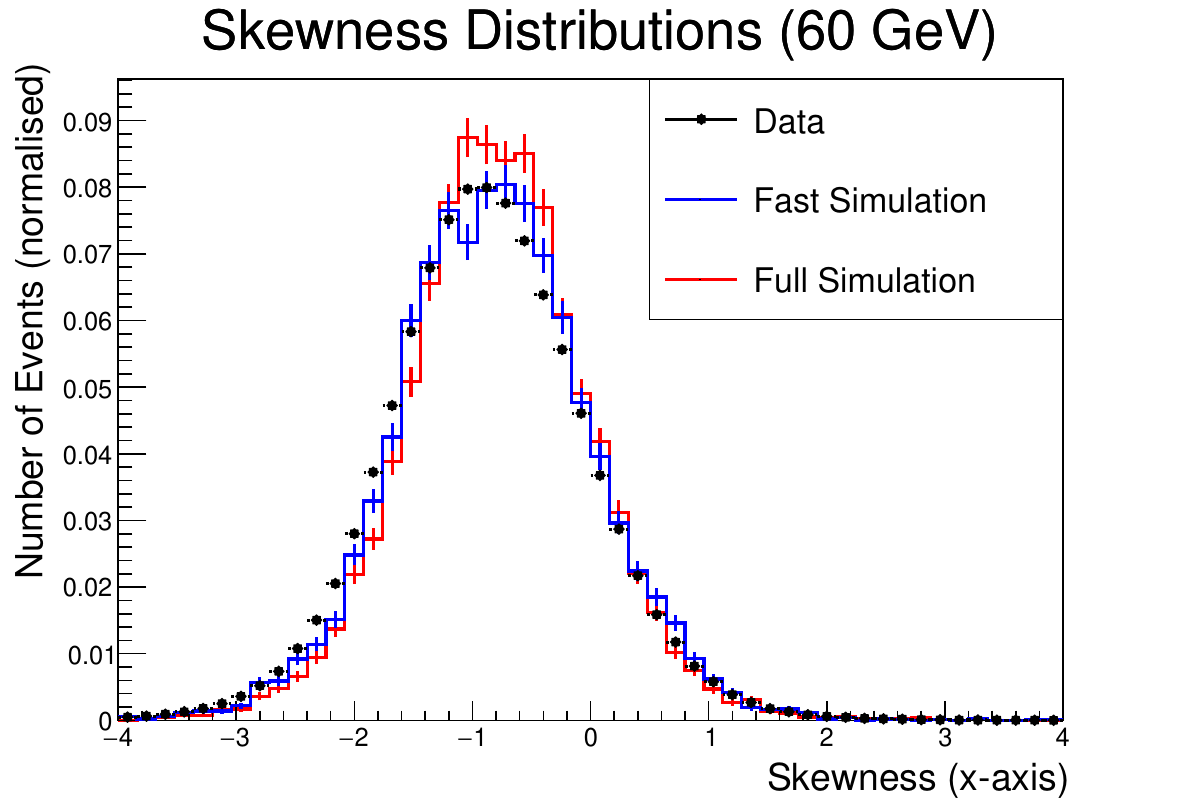}}
    \subfigure[]{\includegraphics[width = 0.49\textwidth]{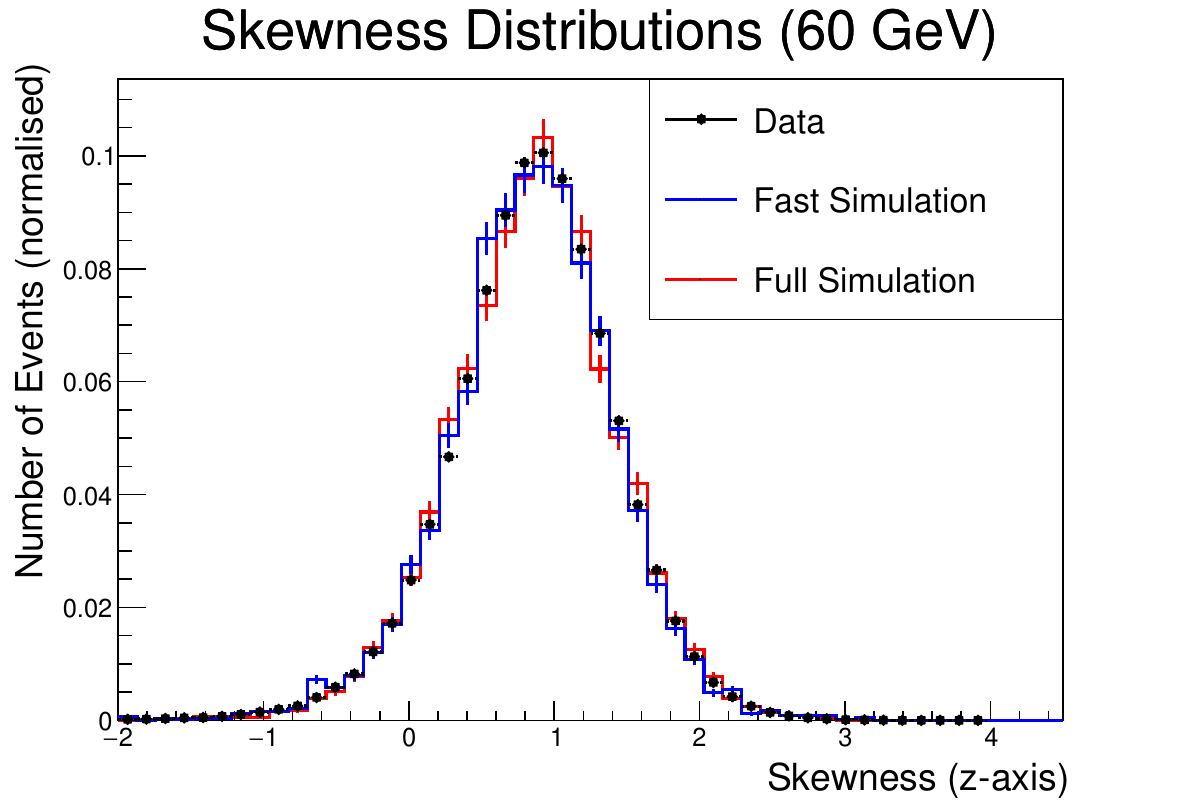}}
    \subfigure[]{\includegraphics[width = 0.49\textwidth]{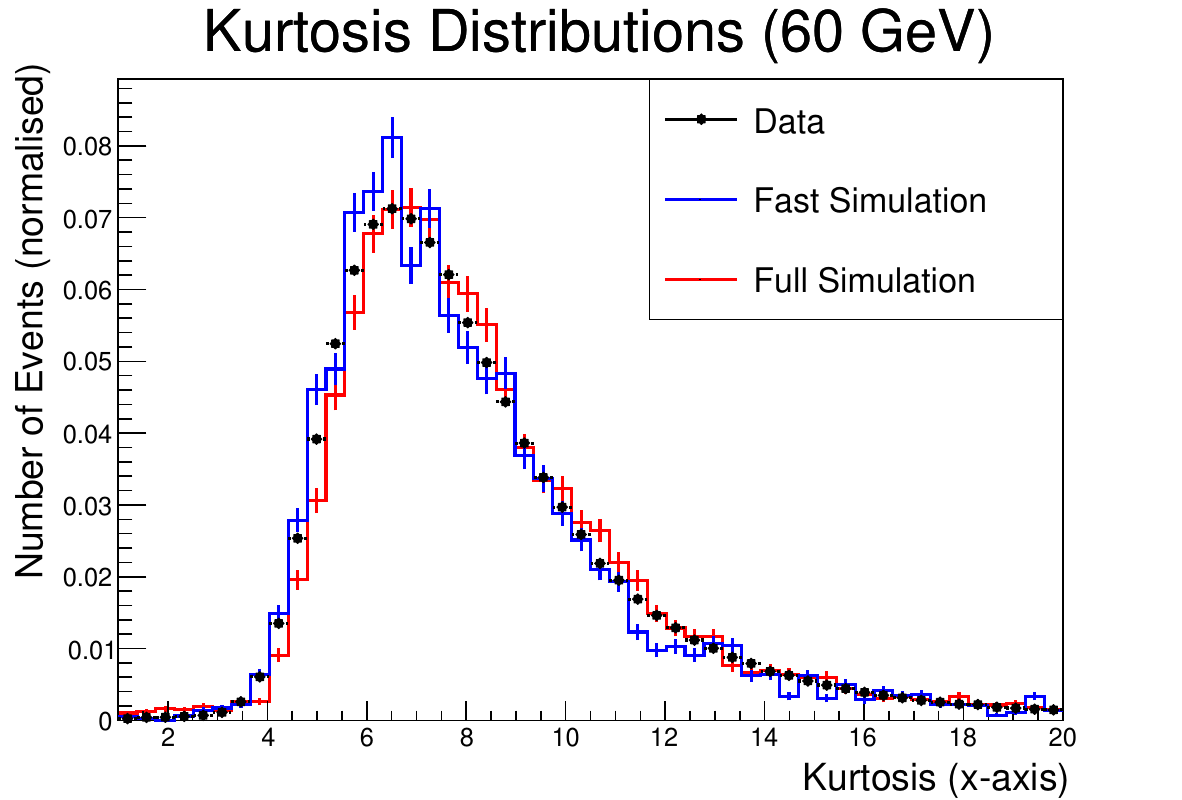}}
    \subfigure[]{\includegraphics[width = 0.49\textwidth]{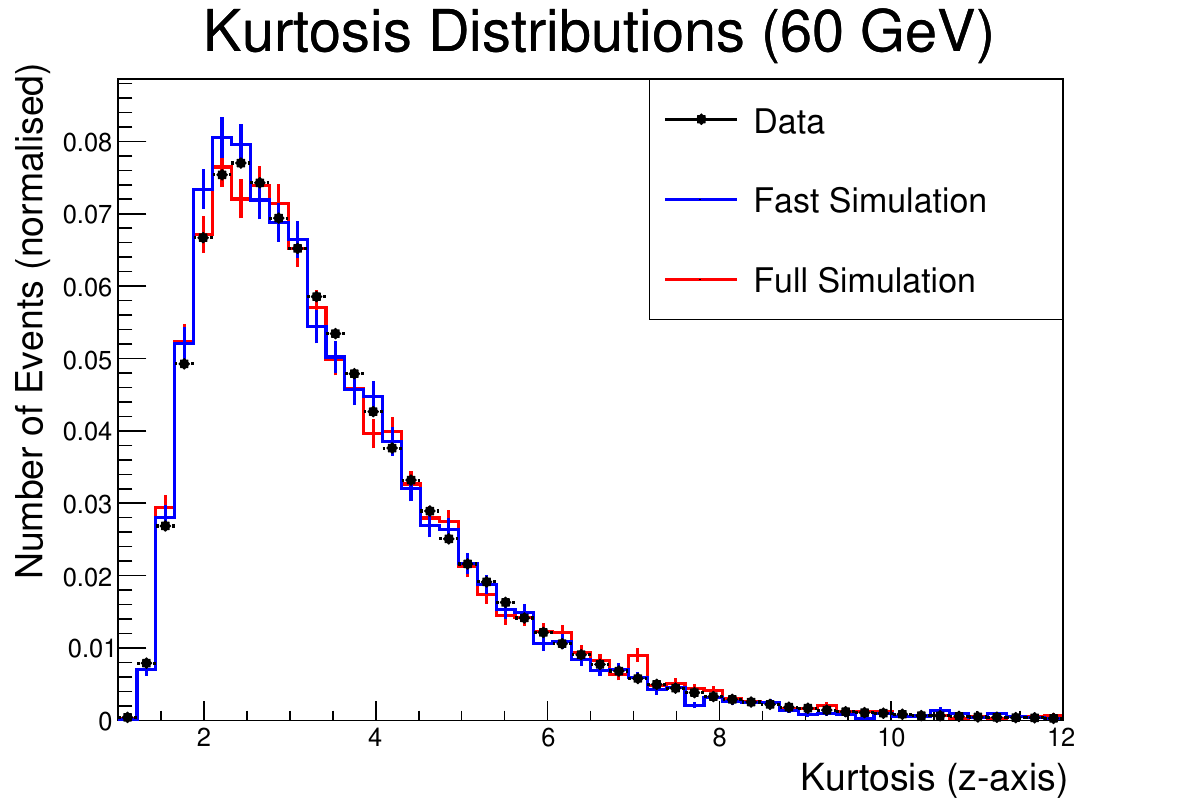}}
    \caption{Distributions of different shower moments for $\SI{60}{\giga\electronvolt}$ pions. The upper row shows histograms for the shower variance, the middle row for the skewness, and the lower row for the kurtosis. Furthermore, the left column depicts all shower moments along the $x$-axis and the right one along the $z$-axis. Black points represent the complete dataset, dark blue curves the fast simulation, and red curves depict full simulation. The fast simulation approach agrees very well with data and even outperforms the full simulation in parts.}
    \label{fig: shower moments for simulation with KDEs 60 GeV}
\end{figure}
\begin{figure}[hp]
    \centering
    \subfigure[]{\includegraphics[width = 0.49\textwidth]{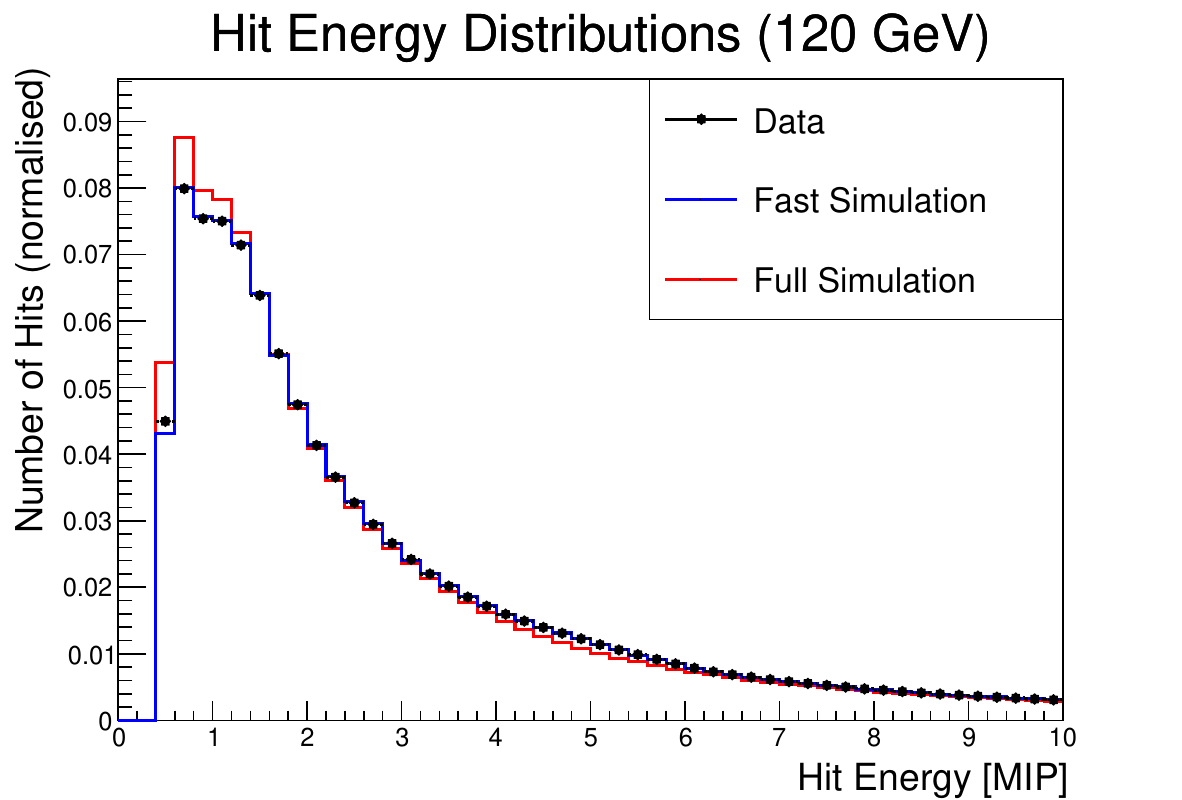}}
    \subfigure[]{\includegraphics[width = 0.49\textwidth]{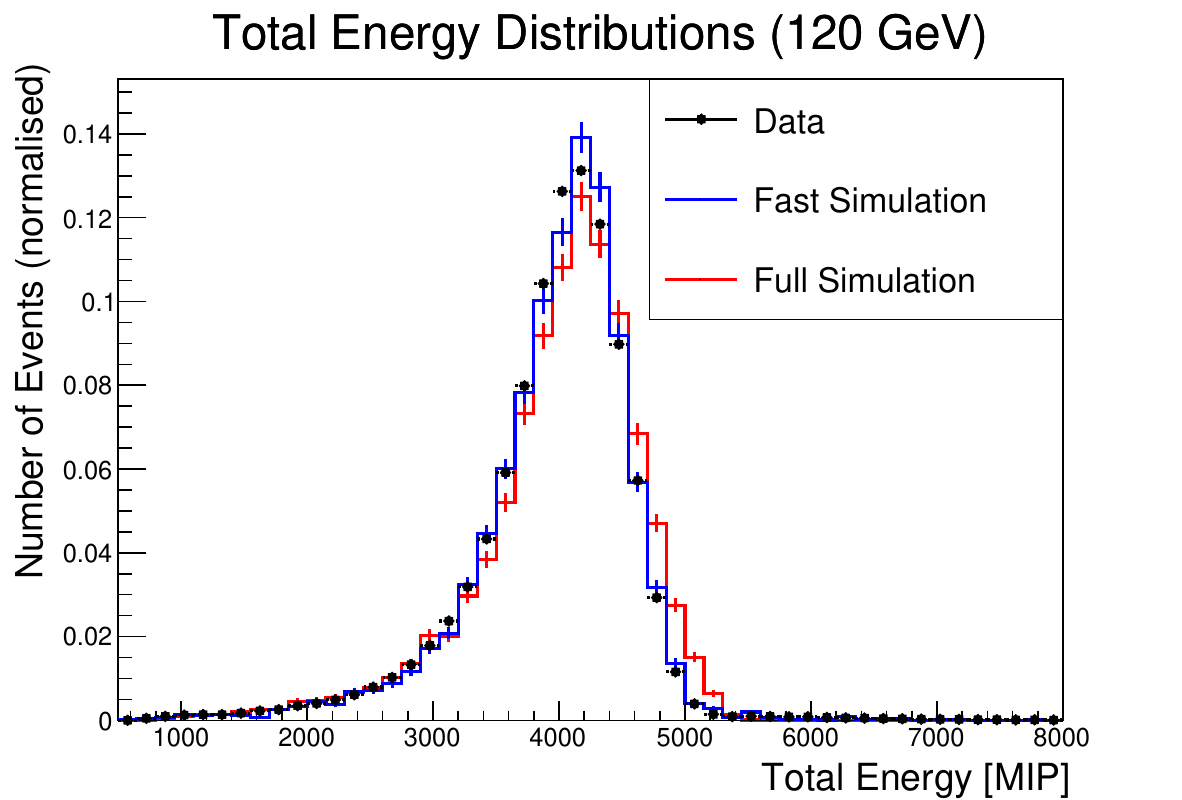}}
    \subfigure[]{\includegraphics[width = 0.49\textwidth]{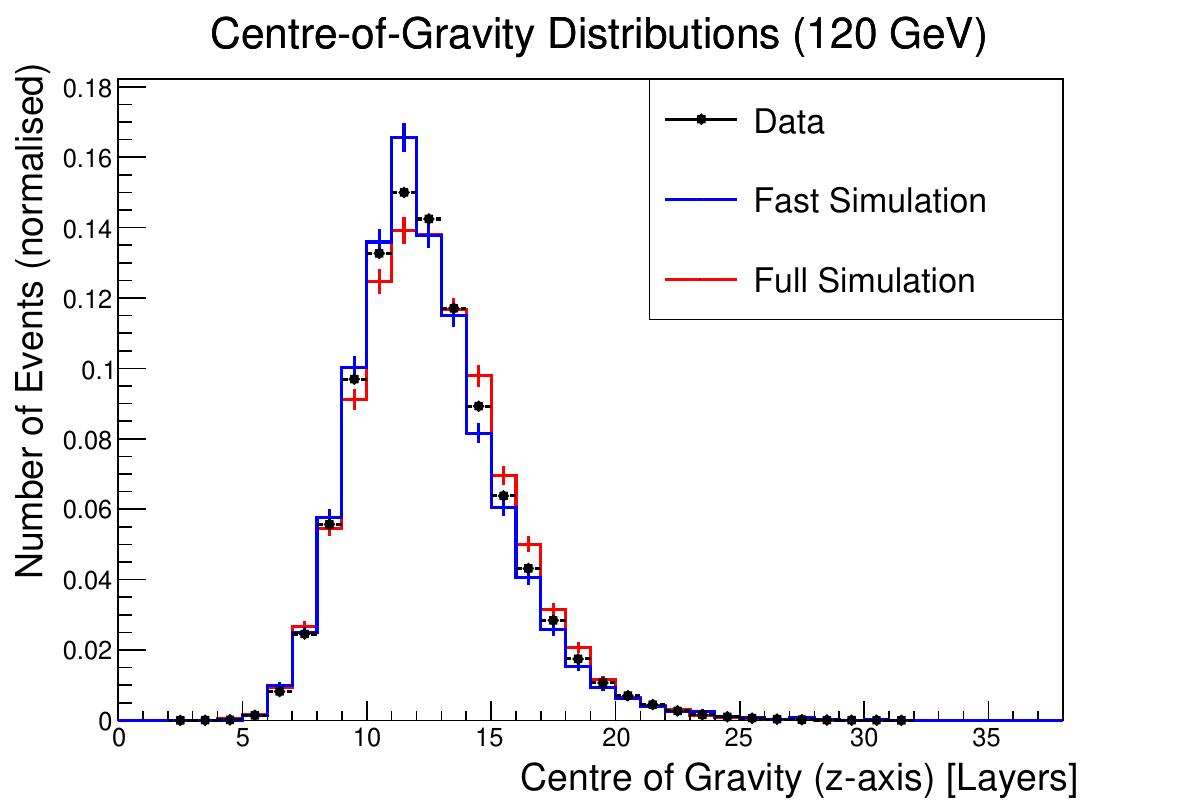}}
    \subfigure[]{\includegraphics[width = 0.49\textwidth]{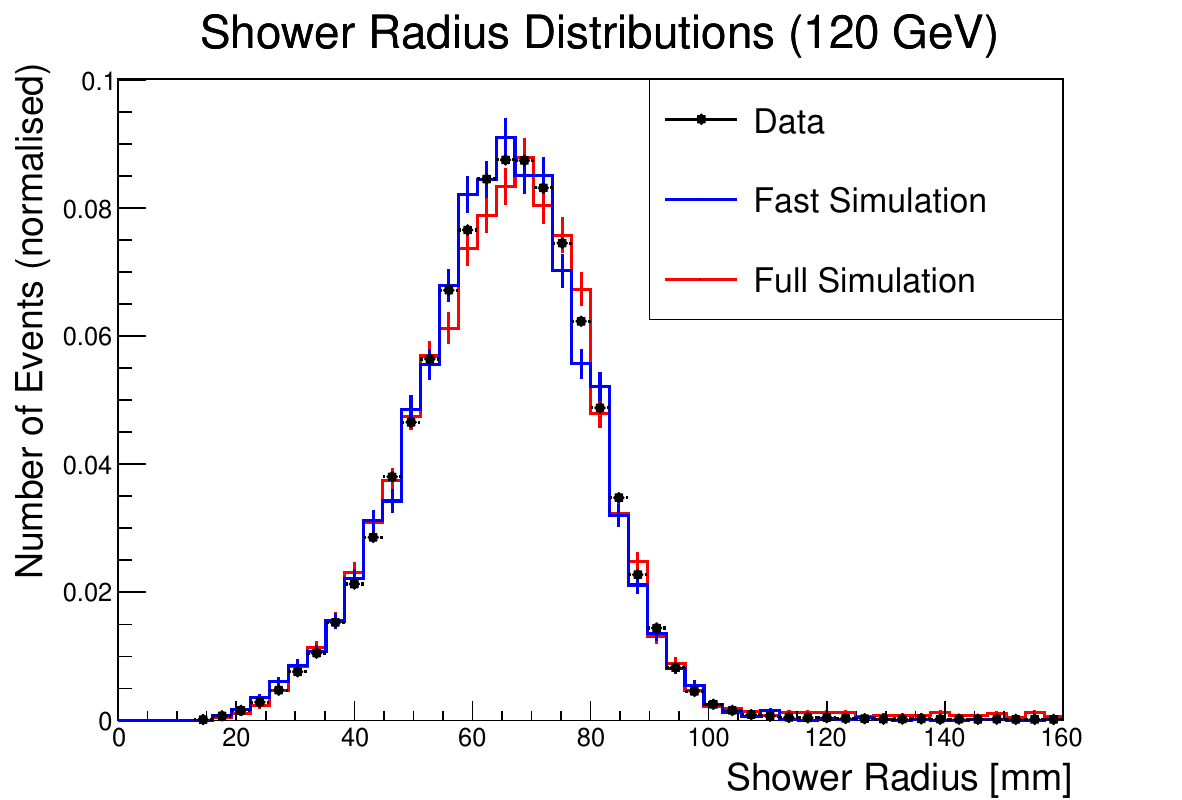}}
    \subfigure[]{\includegraphics[width = 0.49\textwidth]{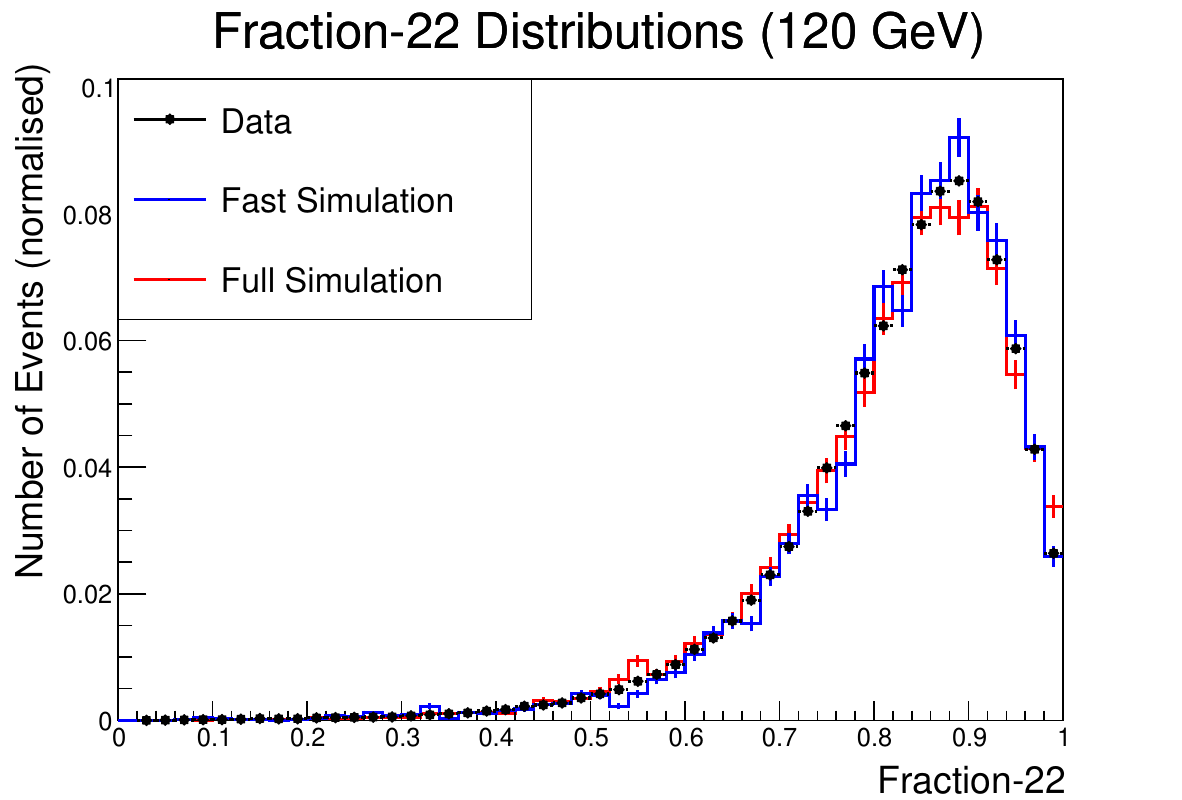}}
    \subfigure[]{\includegraphics[width = 0.49\textwidth]{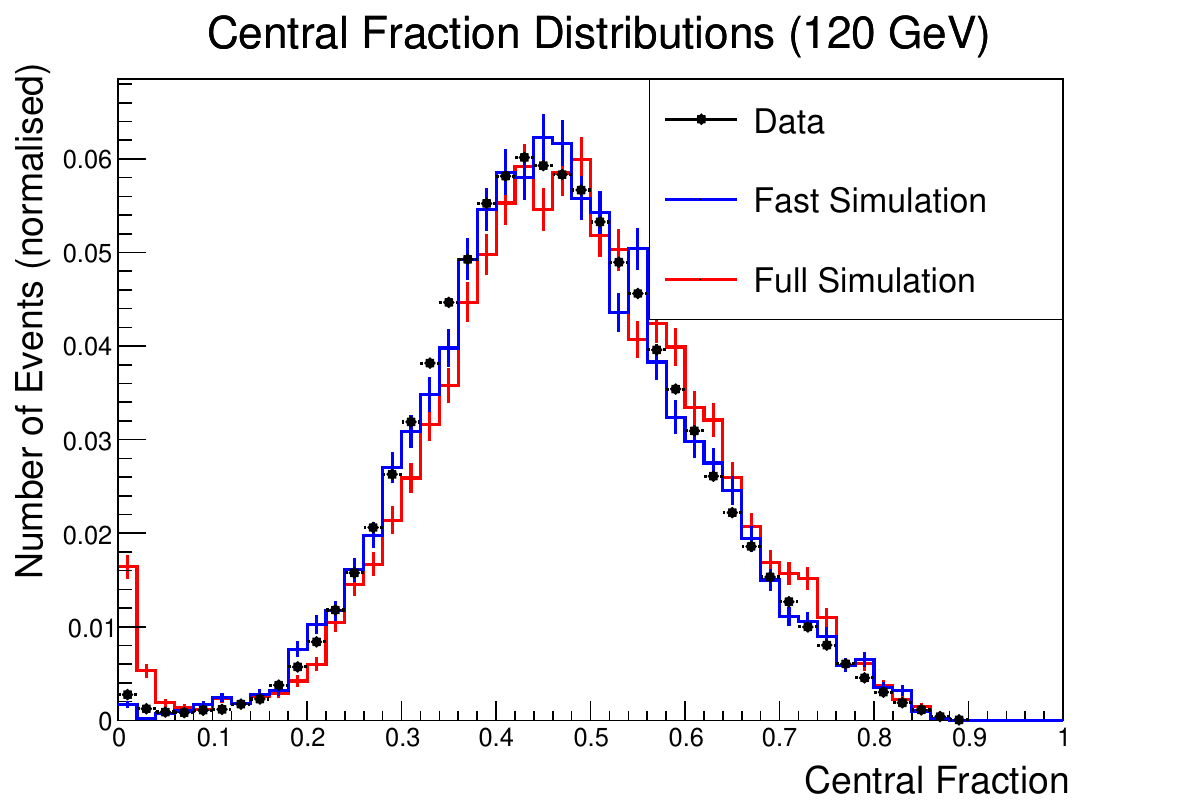}}
    \caption{Distributions of the same kinematic shower variables that have already been shown in Figure \ref{fig: kinematic shower variables for simulation with KDEs 60 GeV}, but for $\SI{120}{\giga\electronvolt}$ pions. The PDFs exhibit the same behaviour that has already been described in Figure \ref{fig: kinematic shower variables for simulation with KDEs 60 GeV}.}
    \label{fig: kinematic shower variables for simulation with KDEs 120 GeV}
\end{figure}
\begin{figure}[hp]
    \centering
    \subfigure[]{\includegraphics[width = 0.49\textwidth]{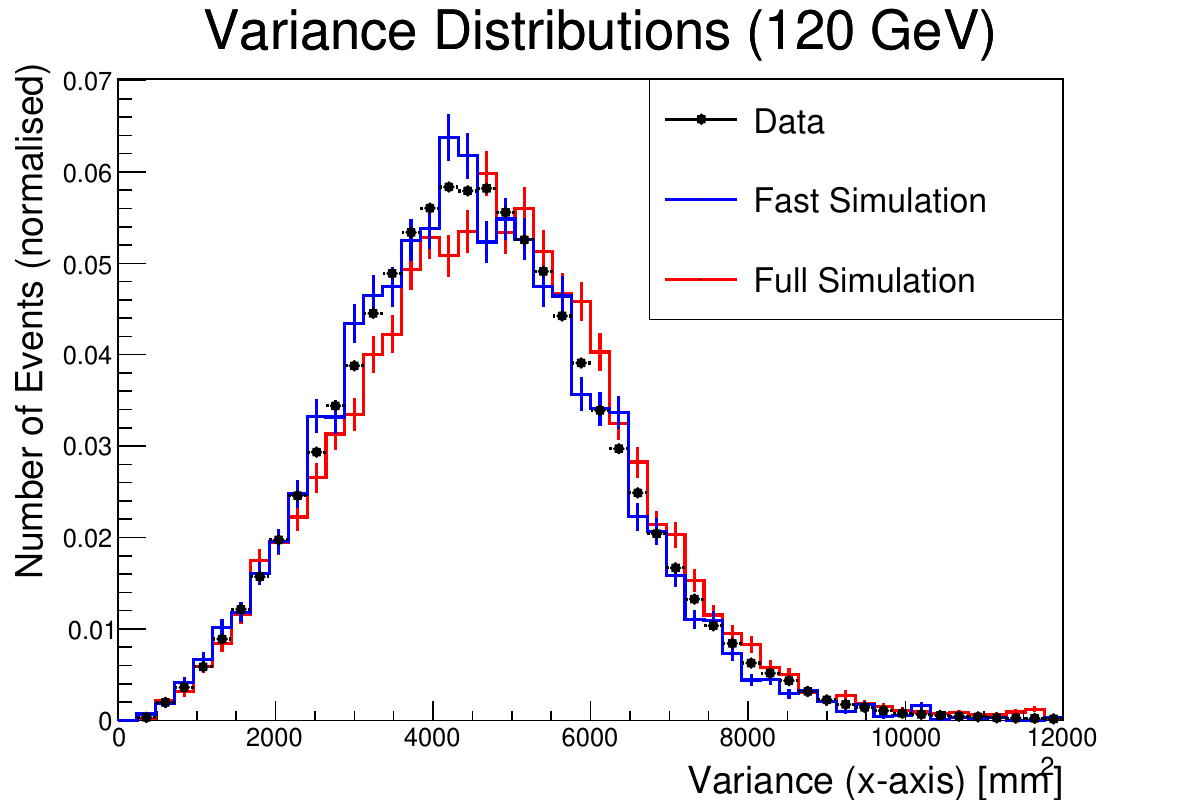}}
    \subfigure[]{\includegraphics[width = 0.49\textwidth]{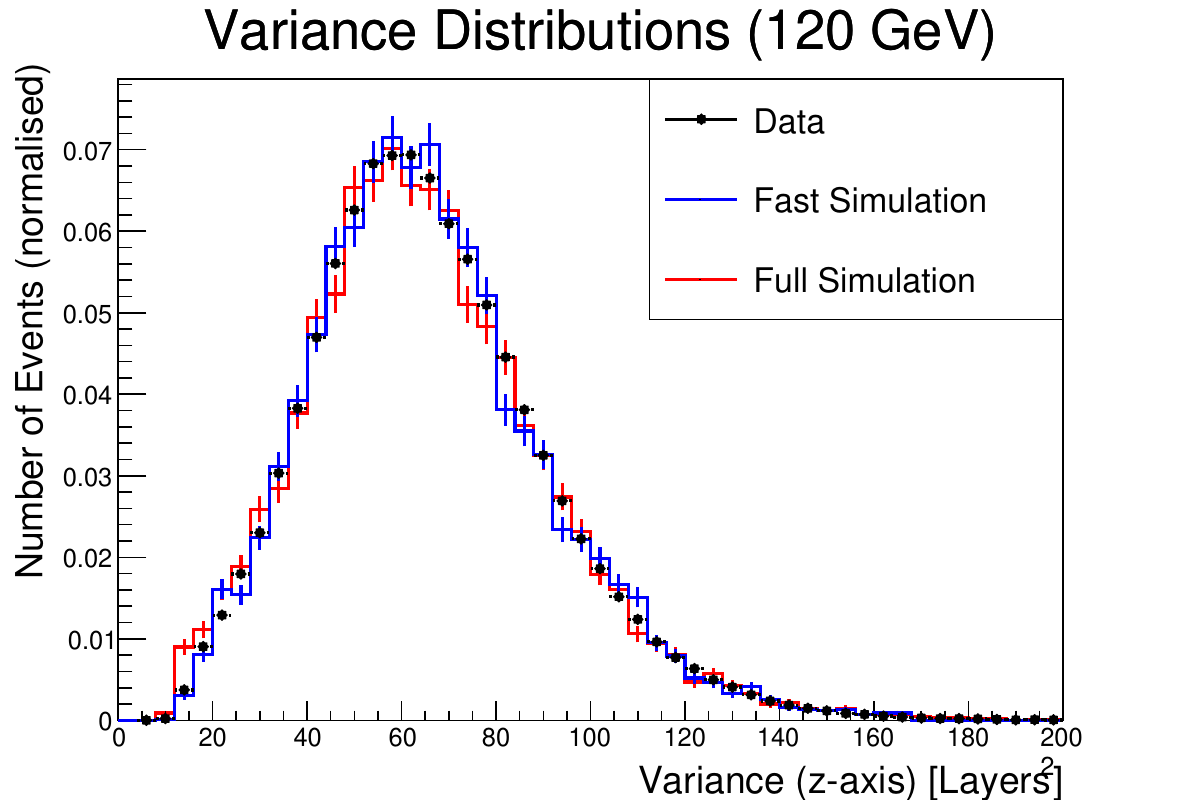}}
    \subfigure[]{\includegraphics[width = 0.49\textwidth]{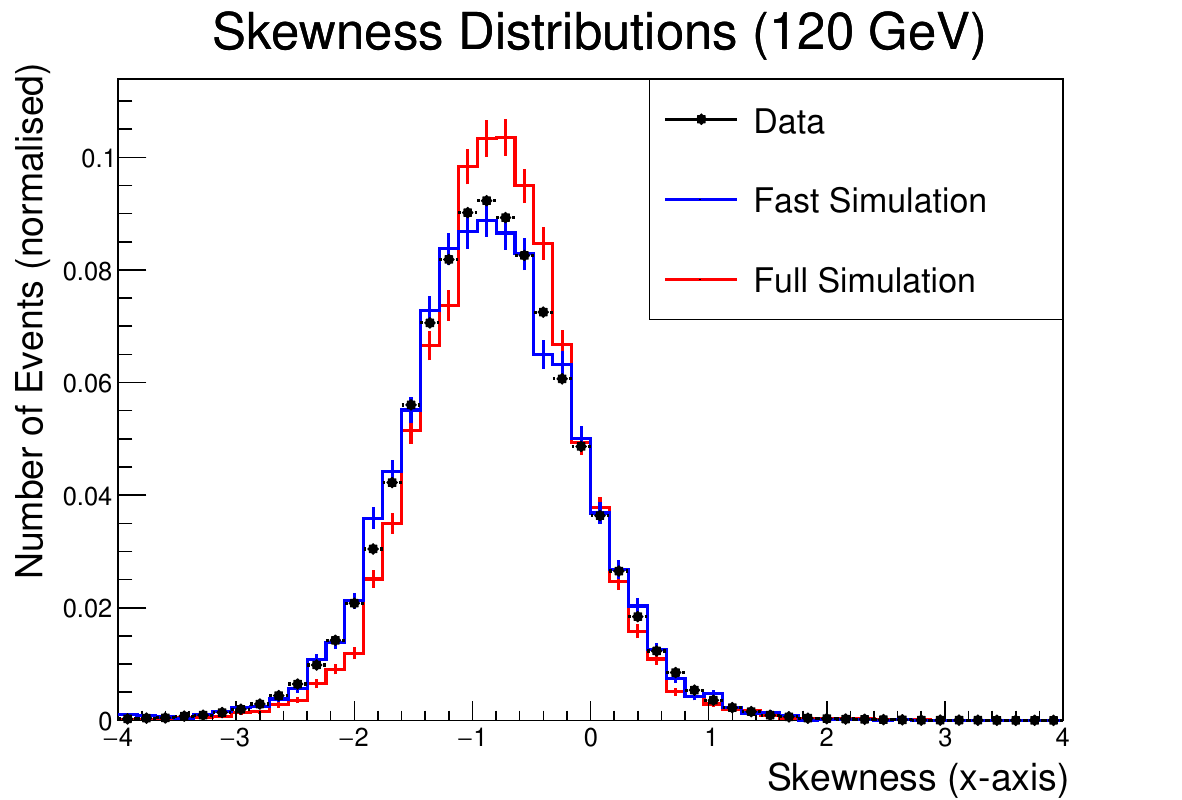}}
    \subfigure[]{\includegraphics[width = 0.49\textwidth]{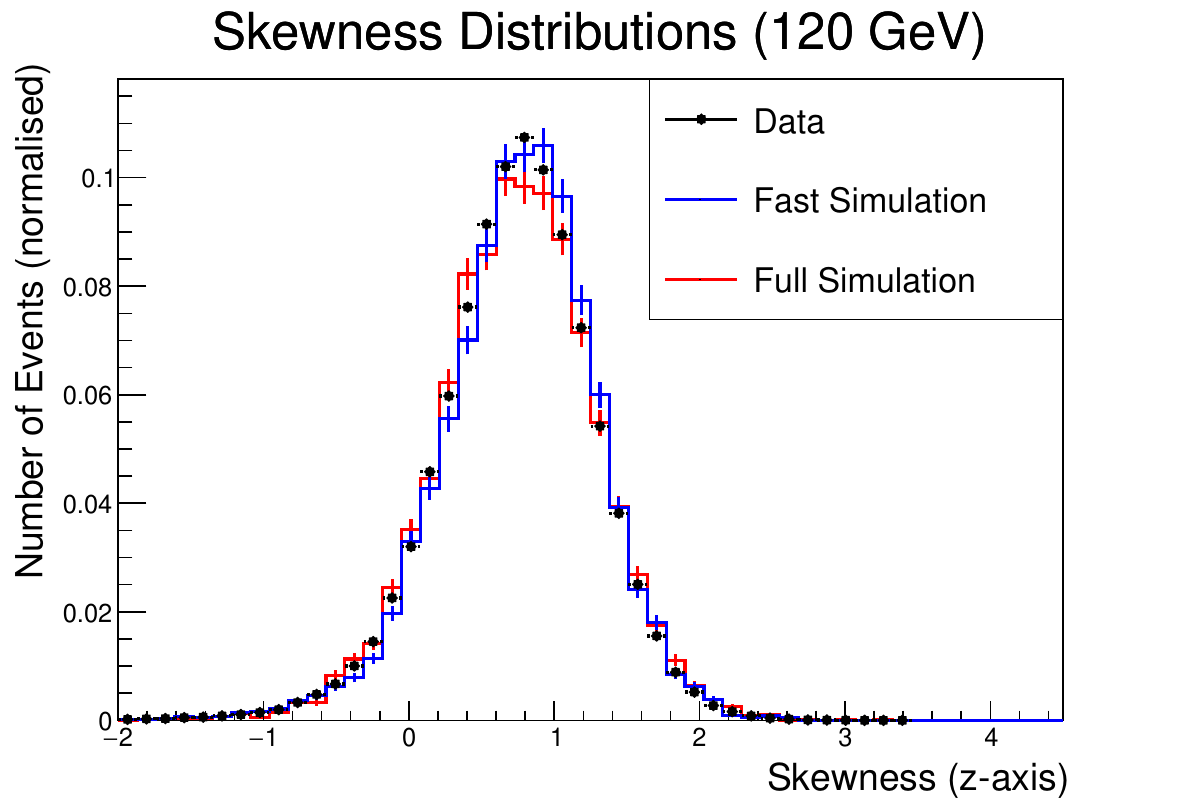}}
    \subfigure[]{\includegraphics[width = 0.49\textwidth]{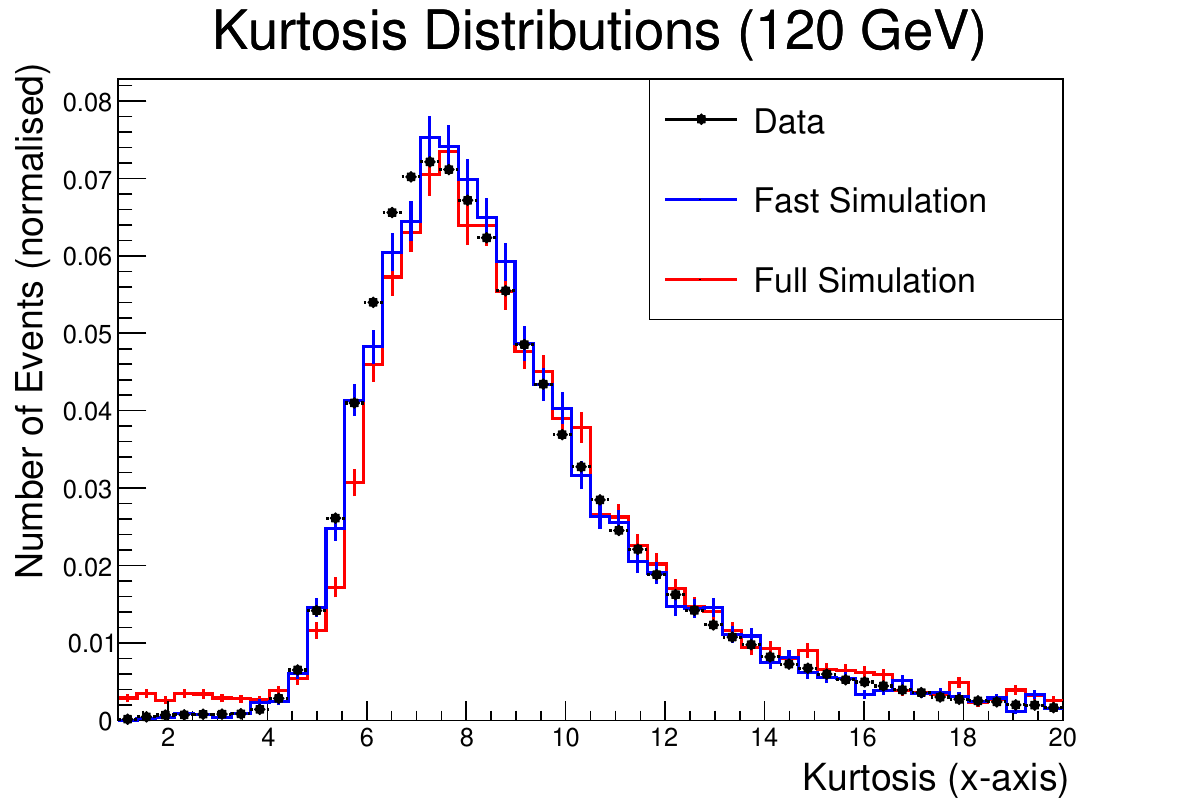}}
    \subfigure[]{\includegraphics[width = 0.49\textwidth]{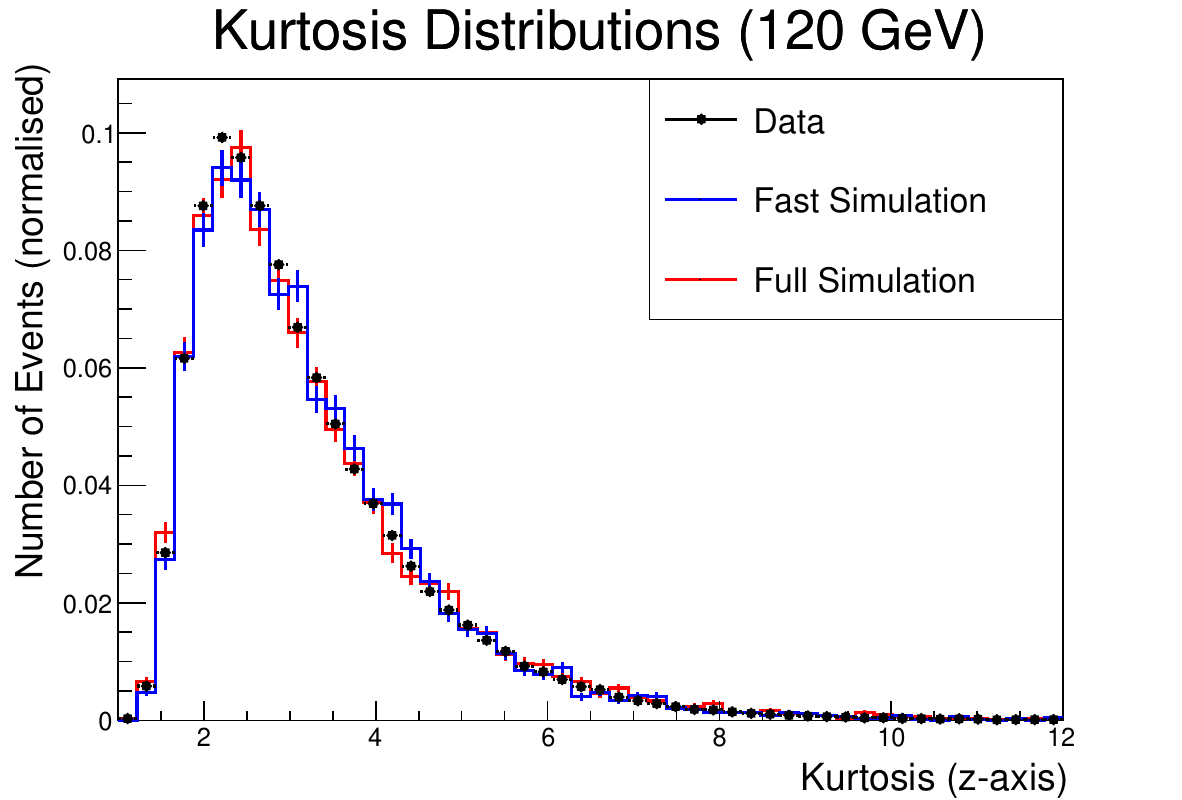}}
    \caption{Distributions of the same shower moments that have already been shown in Figure \ref{fig: shower moments for simulation with KDEs 60 GeV}, but for $\SI{120}{\giga\electronvolt}$ pions. The PDFs exhibit the same behaviour that has already been described in Figure \ref{fig: shower moments for simulation with KDEs 60 GeV}.}
    \label{fig: shower moments for simulation with KDEs 120 GeV}
\end{figure}

Results of the fast simulation modelling these variables are shown in Figures \ref{fig: kinematic shower variables for simulation with KDEs 60 GeV} and \ref{fig: shower moments for simulation with KDEs 60 GeV} for $\SI{60}{\giga\electronvolt}$  and in Figures \ref{fig: kinematic shower variables for simulation with KDEs 120 GeV} and \ref{fig: shower moments for simulation with KDEs 120 GeV} for $\SI{120}{\giga\electronvolt}$ pions. In all cases, the KDE-based fast simulation is able to successfully reproduce the expected kinematic behaviour, as there are no inconsistencies between data and fast simulation. In several distributions, the fast simulation performs even slightly better than the full simulation. This is particularly evident in Figures \ref{fig: kinematic shower variables for simulation with KDEs 60 GeV} and \ref{fig: kinematic shower variables for simulation with KDEs 120 GeV} for the distributions of single hit energies and the total energy. For the former, the height of the peak bin of the full simulation PDF is marginally higher than observed in data. For the latter, the full simulation exhibits a small shift towards larger energy values compared to both data and fast simulation.
\begin{figure}[ht]
    \centering
    \subfigure[]{\includegraphics[width = 0.49\textwidth]{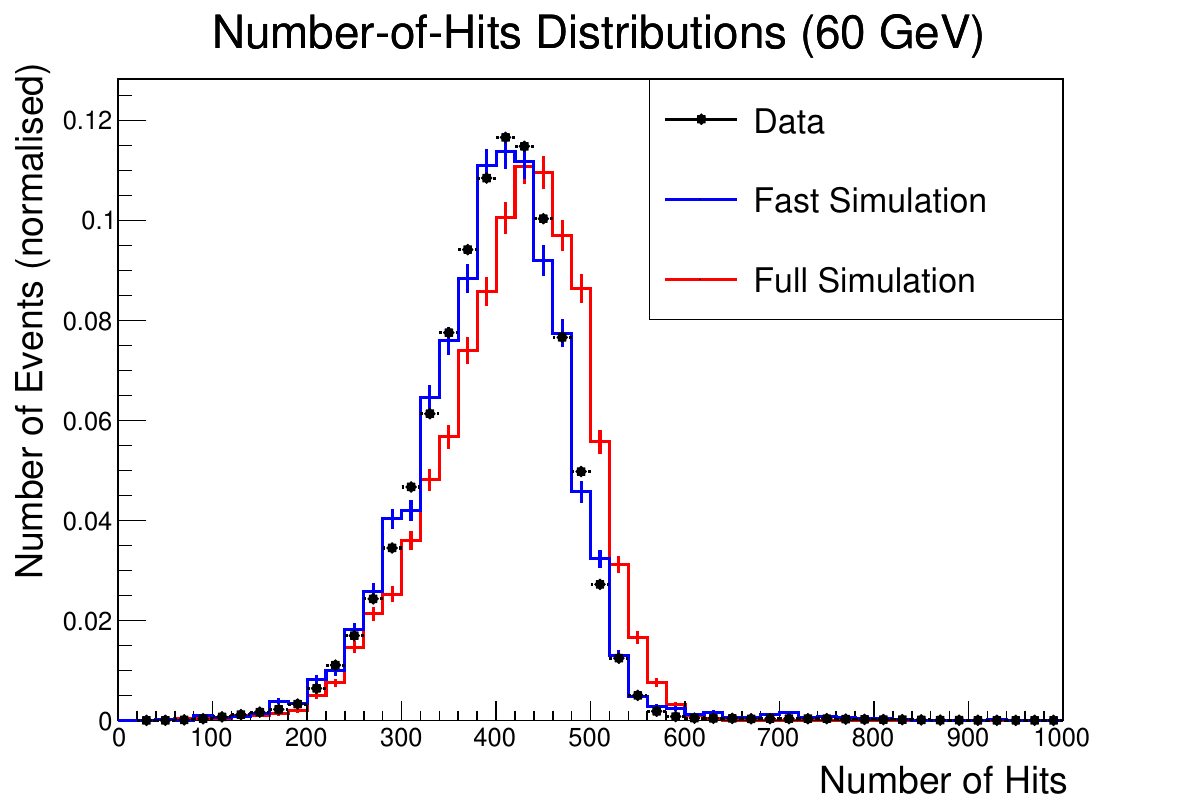}}
    \subfigure[]{\includegraphics[width = 0.49\textwidth]{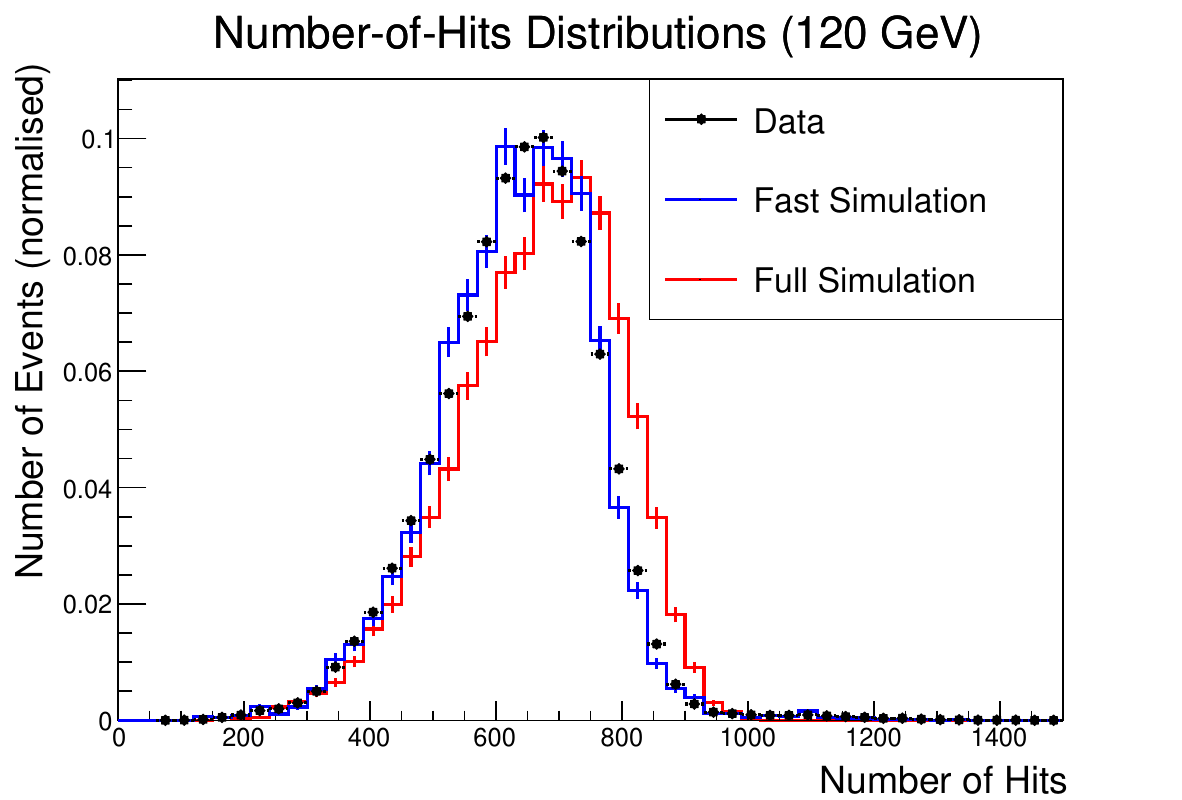}}
    \caption{Distributions for the number of hits per event for $\SI{60}{\giga\electronvolt}$ (left) and $\SI{120}{\giga\electronvolt}$ (right) pions. Data is shown as black points, and the results obtained from fast (full) simulation are depicted as dark blue (red) curves. Both energies show evidence that the fast simulation performs slightly better than the full simulation.}
    \label{fig: number of hits for KDE simulation}
\end{figure}
\par
Finally, Figure \ref{fig: number of hits for KDE simulation} presents distributions of the number of hits per event. Once again, one can observe excellent agreement between data and the fast simulation for both pion energies. Minor disagreement, however, remains between either of these two PDFs and the one obtained from full simulation. In summary, even on hit level, KDEs are a very reliable method of obtaining PDFs to simulate energy distributions, and this simulation method is capable of recreating the kinematic behaviour of a pion shower with high accuracy. Additional simulations performed at other beam energies and compared with the 2018 test beam dataset confirm a similarly high level of agreement.

\subsection{Simulated Correlations between Kinematic Shower Variables}
\label{subsec: simulated correlations between kinematic shower variables}

In Section \ref{subsec: simulated distributions of kinematic shower variables} it has been shown that the fast simulation algorithm is able to recreate distributions of various kinematic shower variables accurately, partly even exceeding those obtained from full simulation. Another important cross check is to investigate whether the fast simulation algorithm also preserves (linear) correlations between the aforementioned variables. Visually, this can be done by plotting two distinct kinematic variables as a pair in a two-dimensional histogram, separately for both data and fast simulation.
\begin{figure}[hp]
    \centering
    \subfigure[]{\includegraphics[width = 0.49\textwidth, page = 1]{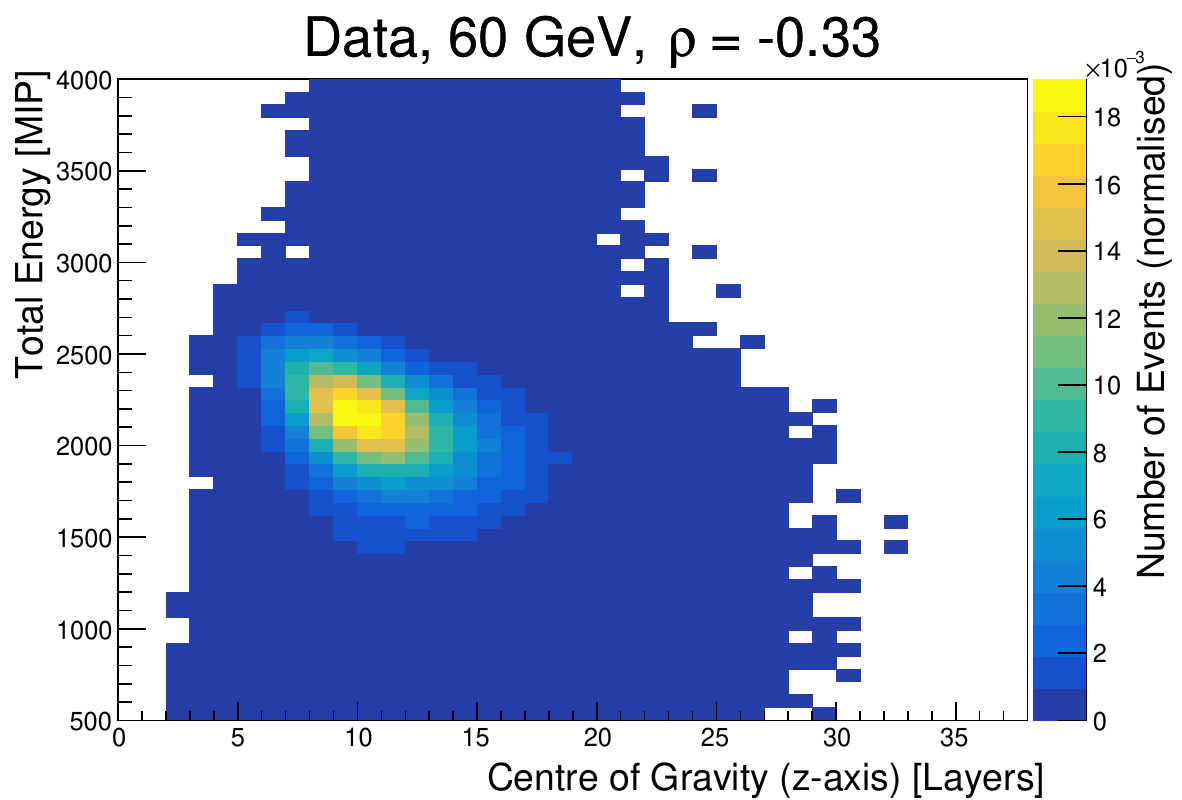}}
    \subfigure[]{\includegraphics[width = 0.49\textwidth, page = 1]{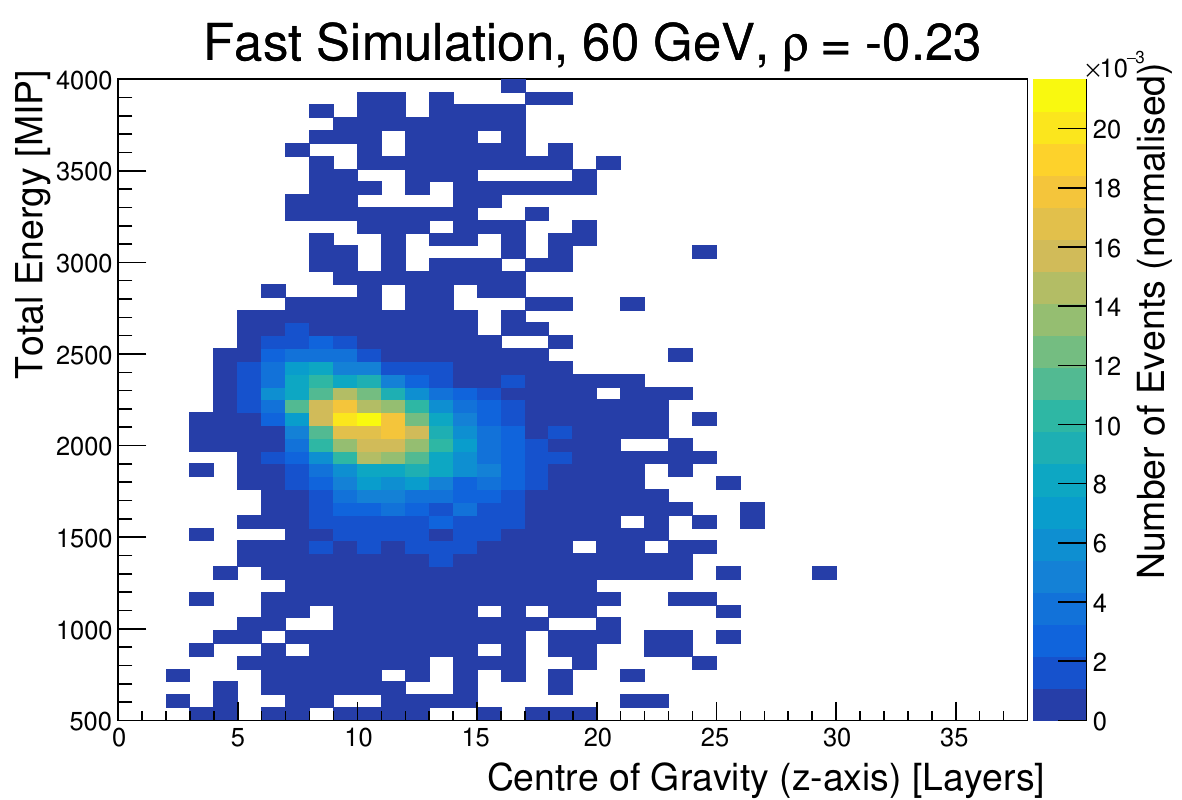}}
    \subfigure[]{\includegraphics[width = 0.49\textwidth, page = 2]{figures/hitlevel/60GeV/CorrelationPlots/KinematicVariables/TotalEnergyCorrelations_Data.pdf}}
    \subfigure[]{\includegraphics[width = 0.49\textwidth, page = 2]{figures/hitlevel/60GeV/CorrelationPlots/KinematicVariables/TotalEnergyCorrelations_FastSimulation.pdf}}
    \subfigure[]{\includegraphics[width = 0.49\textwidth, page = 3]{figures/hitlevel/60GeV/CorrelationPlots/KinematicVariables/TotalEnergyCorrelations_Data.pdf}}
    \subfigure[]{\includegraphics[width = 0.49\textwidth, page = 3]{figures/hitlevel/60GeV/CorrelationPlots/KinematicVariables/TotalEnergyCorrelations_FastSimulation.pdf}}
    \caption{Two-dimensional correlation plots for $\SI{60}{\giga\electronvolt}$ pion showers between the total energy and either the CoG$_{z}$ (top row), the shower radius (middle row), or the central fraction (bottom row). All histograms are shown for data (left column) and fast simulation (right column). For each row, very good agreement is visible between data and fast simulation.}
    \label{fig: correlations total energy vs. kinematic variables 60 GeV}
\end{figure}
\begin{figure}[hp]
    \centering
    \subfigure[]{\includegraphics[width = 0.49\textwidth, page = 1]{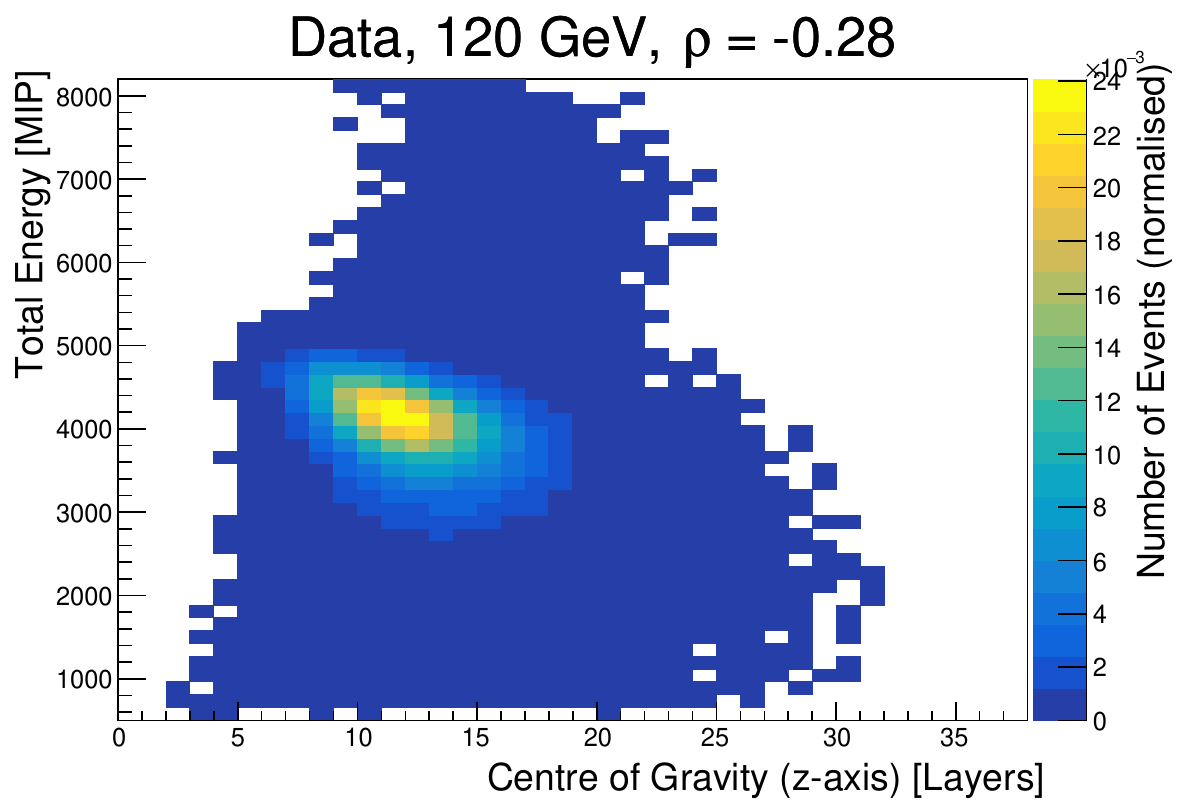}}
    \subfigure[]{\includegraphics[width = 0.49\textwidth, page = 1]{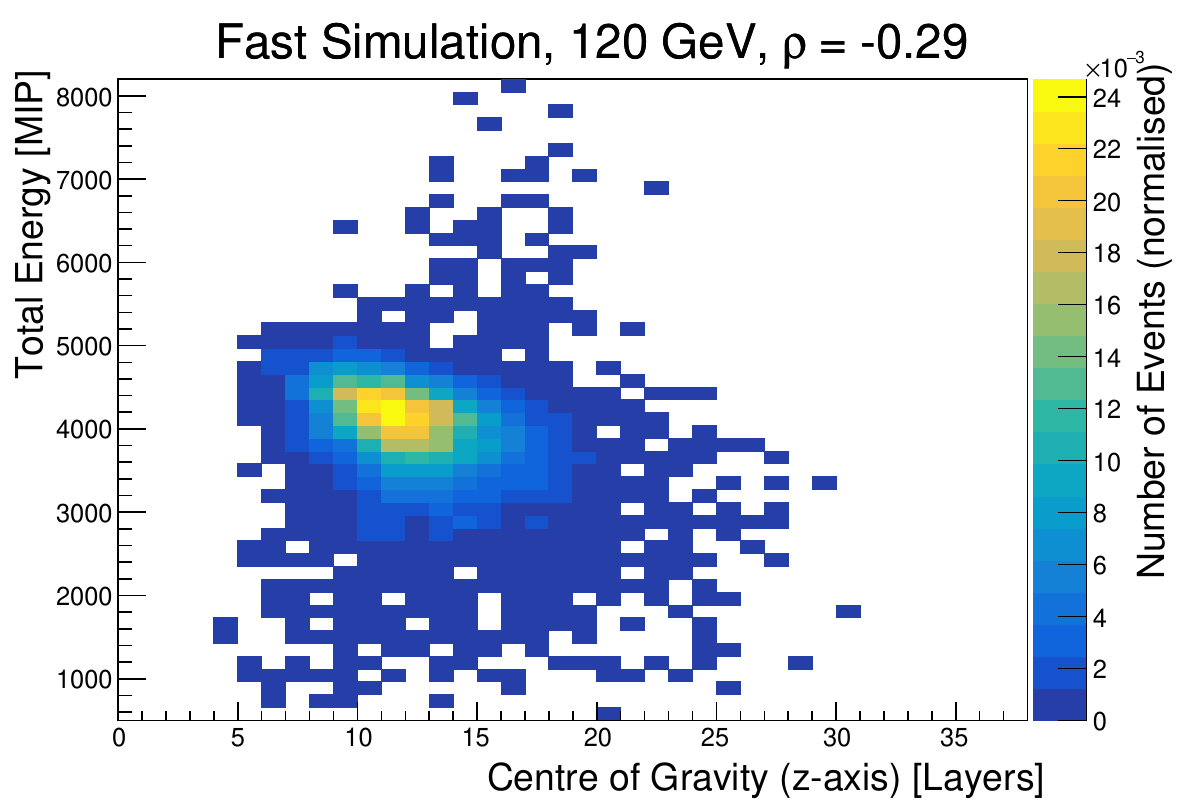}}
    \subfigure[]{\includegraphics[width = 0.49\textwidth, page = 2]{figures/hitlevel/120GeV/CorrelationPlots/KinematicVariables/TotalEnergyCorrelations_Data.pdf}}
    \subfigure[]{\includegraphics[width = 0.49\textwidth, page = 2]{figures/hitlevel/120GeV/CorrelationPlots/KinematicVariables/TotalEnergyCorrelations_FastSimulation.pdf}}
    \subfigure[]{\includegraphics[width = 0.49\textwidth, page = 3]{figures/hitlevel/120GeV/CorrelationPlots/KinematicVariables/TotalEnergyCorrelations_Data.pdf}}
    \subfigure[]{\includegraphics[width = 0.49\textwidth, page = 3]{figures/hitlevel/120GeV/CorrelationPlots/KinematicVariables/TotalEnergyCorrelations_FastSimulation.pdf}}
    \caption{Two-dimensional correlation plots for $\SI{120}{\giga\electronvolt}$ pion showers between the total energy and either the CoG$_{z}$ (top row), the shower radius (middle row), or the central fraction (bottom row). All histograms are shown for data (left column) and fast simulation (right column). For each row, very good agreement is visible between data and fast simulation.}
    \label{fig: correlations total energy vs. kinematic variables 120 GeV}
\end{figure}
\par
Comparisons between simulated correlation factors and those obtained from data are shown in Figures \ref{fig: correlations total energy vs. kinematic variables 60 GeV} and \ref{fig: correlations total energy vs. kinematic variables 120 GeV} for $\SI{60}{\giga\electronvolt}$ and $\SI{120}{\giga\electronvolt}$, respectively. These show examples of correlation plots between the total energy and either the longitudinal CoG, the mean shower radius, or the central fraction for both data and fast simulation. For each pair of variables, the two corresponding plots exhibit a high degree of visual similarity. A slight anticorrelation is observed between the total energy and both the CoG$_{z}$ as well as the shower radius, while, on the other hand, the total energy and the central fraction are positively correlated. Overall, the correlation factors visible in the panel titles only differ by, at most, $\SI{0.1}{}$ and $\SI{0.01}{}$ for $\SI{60}{\giga\electronvolt}$ and $\SI{120}{\giga\electronvolt}$, respectively.
\par
\begin{figure}[ht]
    \centering
    \includegraphics[width = 1\textwidth]{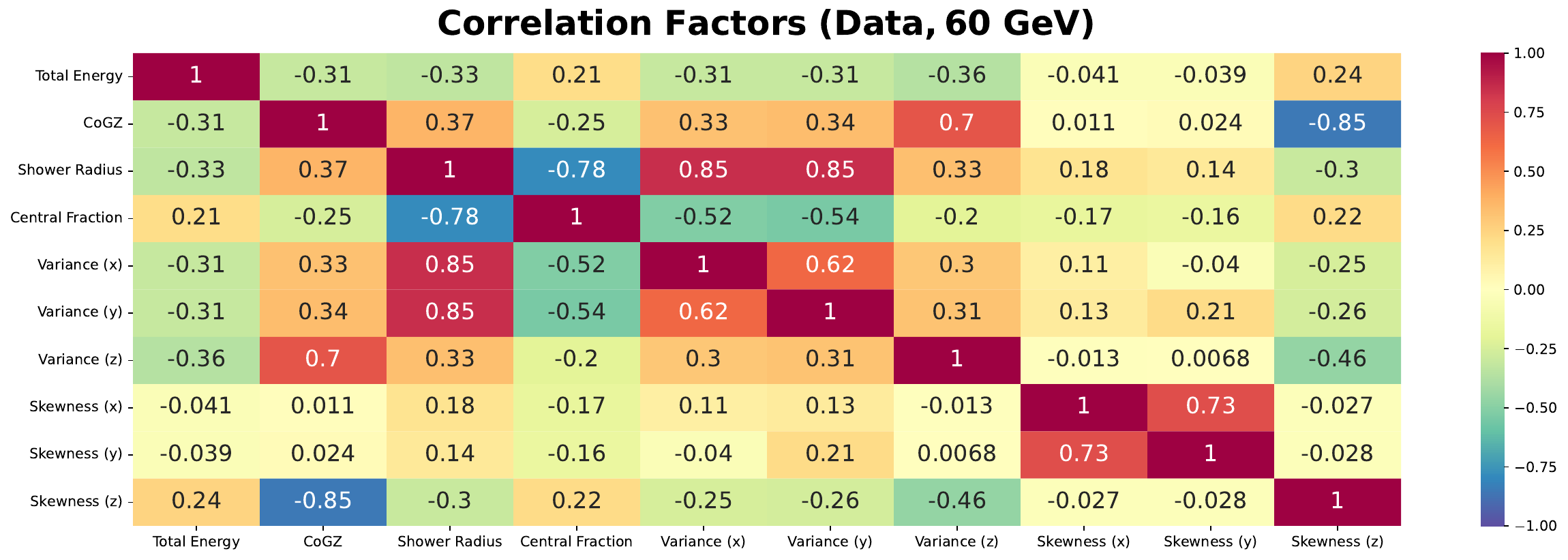}
    \includegraphics[width = 1\textwidth]{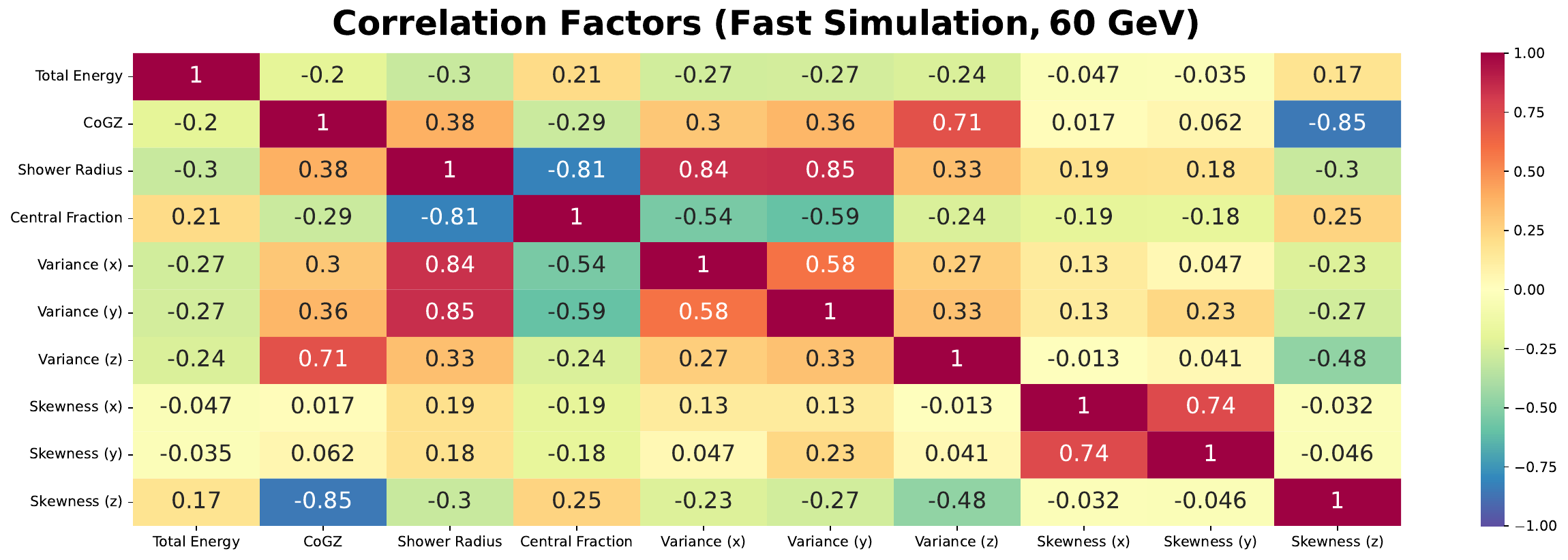}
    \caption{Correlation matrices for kinematic shower variables obtained from data (upper plot) and fast simulation (lower plot) for $\SI{60}{\giga\electronvolt}$ pions. Dark red colouring represents strong correlation between two variables, whereas dark blue indicates strong anticorrelation. The matrices are symmetric about their diagonals and agree very well with each other.}
    \label{fig: correlation matrices data and simulation 60 GeV}
\end{figure}
\begin{figure}[ht]
    \centering
    \includegraphics[width = 1\textwidth]{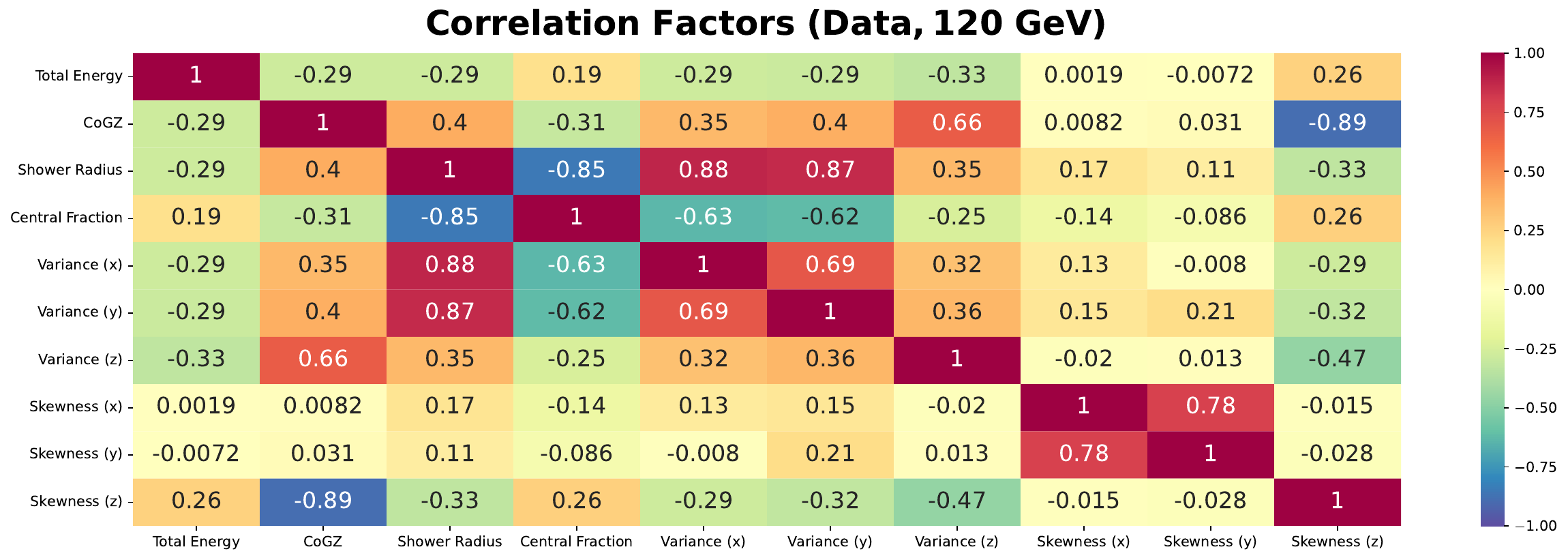}
    \includegraphics[width = 1\textwidth]{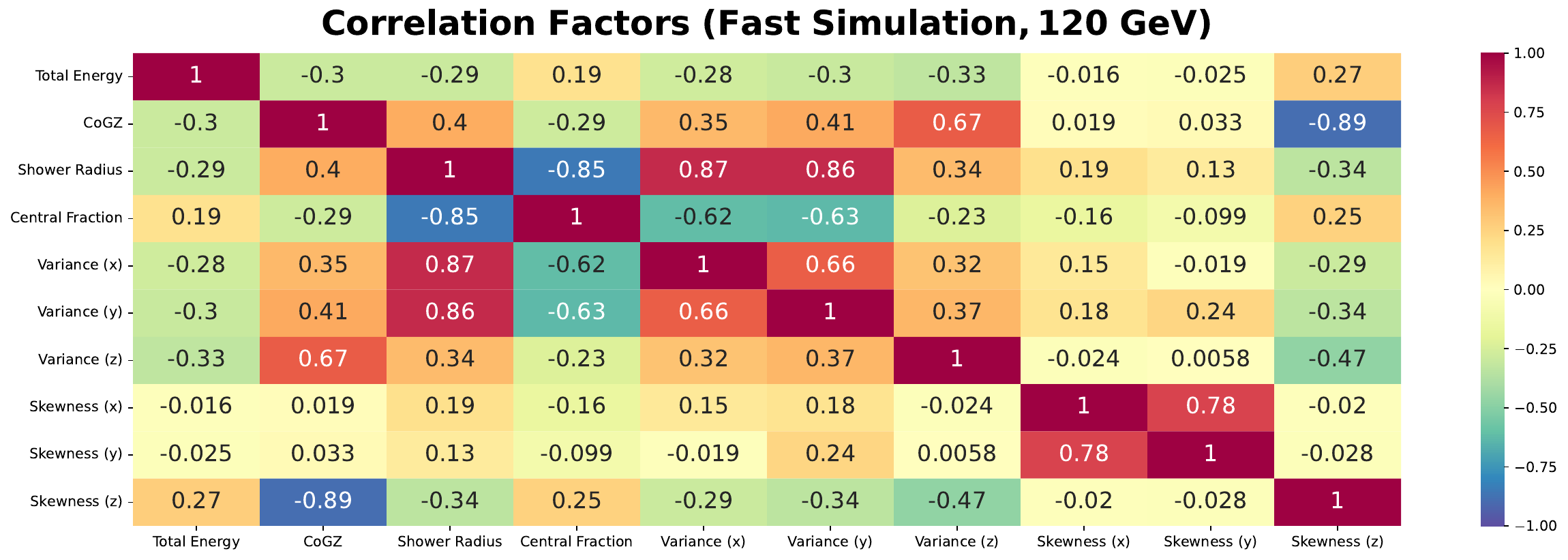}
    \caption{Correlation matrices for kinematic shower variables obtained from data (upper plot) and fast simulation (lower plot) for $\SI{60}{\giga\electronvolt}$ pions. Dark red colouring represents strong correlation between two variables, whereas dark blue indicates strong anticorrelation. The matrices are symmetric about their diagonals and agree very well with each other.}
    \label{fig: correlation matrices data and simulation 120 GeV}
\end{figure}

In addition to the previously discussed plots, Figures \ref{fig: correlation matrices data and simulation 60 GeV} and \ref{fig: correlation matrices data and simulation 120 GeV} present correlation matrices for data and fast simulation, respectively. These matrices summarise the correlation factors derived from Figures \ref{fig: correlations total energy vs. kinematic variables 60 GeV} to \ref{fig: correlations total energy vs. kinematic variables 120 GeV}, also for variable pairs that have not been shown thus far. In this representation, red (blue) colouring indicates strong (anti-)correlation. Correlation factors are shown for the total energy, the longitudinal CoG, the mean shower radius, the central fraction, as well as the variance and skewness in all three spatial dimensions. The comparison between data and fast simulation reveals very good agreement across all variables, with the largest observed correlation difference being $\SI{0.11}{}$ at $\SI{60}{\giga\electronvolt}$. A similar level of consistency is also observed for other test beam energies. In summary, the KDE-based fast simulation algorithm not only reproduces kinematic distributions accurately, but also preserves correlation factors between kinematic shower variables at hit level.
\par
\begin{figure}[ht]
    \centering
    \subfigure[]{\includegraphics[width = 0.49\textwidth]{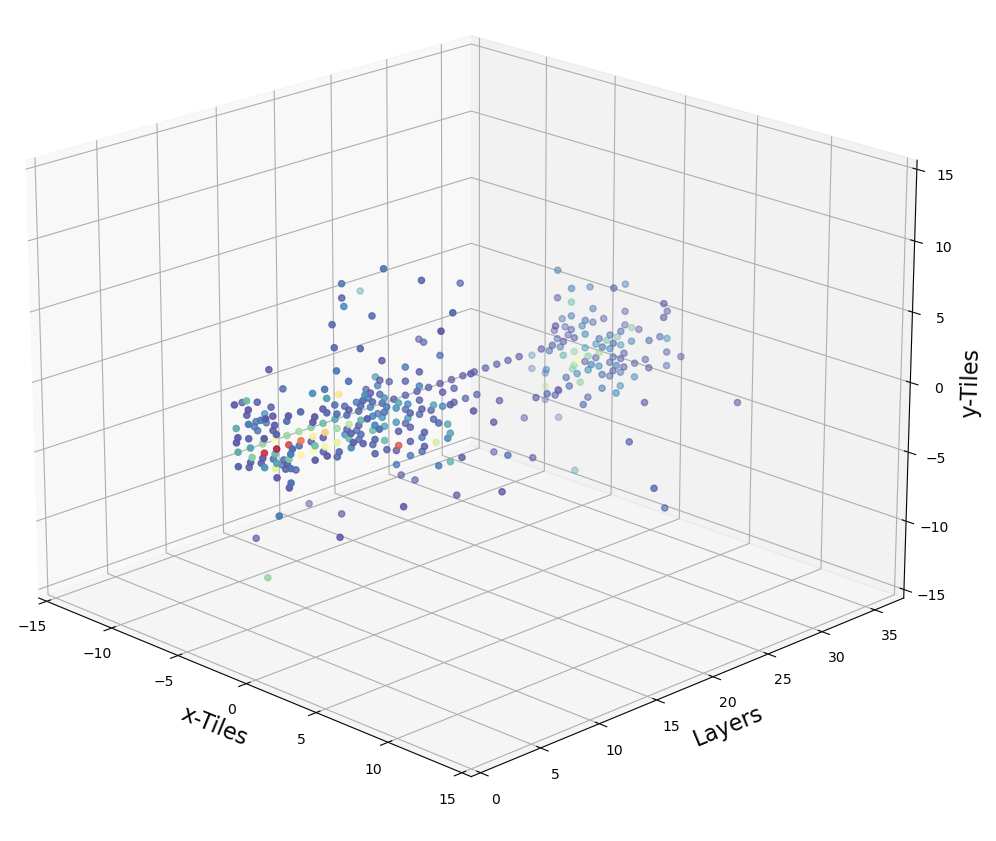}}
    \subfigure[]{\includegraphics[width = 0.49\textwidth]{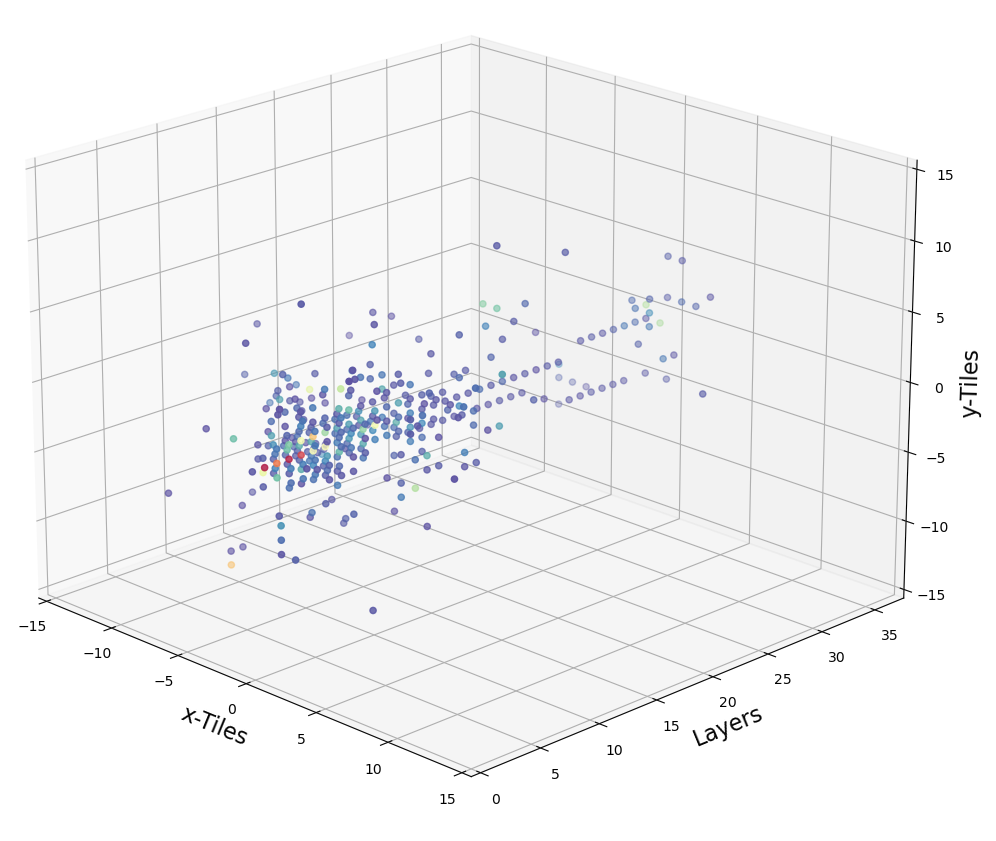}}
    \caption{Three-dimensional depictions of $\SI{60}{\giga\electronvolt}$ pion showers obtained from data (left) and fast simulation (right). The axes labelled as ``$x$-Tiles'' and ``$y$-Tiles'' represent the $x$- and $y$-axis, respectively, of the \ahcal; the one labelled as ``Layers'' points along the $z$-direction of the detector. Each point represents an individual hit with blue (red) colouring corresponding to low-energetic (high-energetic) ones. One can observe track-like structures as well as small electromagnetic subshowers within the whole simulated pion shower.}
    \label{fig: 3D shower plots}
\end{figure}

Finally, Figure \ref{fig: 3D shower plots} depicts two examples of how $\SI{60}{\giga\electronvolt}$ pion showers obtained from data and fast simulation, respectively, look like in a three-dimensional calorimeter. In both plots, each dot represents either a detected or simulated hit with colouring indicating the hit energy. The overall shower structures are remarkably similar and exhibit no significant visual differences between data and fast simulation. Note that the fast simulation algorithm is even capable of simulating single tracks and small, electromagnetic clusters within the whole simulated hadron shower, as one would observe them in real calorimeters.

\subsection{Computational Requirements}
\label{subsec: computational requirements KDEs}

The investigation presented in this section was carried out on a \textit{Dell Precision 7865 Tower} equipped with an \textit{AMD Ryzen Threadripper PRO 5965WX 24-Cores} processor. From the test beam dataset, a subset comprising $\SI{10000}{}$ events has been used as input for the KDE, from which $\SI{10000}{}$ events were randomly generated. Each dataset required $\SI{2.92}{\giga\byte}$ of random-access memory, resulting in a total of twice the size, i.e. $\SI{5.84}{\giga\byte}$. This procedure was repeated ten times in order to obtain a statistically significant simulated dataset.
\par
The average computing times for generating one and $\SI{10000}{}$ events, respectively, are listed in Table \ref{tab: computing times for hit level simulation} for $\SI{60}{\giga\electronvolt}$ pions for both fast and full simulation. On average, the fast simulation requires a few hundreds of seconds to simulate $\SI{10000}{}$ events, which corresponds to an average computation time per event of roughly $\SI{23}{\milli\second}$. Most of the computing time, in this case, is required to read the input dataset and write the simulated sample to disk. In comparison, simulating, digitising, and reconstructing $\SI{10000}{}$ events via full simulation requires multiple days, resulting in an average computation time per event of approximately $\SI{42}{\second}$. The fast simulation, therefore, clearly excels the full simulation by multiple orders of magnitude and achieves an average speed-up factor of at least $\mathcal{O}(\SI{1000}{})$. Comparable performance has also been observed for other pion energies.
\begin{table}[ht]
    \centering
    \caption{Average computing times for the simulation of $\SI{60}{\giga\electronvolt}$ pion showers. The two rows show the computing times for fast and full simulation, respectively. Column 2 and 3 show how much time is required for simulating $\SI{10000}{}$ events and a single event, respectively, on average.}
    \begin{tabular}{|c|c|c|}
        \hline
        \multirow{2}{*}{Simulation} & \multicolumn{2}{c|}{Average Computing Time}\\
        \cline{2-3}
        & $\SI{10000}{}$ events & One event\\
        \hline
        \hline
        Fast & $\SI{227}{\second}$ & $\SI{22.7}{\milli\second}$\\
        Full & $\SI{4.81}{\day}$ & $\SI{41.5}{\second}$\\
        \hline
    \end{tabular}
    \label{tab: computing times for hit level simulation}
\end{table}

%% file: chapters/interpolation.tex
\section{Interpolation Studies of Hit Energy Distributions using Kernel Density Estimators}
\label{sec: interpolation studies of hit energy distributions using kernel density estimators}

Since the simulation presented in Section \ref{sec: simulation of hit energy distributions using kernel density estimators} showed very good agreement with the data, interpolations of simulated pion showers have also been investigated. More precisely, the hit energy distributions of individual calorimeter tiles have been interpolated between different initial pion energies. Interpolations play a crucial role for this data-driven fast simulation because it allows for the prediction of pion shower behaviour at arbitrary beam energies. Otherwise, without the interpolation, the fast simulation would be confined to the energies of the test beam dataset. However, due to the limited size of the dataset this simulation is based upon, the interpolation algorithm could only be tested and validated using the available pion energies from the test beam campaign.
\par
This section presents the results of how the fast simulation algorithm introduced in Section \ref{sec: kernel density estimators for pion showers} is used to interpolate simulated hit energy distributions of single pion showers between various initial energies. The exact method and the mathematical procedures behind the interpolation are first introduced in Section \ref{subsec: mathematical background of interpolations}. After that, interpolated kinematic distributions are shown in Section \ref{subsec: kinematic distributions of interpolated individual hit energies} and compared with kinematic distributions that were obtained from both data and full simulation. Finally, Section \ref{subsec: interpolated correlations between kinematic shower variables} concludes this chapter by comparing the correlations of shower variables at interpolated energies with those obtained from data, followed by a short performance estimation of the interpolation algorithm.

\subsection{Algorithmic Approach for Energy Interpolations}
\label{subsec: mathematical background of interpolations}

The objective of an interpolation is not only to predict distributions of hit energies correctly, but also to preserve the (anti-)correlations between kinematic shower variables as precisely as possible. For this reason, the interpolation was done in the following way.
\par
\begin{figure}[hp]
    \centering
    \includegraphics[width = 1\textwidth]{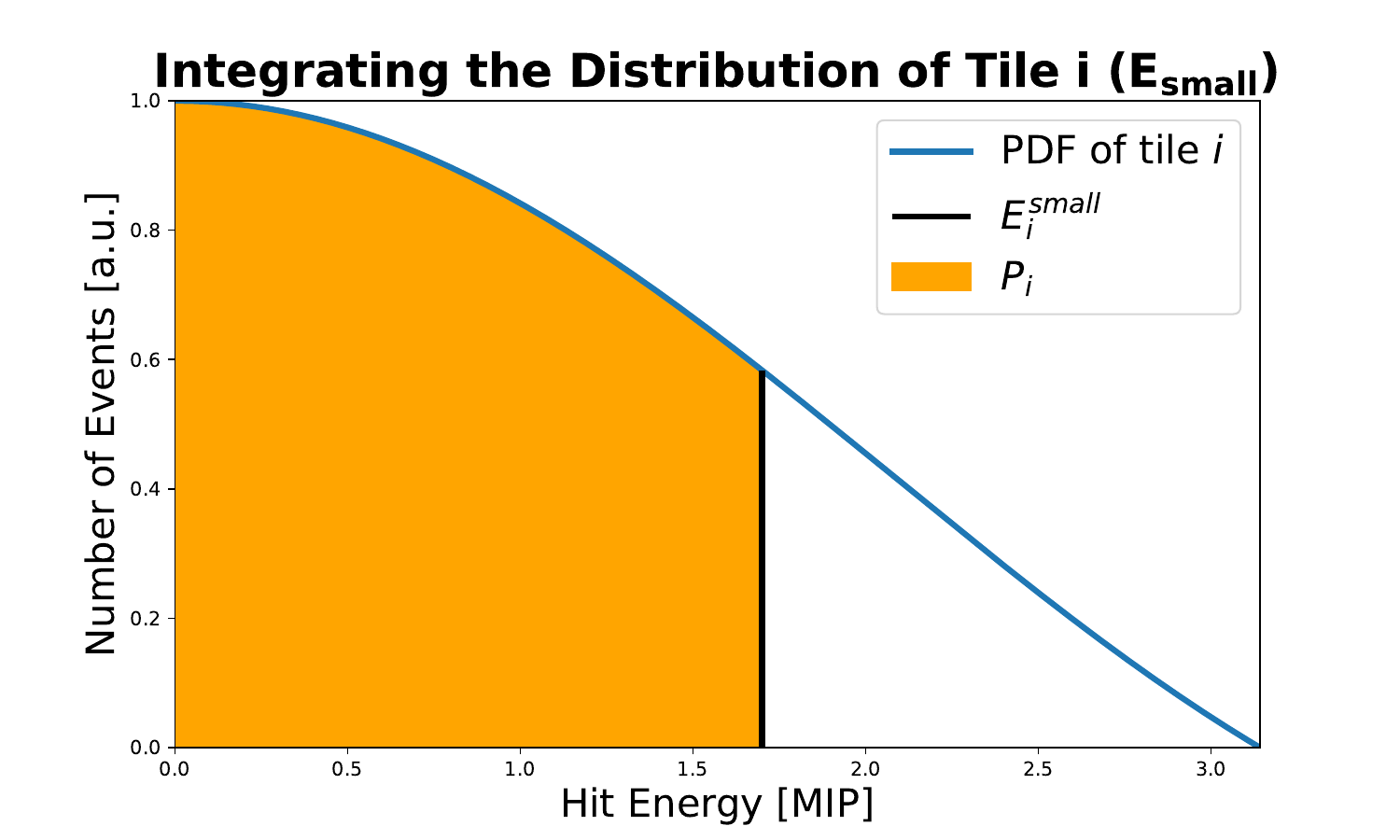}
    \includegraphics[width = 1\textwidth]{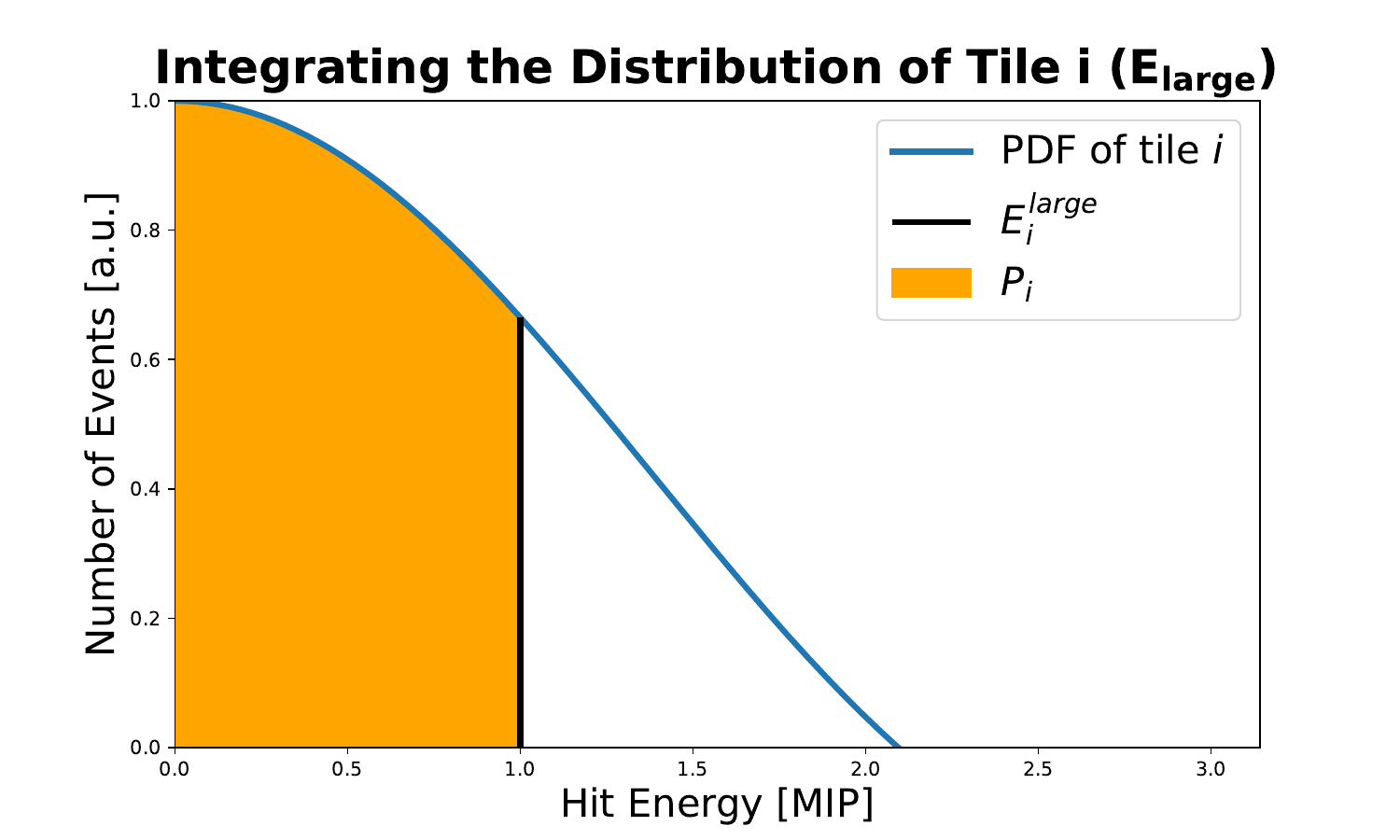}
    \caption{Illustrative example of integrating hit energy distributions. The upper plot corresponds to integrating the PDF of the hit energy for tile $i$ at $E^{\text{small}}$ from the smallest bin to $E^{\text{small}}_{i}$. The lower plot, on the other hand, also corresponds to integrating the PDF of the hit energy for tile $i$, but at $E^{\text{large}}$ until both orange-shaded areas are equal. This is the case at the upper bound $E^{\text{large}}_{i}$.}
    \label{fig: example PDFs small and large}
\end{figure}

Interpolating to a target energy, $E^{\text{int}}$, requires the cumulative hit energy distributions of two different reference pion energies, for example $E^{\text{small}}$ and $E^{\text{large}}$, chosen such that $E^{\text{small}}<E^{\text{int}}<E^{\text{large}}$. An explicit example could be $E^{\text{small}}=\SI{40}{\giga\electronvolt}$ and $E^{\text{large}}=\SI{80}{\giga\electronvolt}$ as reference energies and $E^{\text{int}}=\SI{60}{\giga\electronvolt}$ as target energy for a dataset that does not include $\SI{60}{\giga\electronvolt}$ pion data. Given the chosen CoG cuts, there are in total $N=31\times31\times38=\SI{36518}{}$ hit energy distributions\footnote{These hit energy distributions should not be confused with the one introduced in Section \ref{subsec: simulated distributions of kinematic shower variables} which shows the hit energy distribution of the entire detector, not that of a single tile. For the entire detector, one event contributes $\SI{36518}{}$ hit energies to its hit energy distribution, whereas on a tile-by-tile basis, only the $i$th hit energy of an event is added to the hit energy distribution of the $i$th calorimeter tile. For $n$ events, the former thus contains $n\times\SI{36518}{}$ entries, while the hit energy distributions of each tile only contain $n$ entries.} for each of $E^{\text{small}}$ and $E^{\text{large}}$. The interpolation procedure begins by generating a single event for $E^{\text{small}}$ using KDEs, resulting in hit energies $E^{\text{small}}_{0}, E^{\text{small}}_{1}, ..., E^{\text{small}}_{N-1}$. For each tile $i$, the value of the cumulative hit energy PDF at $E^{\text{small}}_{i}$ is then evaluated, which yields a set of $N$ real numbers between zero and one (because the PDFs are normalised to unity): $A^{\text{small}}_{0}, A^{\text{small}}_{1}, ..., A^{\text{small}}_{N-1}$. This step corresponds to integrating the hit energy PDF of tile $i$ from the left (the smallest bin) to $E^{\text{small}}_{i}$ as illustratively shown in the upper plot in Figure \ref{fig: example PDFs small and large}.
\par
Next, the cumulative hit energy distributions corresponding to $E^{\text{large}}$ are considered. For every tile $i$, one identifies the bin of the cumulative PDF whose height equals $A^{\text{small}}_{i}$. This is the case at some hit energy $E^{\text{large}}_{i}$. In other words, this step corresponds to integrating the hit energy distribution for tile $i$ at $E^{\text{large}}$ from the smallest bin until the cumulative area equals $A^{\text{small}}_{i}$, which is the case at $E^{\text{large}}_{i}$. Repeating this procedure for all tiles yields another set of $N$ hit energies: $E^{\text{large}}_{0}, E^{\text{large}}_{1}, ..., E^{\text{large}}_{N-1}$. The lower plot of Figure \ref{fig: example PDFs small and large} shows this process schematically.
\par
The final step of the interpolation involves combining the two sets of hit energies, $E^{\text{small}}_{i}$ and $E^{\text{large}}_{i}$, by applying appropriate weights. This yields one interpolated set of hit energies:
\begin{equation}
    E^{\text{int}}_{i}=w^{\text{small}}E^{\text{small}}_{i}+w^{\text{large}}E^{\text{large}}_{i}\, .
\end{equation}
Both weights are determined by the relative distances of $E^{\text{small}}$ and $E^{\text{large}}$ to the target energy, $E^{\text{int}}$, via
\begin{equation}\label{eq: weight interpolation definition}
    w_{i}=1-\frac{|E^{\text{int}}-E_{i}|}{E^{\text{large}}-E^{\text{small}}}\, ,
\end{equation}
where the index $i$ is either ``small'' or ``large''. By construction, both weights satisfy
\begin{equation}
    w^{\text{small}}+w^{\text{large}}=1\, .
\end{equation}
Equation (\ref{eq: weight interpolation definition}) ensures that the relative contributions of $E^{\text{small}}$ and $E^{\text{large}}$ to the final results reflect their proximity to $E^{\text{int}}$: the closer $E_{i}$ is to $E^{\text{int}}$, the greater its influence on the interpolated events, and vice versa.
\par
The whole interpolation algorithm is then repeated vice versa with the superscripts ``small'' and ``large'' interchanged, which results in another independently simulated, interpolated event. The correlations are, therefore, based $\SI{50}{\percent}$ of the time on $E^{\text{small}}$ and $\SI{50}{\percent}$ of the time on $E^{\text{large}}$. This is done in order to mitigate any potential biases in the final, interpolated distributions that might be introduced by always starting from $E^{\text{small}}$. Combining the results from both directions helps to compensate for systematic tendencies in the interpolation. For example, if interpolating from $E^{\text{small}}$ tends to underestimate, say, the number of hits per shower, then interpolating from $E^{\text{large}}$ will likely overestimate it. In weighting and adding both values afterwards, any over- or underestimation will be ``balanced out'', giving the correct prediction in the end. Repeating this procedure multiple times yields a statistically significant sample of simulated, interpolated events whose kinematic behaviour can be studied, as well as the correlations between its kinematic variables.

\subsection{Kinematic Distributions of Interpolated Individual Hit Energies}
\label{subsec: kinematic distributions of interpolated individual hit energies}

Two interpolations studies have been performed for this investigation, both based on three neighbouring and equidistant initial energies. The first one was conducted for $E^{\text{small}}=\SI{40}{\giga\electronvolt}$, $E^{\text{int}}=\SI{60}{\giga\electronvolt}$, and $E^{\text{large}}=\SI{80}{\giga\electronvolt}$, while the second one used $E^{\text{small}}=\SI{80}{\giga\electronvolt}$, $E^{\text{int}}=\SI{120}{\giga\electronvolt}$, and $E^{\text{large}}=\SI{160}{\giga\electronvolt}$. The procedure presented in the previous section was repeated $\SI{100000}{}$ times for each case, resulting in interpolated datasets of twice the size, i.e. $\SI{200000}{}$ events in total for each interpolated pion energy.
\par
\begin{figure}[hp]
    \centering
    \subfigure[]{\includegraphics[width = 0.49\textwidth]{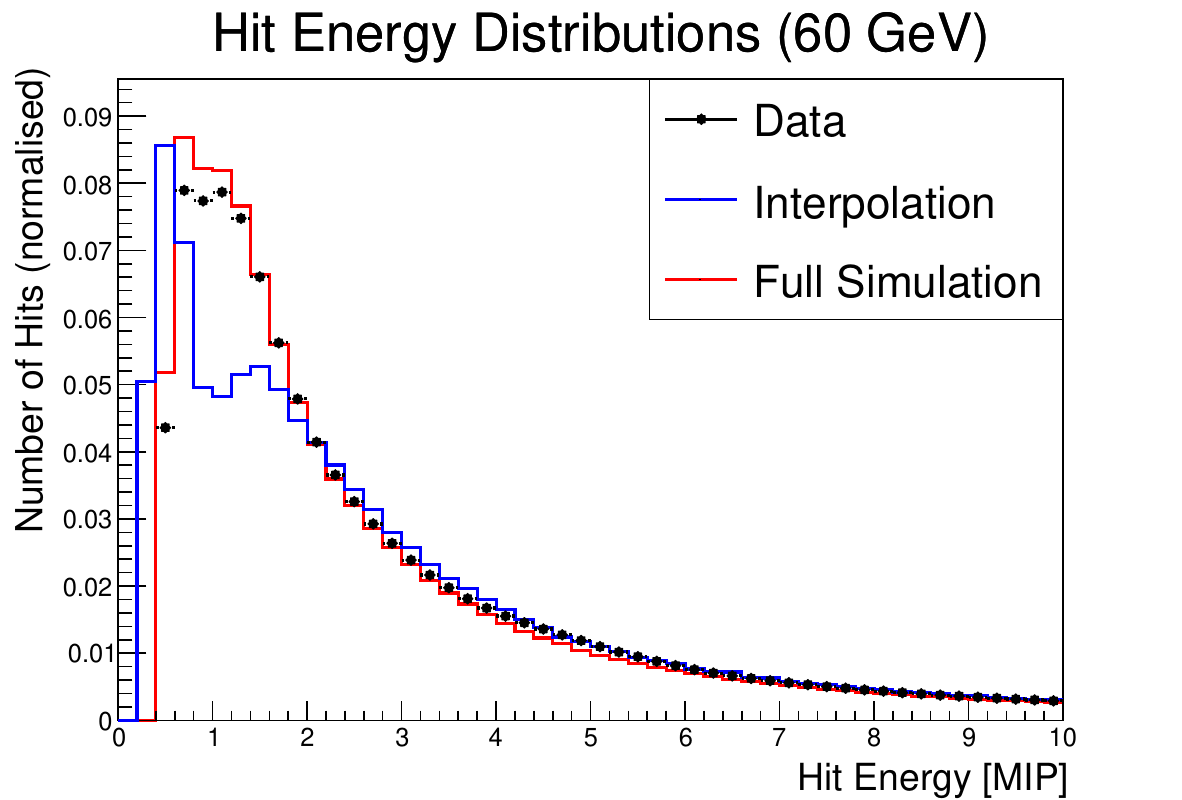}}
    \subfigure[]{\includegraphics[width = 0.49\textwidth]{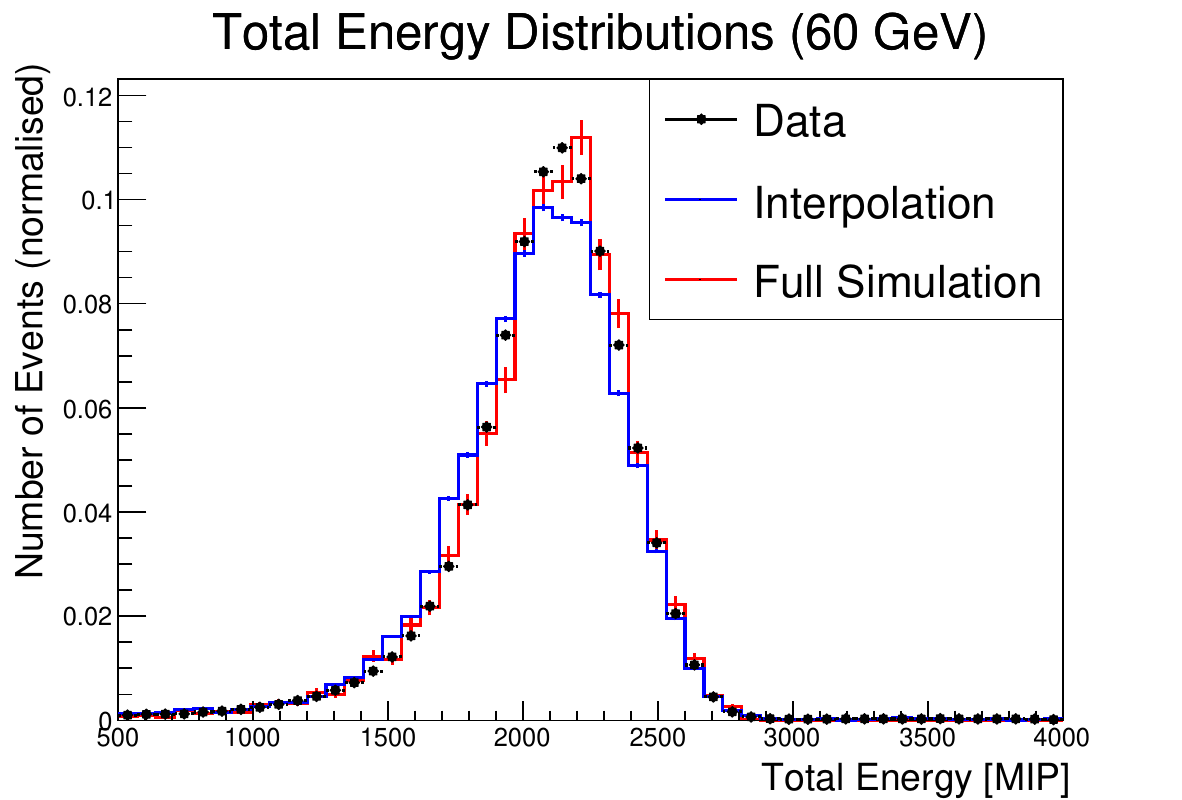}}
    \subfigure[]{\includegraphics[width = 0.49\textwidth]{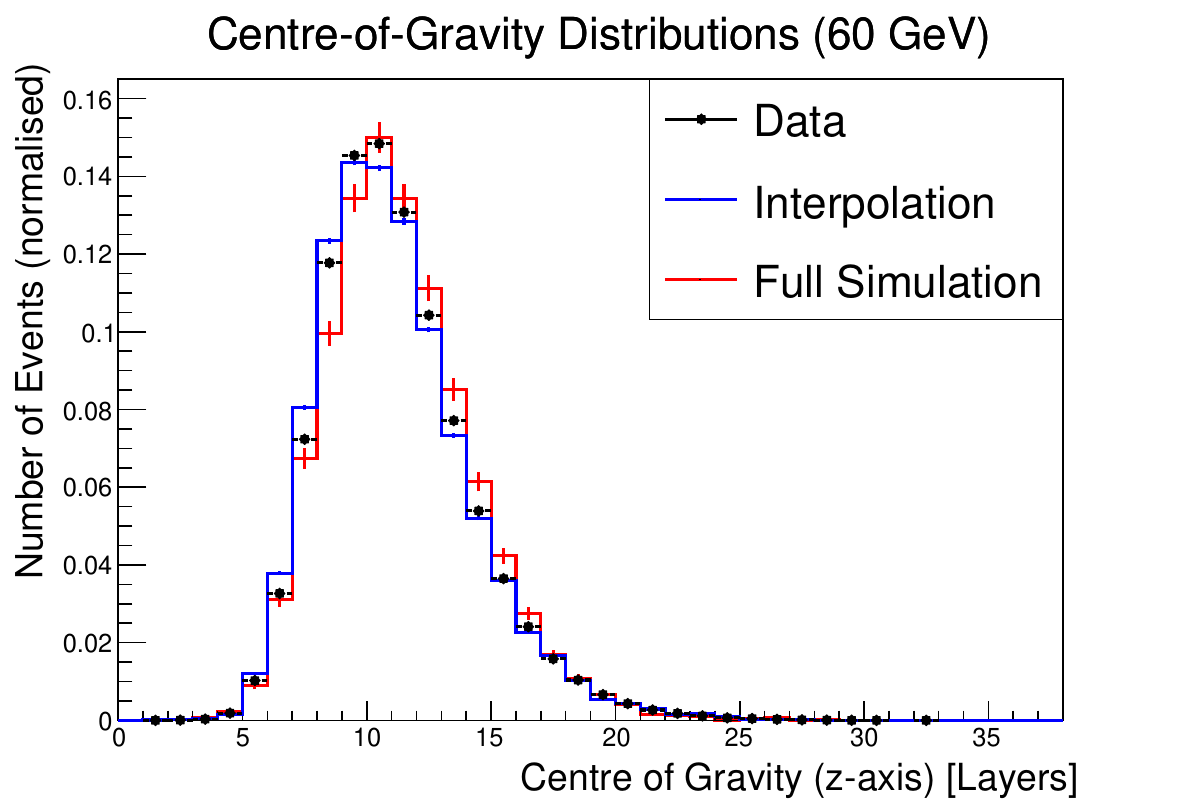}}
    \subfigure[]{\includegraphics[width = 0.49\textwidth]{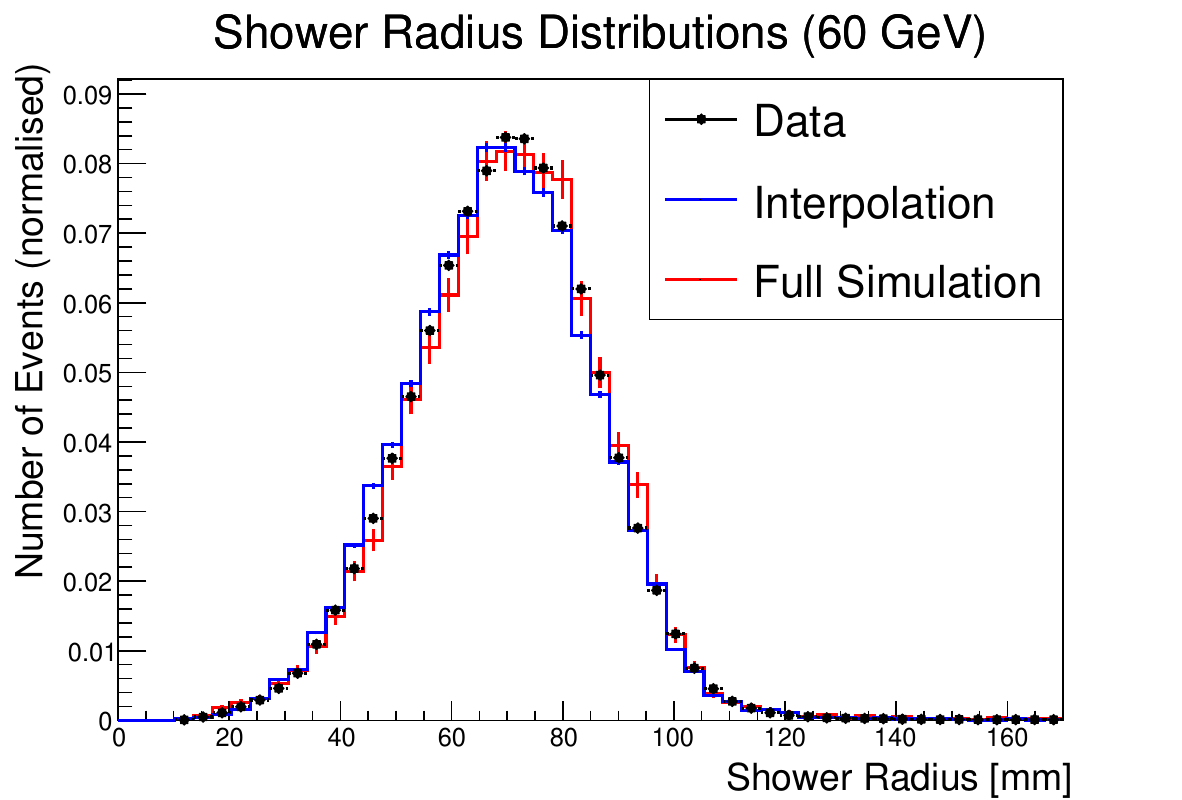}}
    \subfigure[]{\includegraphics[width = 0.49\textwidth]{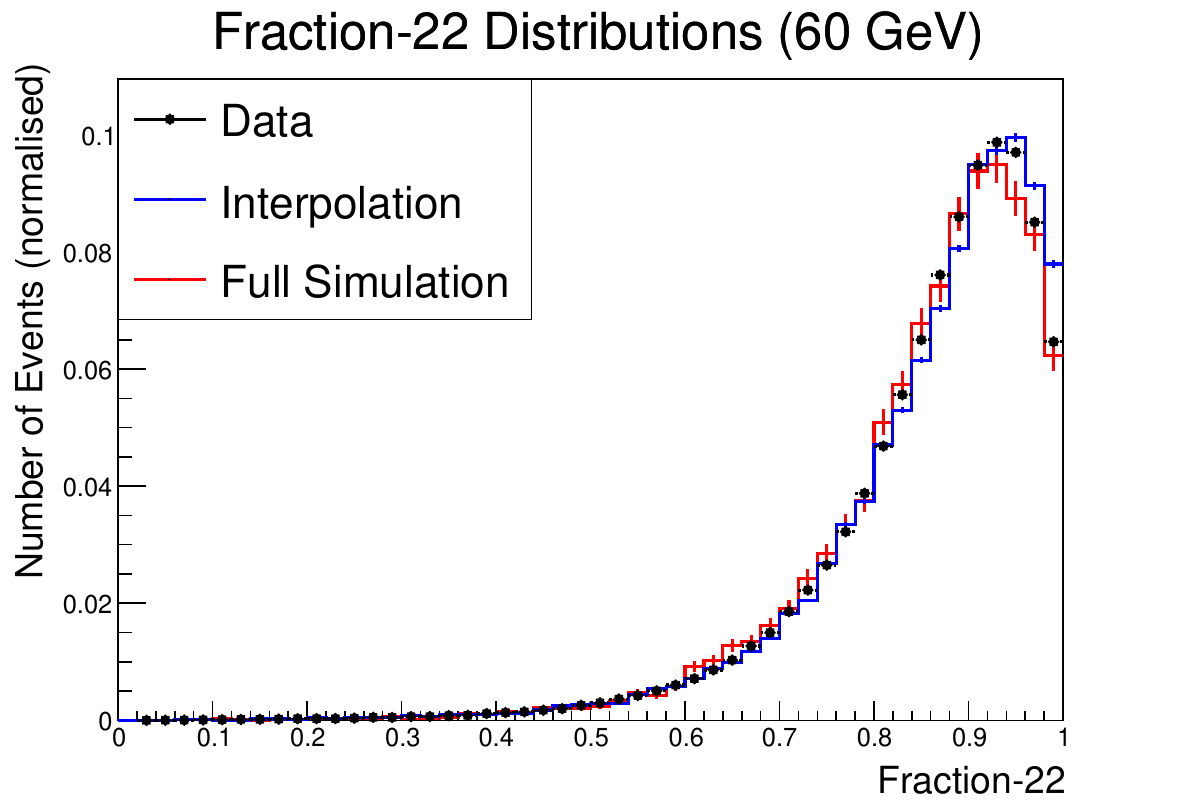}}
    \subfigure[]{\includegraphics[width = 0.49\textwidth]{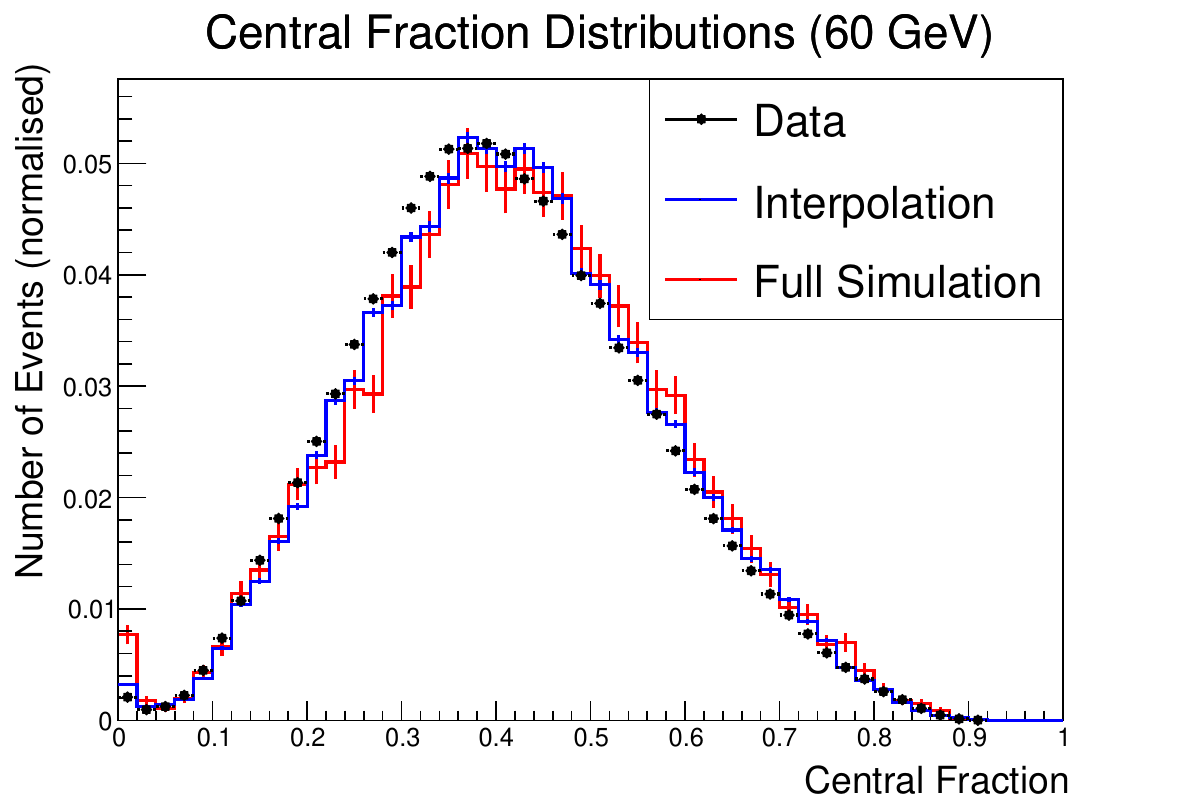}}
    \caption{Distributions of different kinematic shower variables for $\SI{60}{\giga\electronvolt}$ pions. Histograms are shown for (a) the hit energy, (b) the total energy, (c) the CoG along the $z$-axis, (d) the mean shower radius, (e) the energy fraction within the first $22$ layers, and (f) the energy fraction within a cylinder of radius $\SI{30}{\milli\meter}$. Black points represent the complete dataset, dark blue the interpolation, and red curves depict the full simulation. The interpolation agrees very well with data, except for the hit energy and total energy distributions where (small) deviations are visible.}
    \label{fig: kinematic shower variables for equidistant interpolation 60 GeV 1}
\end{figure}
\begin{figure}[hp]
    \centering
    \subfigure[]{\includegraphics[width = 0.49\textwidth]{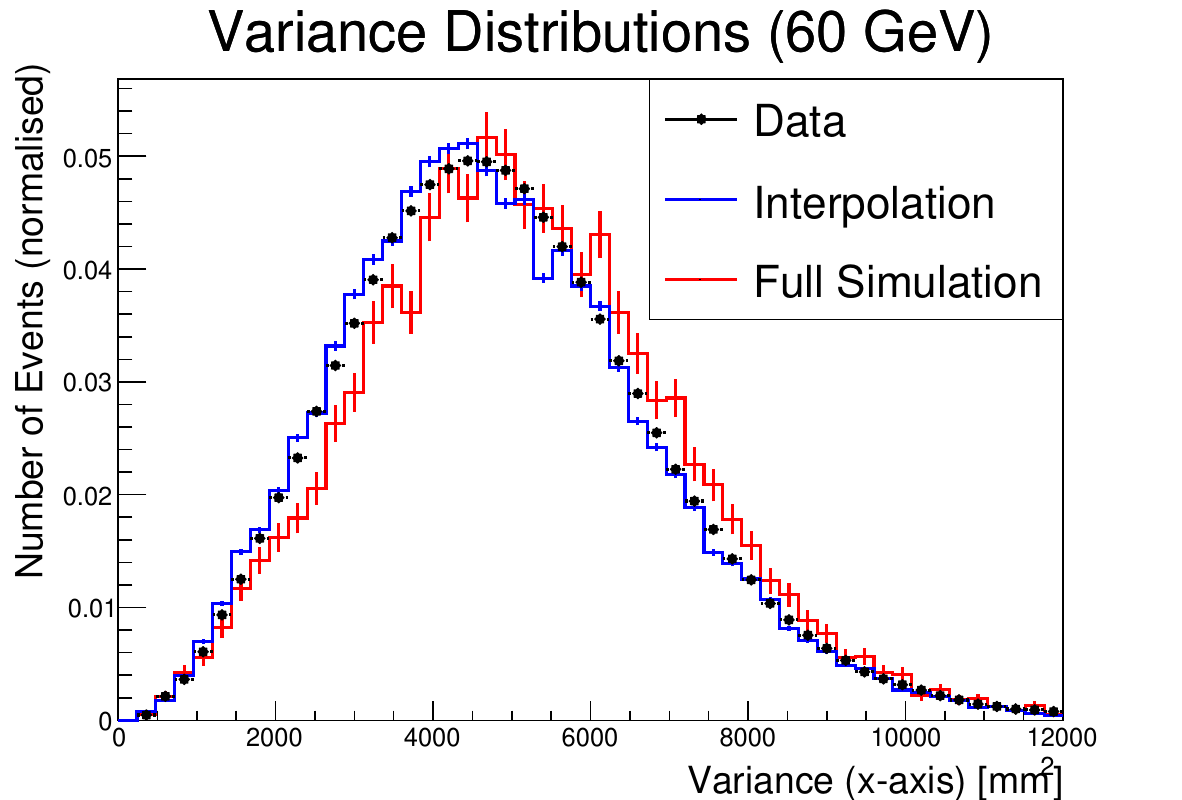}}
    \subfigure[]{\includegraphics[width = 0.49\textwidth]{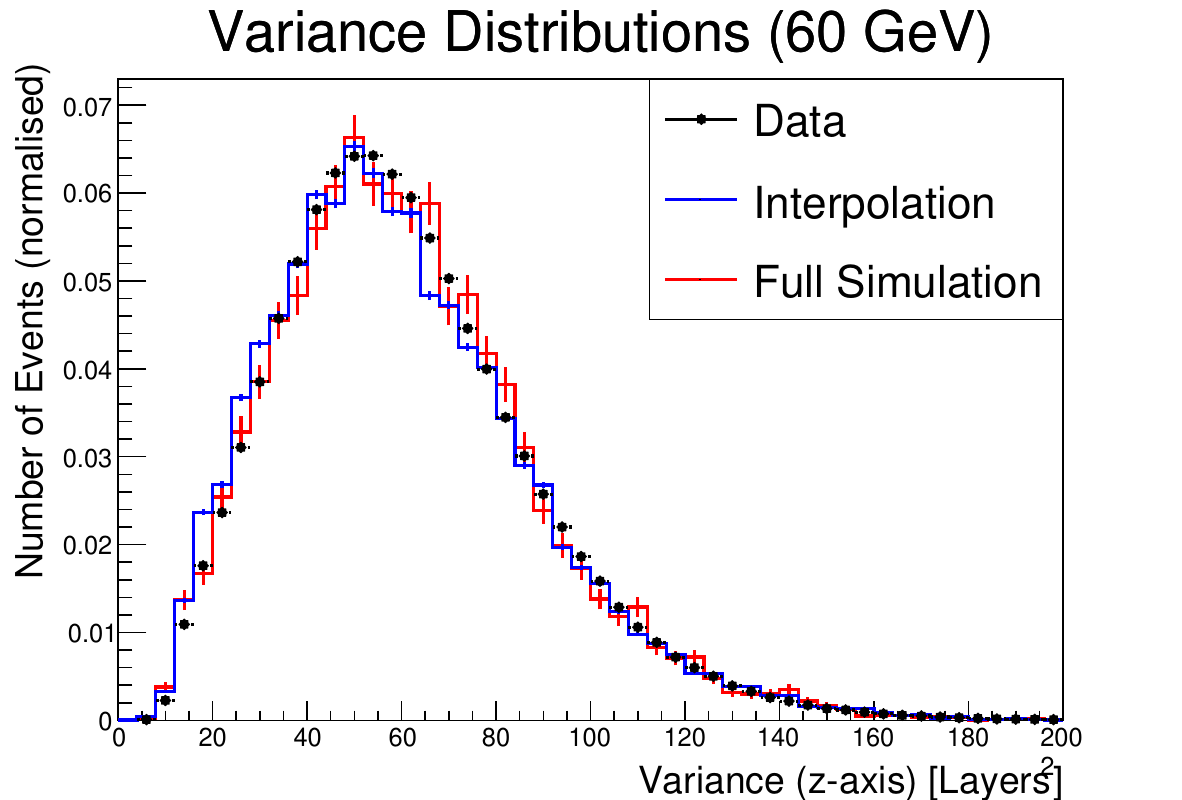}}
    \subfigure[]{\includegraphics[width = 0.49\textwidth]{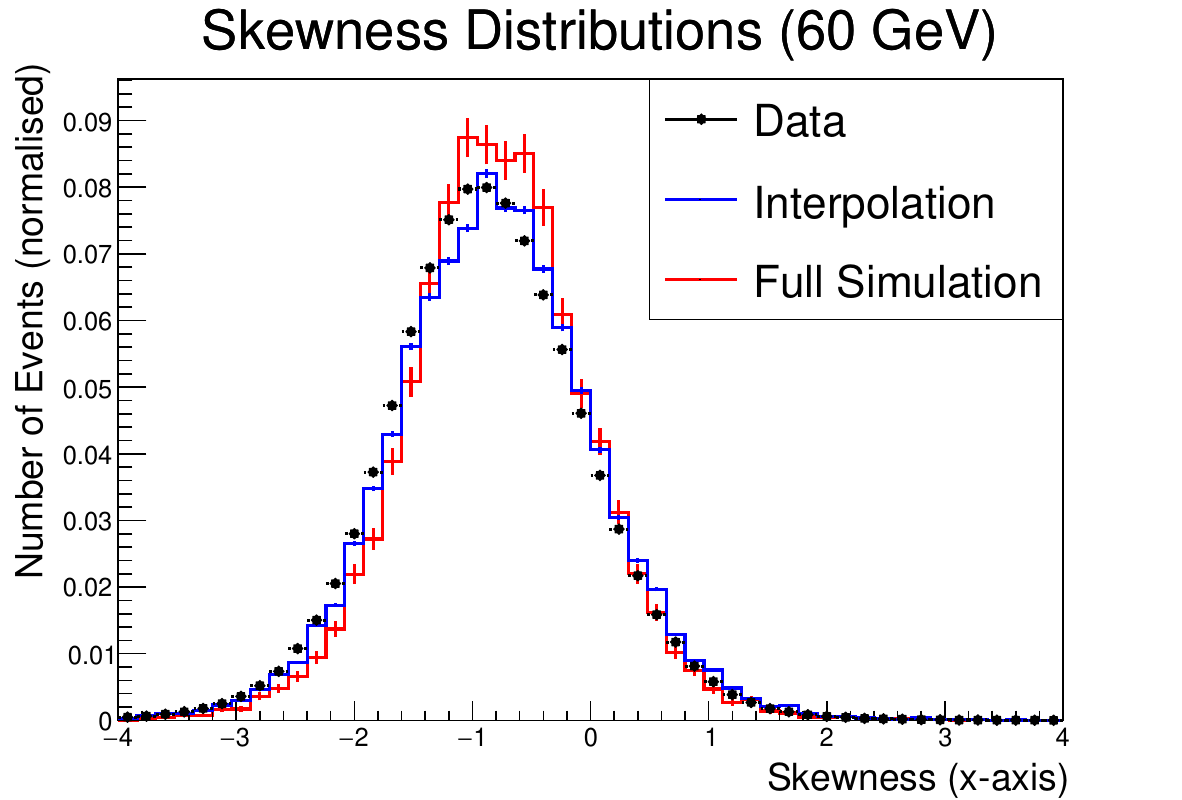}}
    \subfigure[]{\includegraphics[width = 0.49\textwidth]{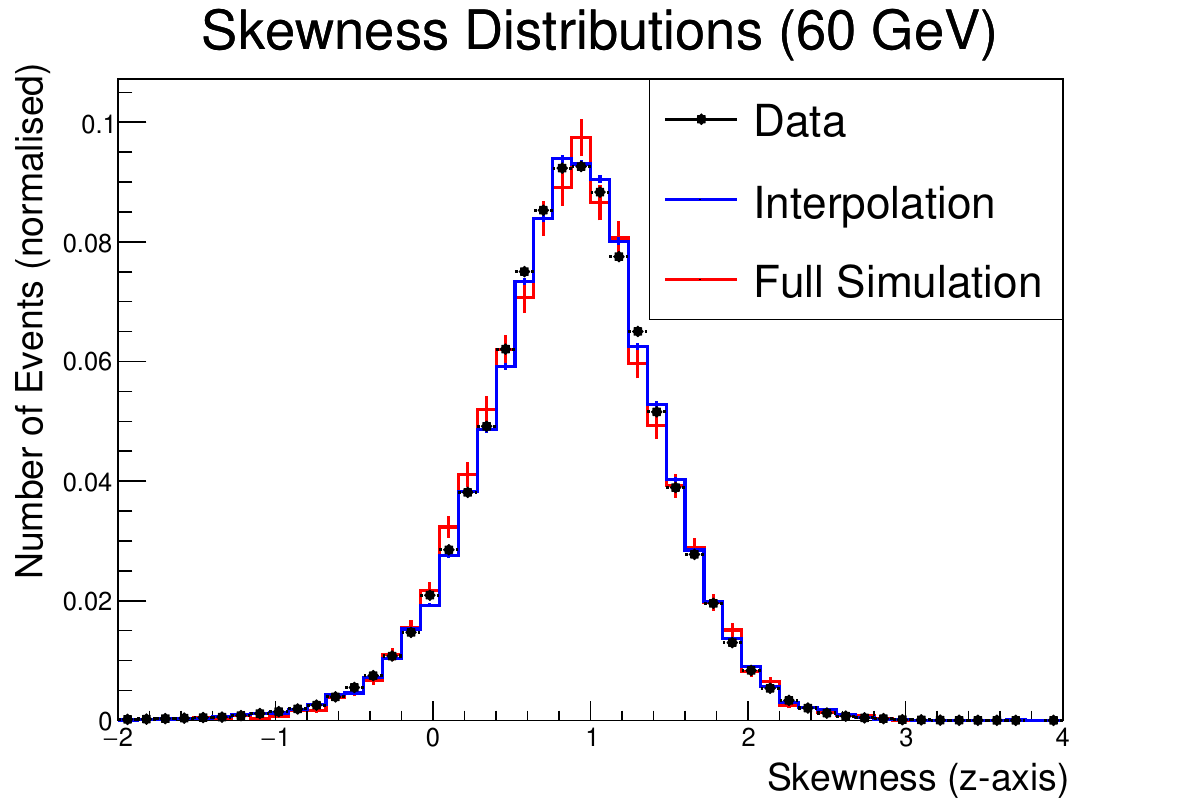}}
    \subfigure[]{\includegraphics[width = 0.49\textwidth]{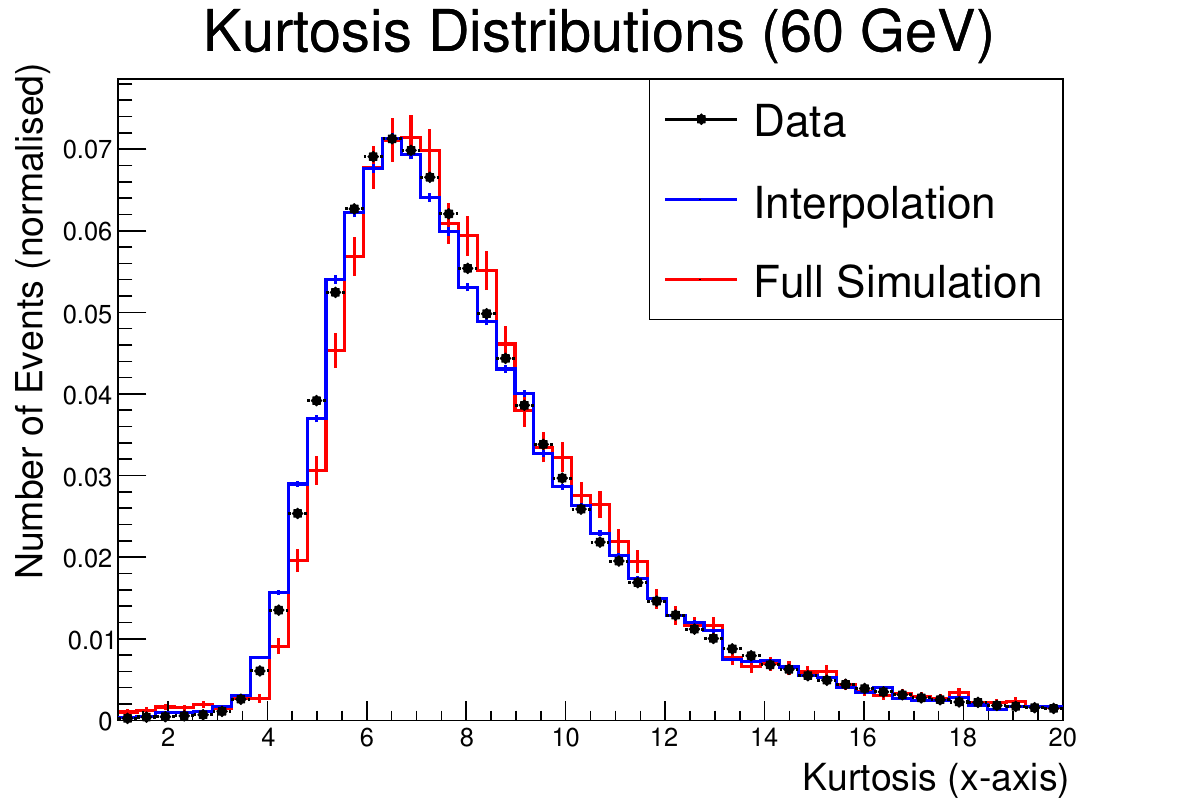}}
    \subfigure[]{\includegraphics[width = 0.49\textwidth]{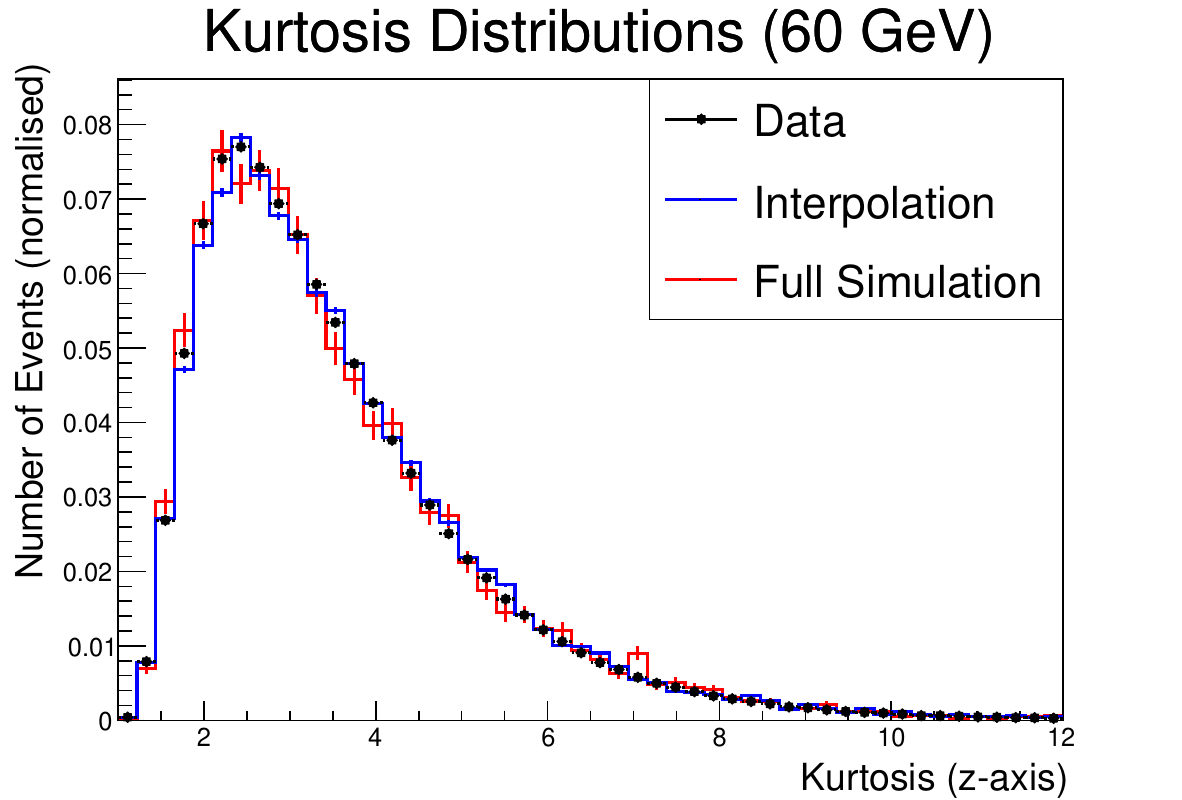}}
    \caption{Distributions of different shower moments for $\SI{60}{\giga\electronvolt}$ pions. The upper row shows histograms for the shower variance, the middle row for the skewness, and the lower row for the kurtosis. Furthermore, the left column depicts all shower moments along the $x$-axis and the right one along the $z$-axis. Black points represent the complete dataset, dark blue the one obtained from interpolation, and red curves depict the full simulation. The interpolation agrees very well with data.}
    \label{fig: kinematic shower variables for equidistant interpolation 60 GeV 2}
\end{figure}
\begin{figure}[hp]
    \centering
    \subfigure[]{\includegraphics[width = 0.49\textwidth]{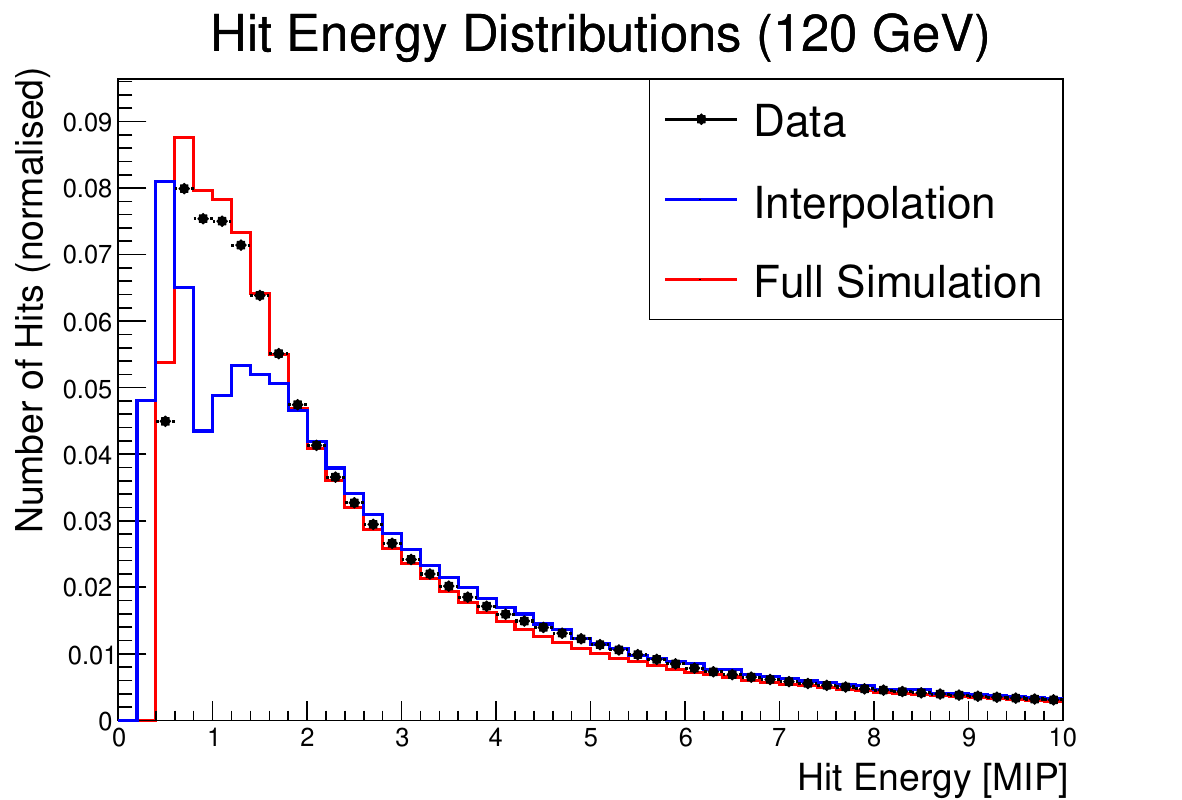}}
    \subfigure[]{\includegraphics[width = 0.49\textwidth]{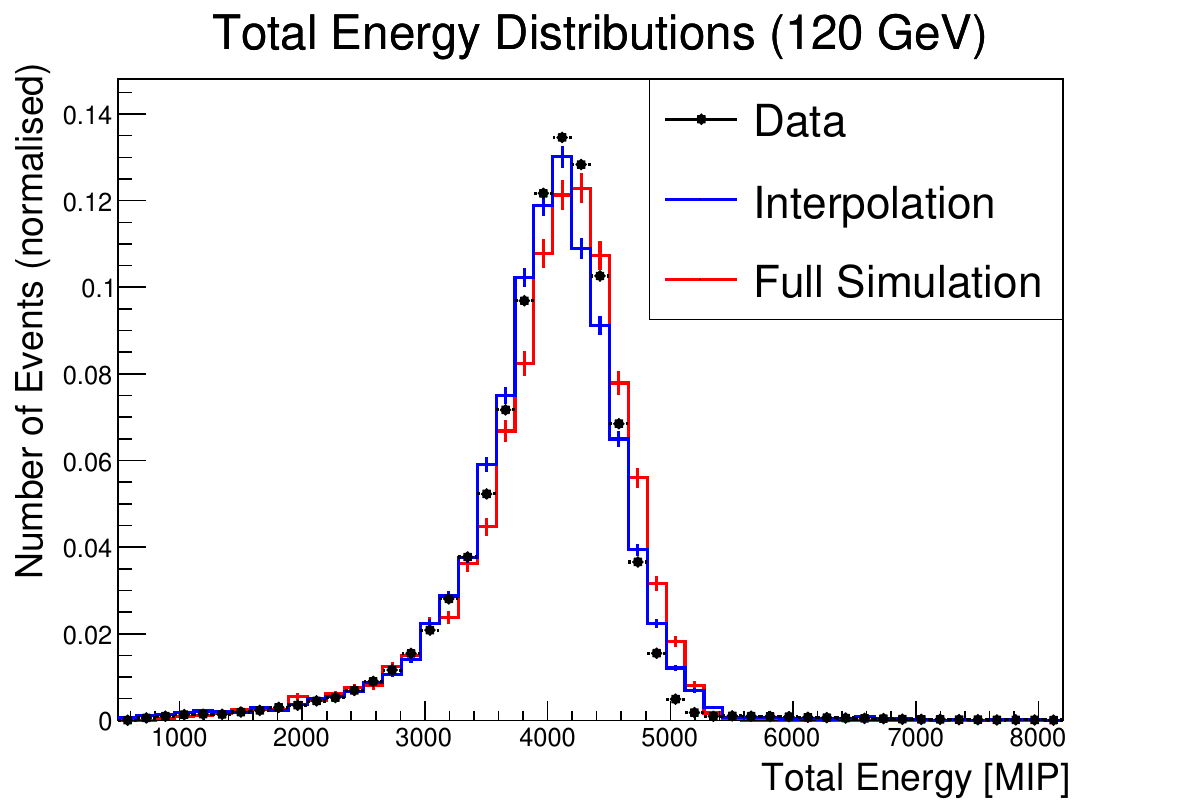}}
    \subfigure[]{\includegraphics[width = 0.49\textwidth]{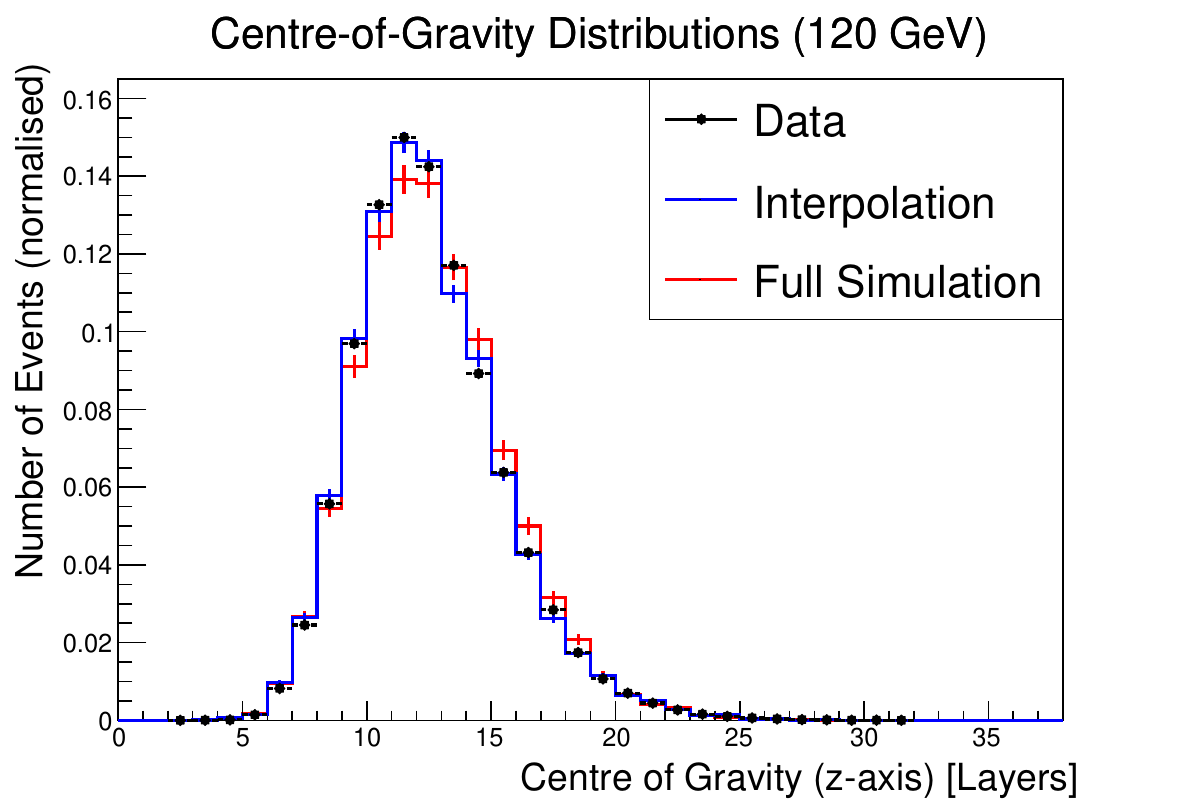}}
    \subfigure[]{\includegraphics[width = 0.49\textwidth]{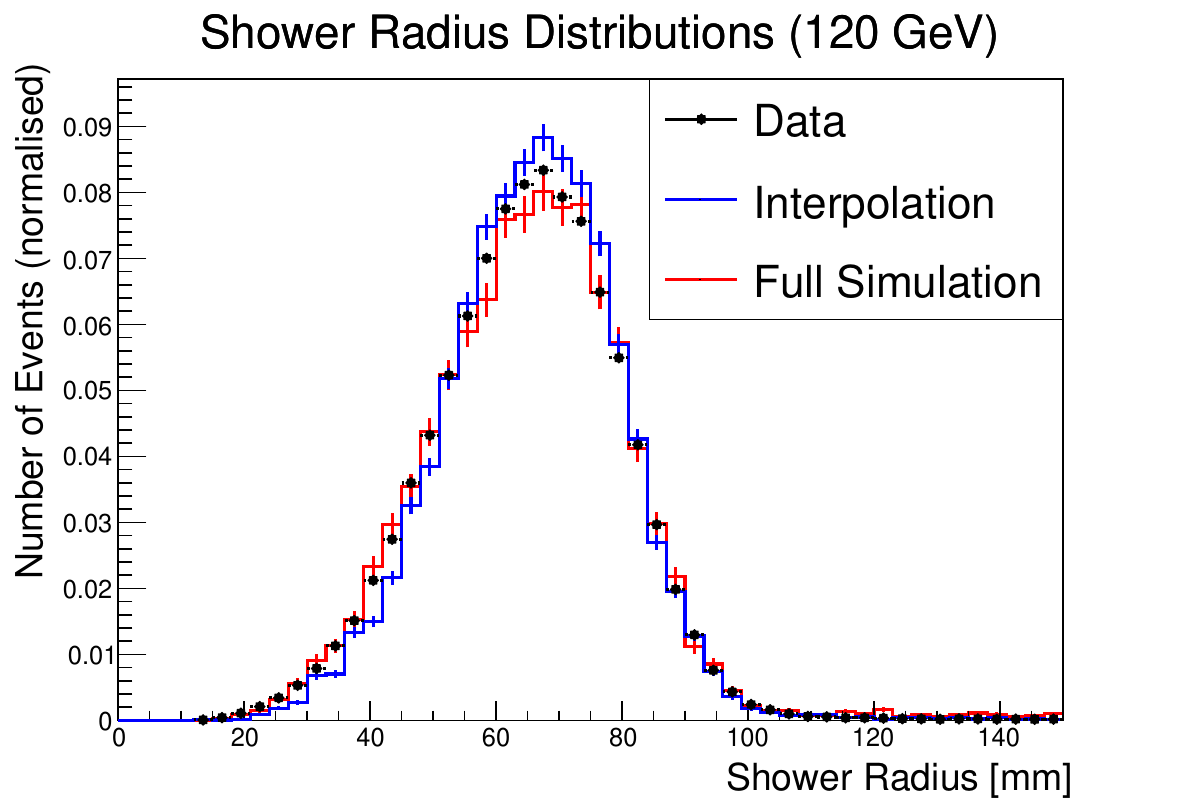}}
    \subfigure[]{\includegraphics[width = 0.49\textwidth]{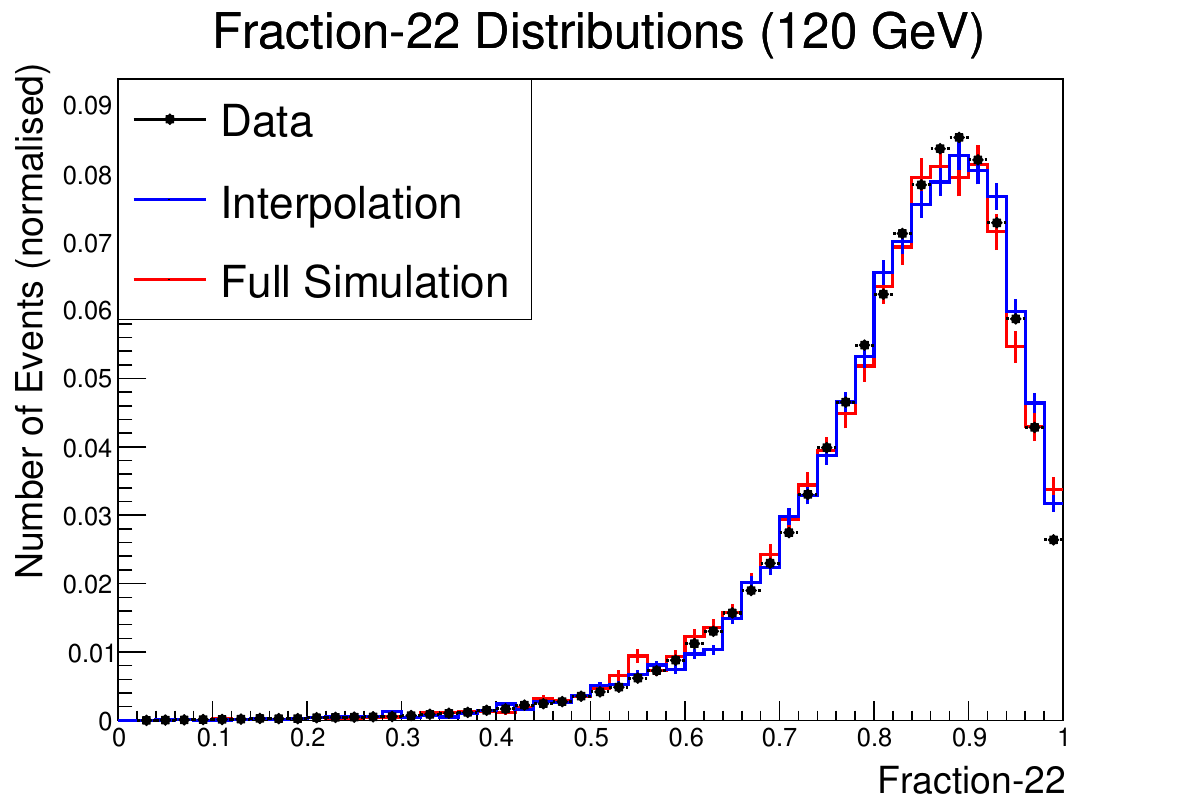}}
    \subfigure[]{\includegraphics[width = 0.49\textwidth]{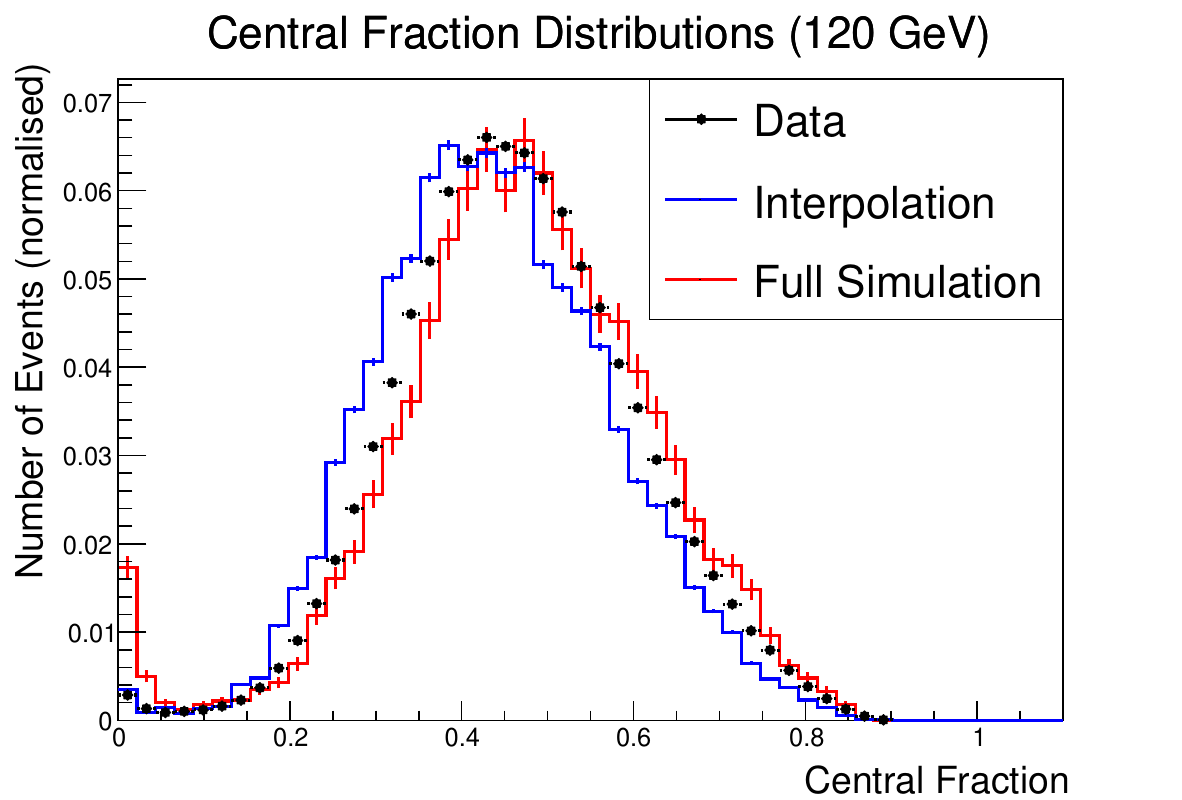}}
    \caption{Distributions of different kinematic shower variables for $\SI{120}{\giga\electronvolt}$ pions. The kinematic variables are the same as shown in Figure \ref{fig: kinematic shower variables for equidistant interpolation 60 GeV 1} and also agree very well with the $\SI{120}{\giga\electronvolt}$ pion dataset, except for the hit energy and the central fraction distributions.}
    \label{fig: kinematic shower variables for equidistant interpolation 120 GeV 1}
\end{figure}
\begin{figure}[hp]
    \centering
    \subfigure[]{\includegraphics[width = 0.49\textwidth]{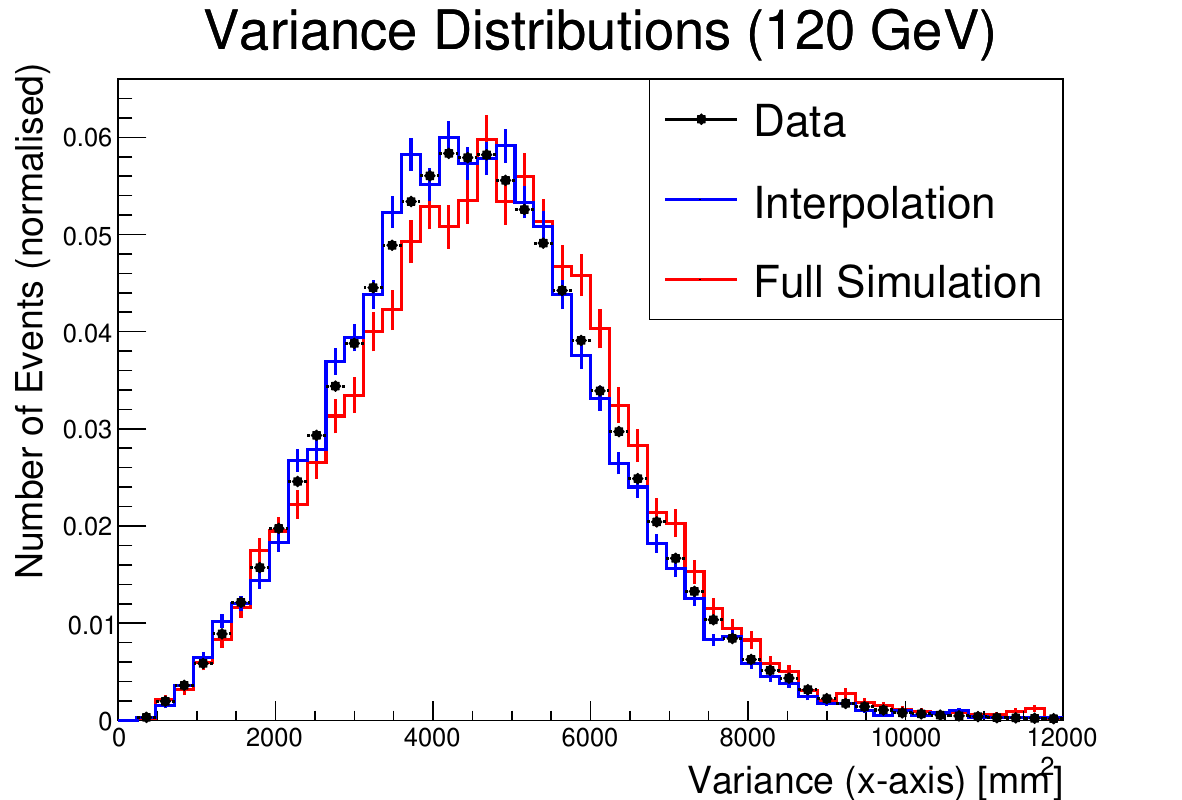}}
    \subfigure[]{\includegraphics[width = 0.49\textwidth]{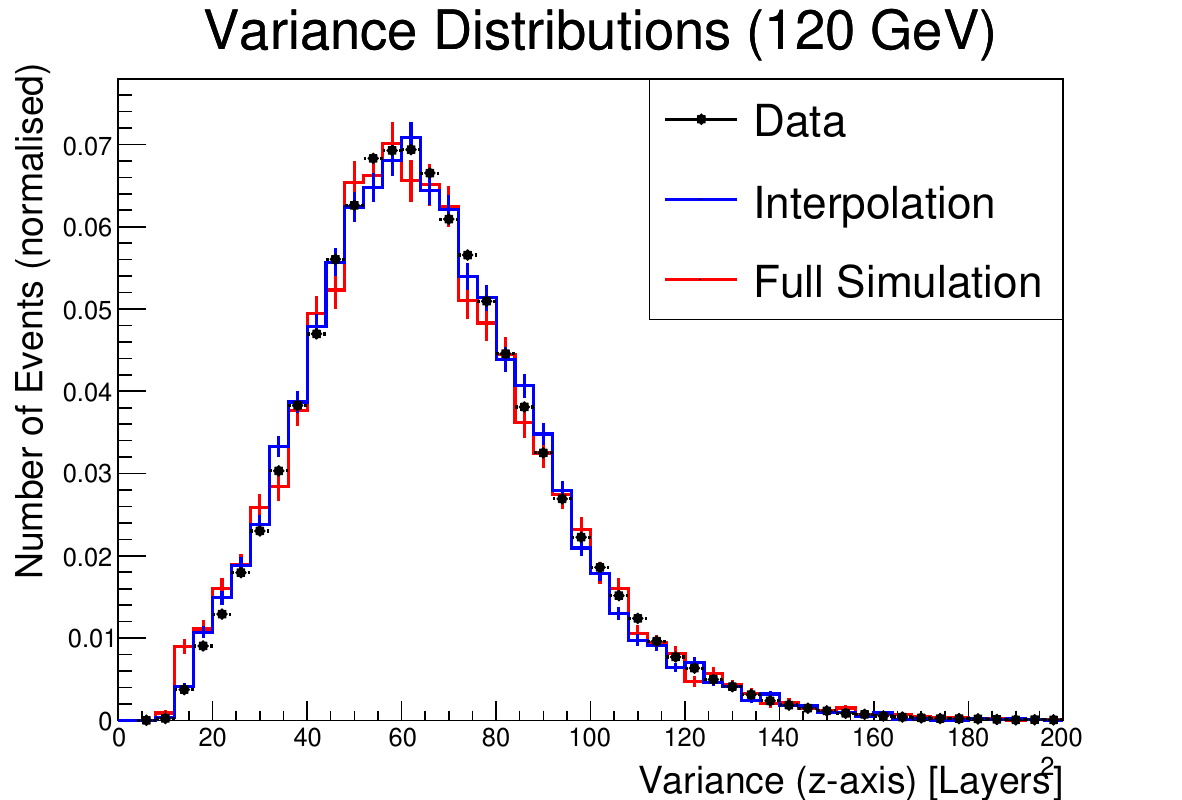}}
    \subfigure[]{\includegraphics[width = 0.49\textwidth]{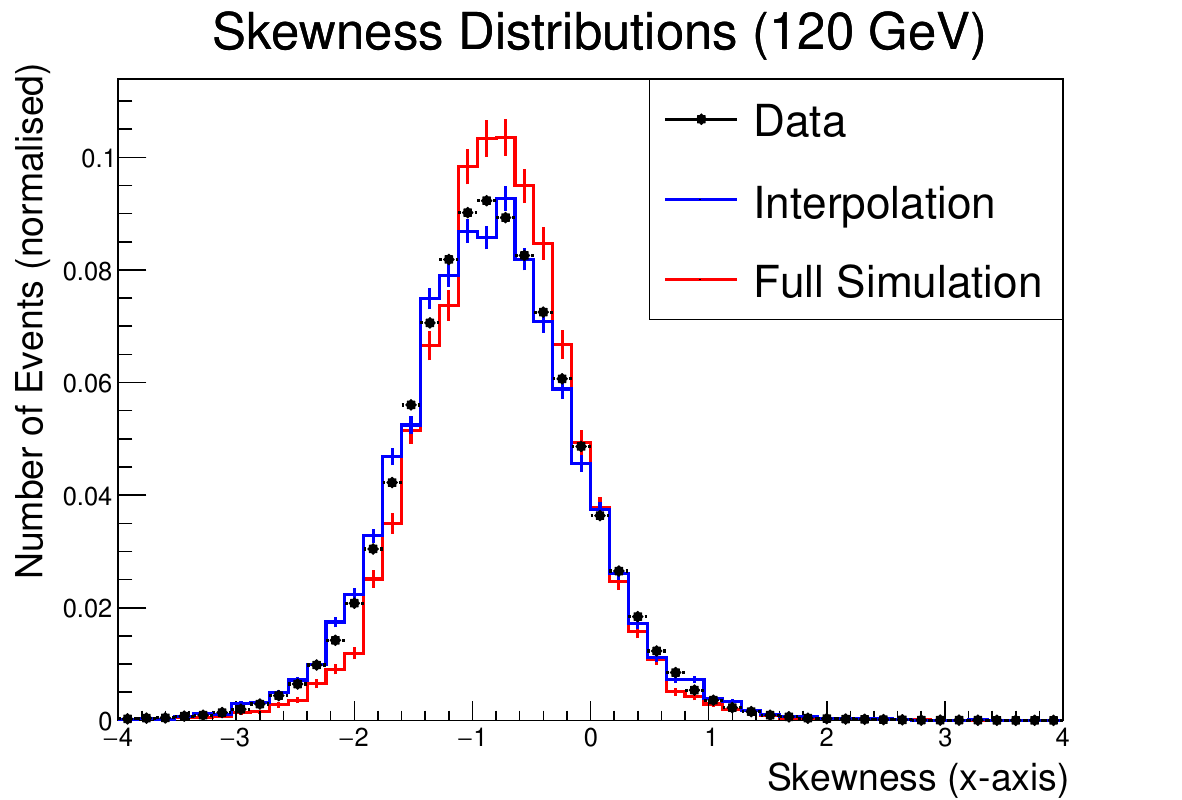}}
    \subfigure[]{\includegraphics[width = 0.49\textwidth]{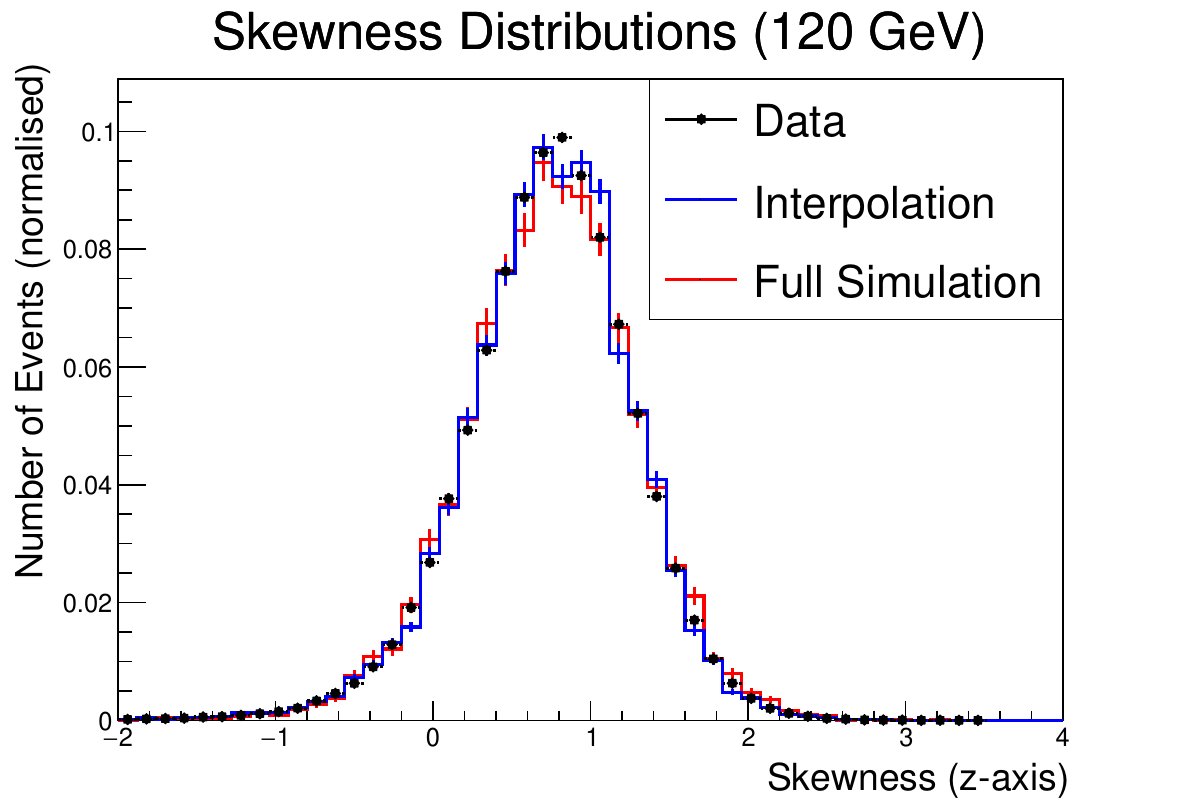}}
    \subfigure[]{\includegraphics[width = 0.49\textwidth]{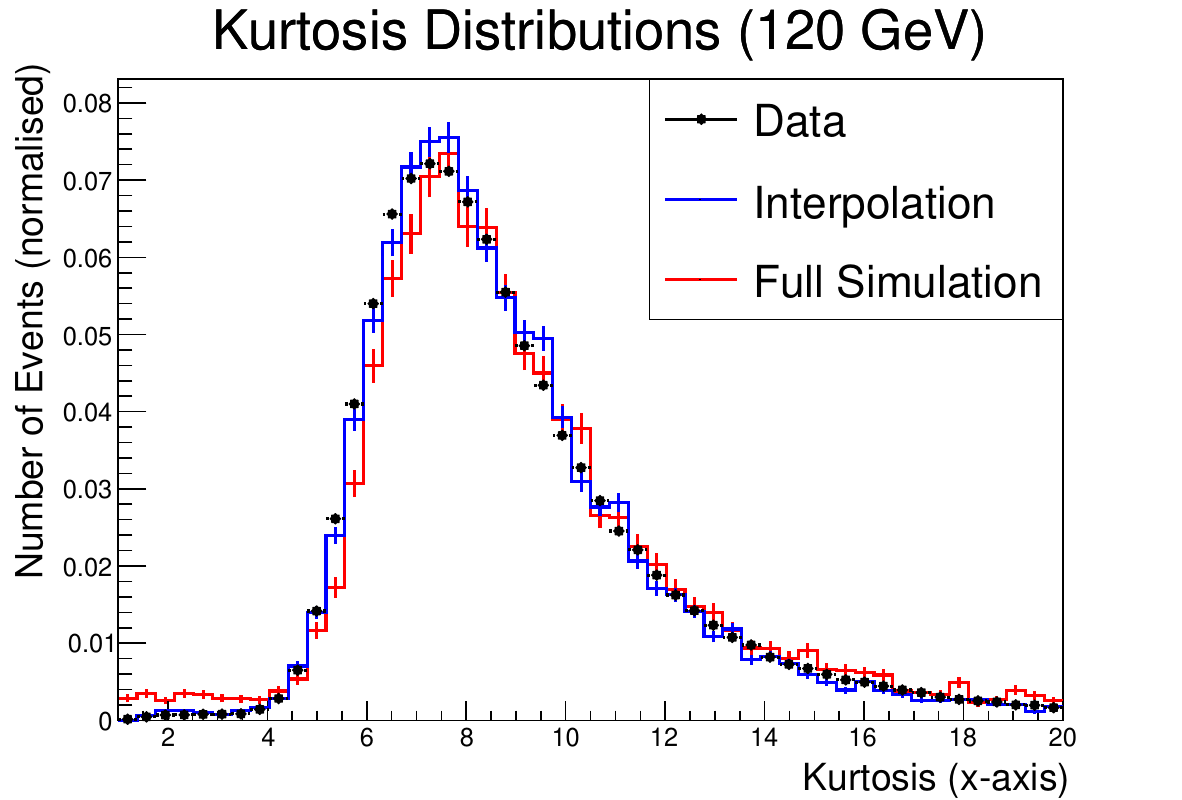}}
    \subfigure[]{\includegraphics[width = 0.49\textwidth]{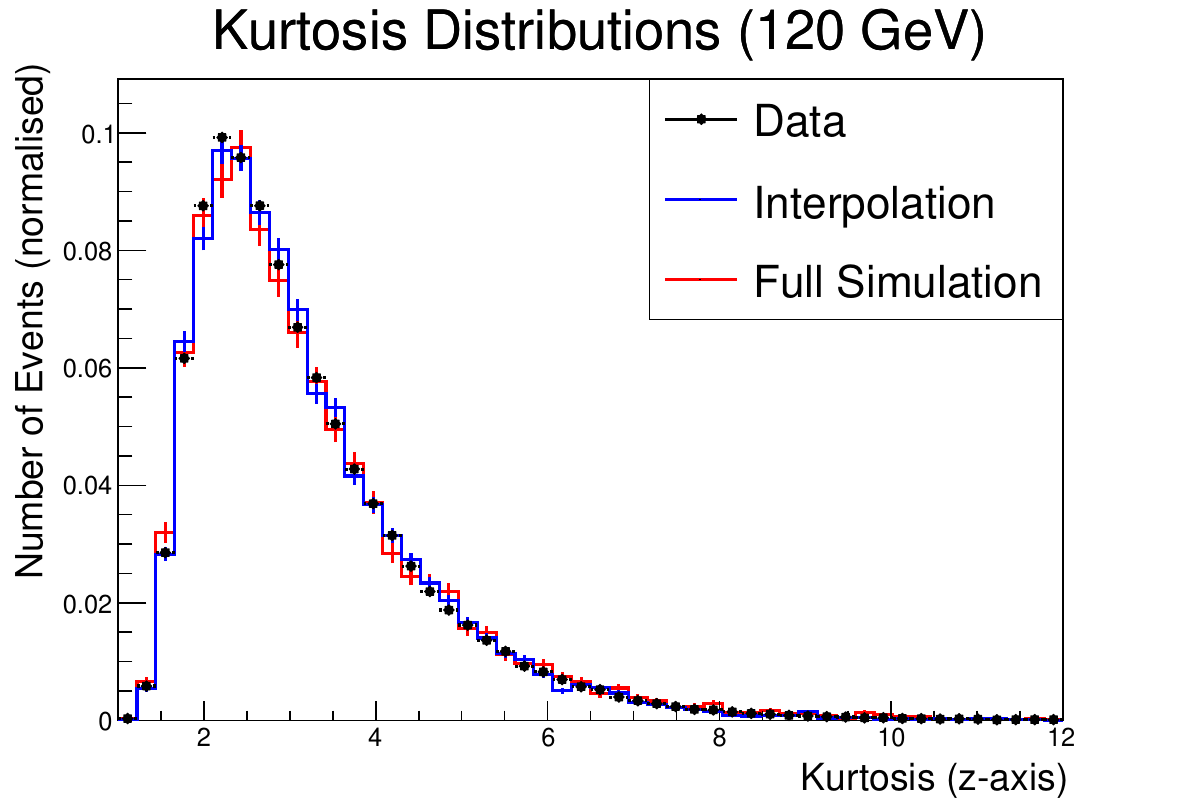}}
    \caption{Distributions of different shower moments for $\SI{120}{\giga\electronvolt}$ pions. The moments are the same as shown in Figure \ref{fig: kinematic shower variables for equidistant interpolation 60 GeV 2} and also agree very well with the $\SI{120}{\giga\electronvolt}$ pion dataset.}
    \label{fig: kinematic shower variables for equidistant interpolation 120 GeV 2}
\end{figure}

From the interpolated datasets described above, distributions of kinematic shower variables were computed which are shown in Figures \ref{fig: kinematic shower variables for equidistant interpolation 60 GeV 1} and \ref{fig: kinematic shower variables for equidistant interpolation 60 GeV 2} for $\SI{60}{\giga\electronvolt}$, as well as in Figures \ref{fig: kinematic shower variables for equidistant interpolation 120 GeV 1} and \ref{fig: kinematic shower variables for equidistant interpolation 120 GeV 2} for $\SI{120}{\giga\electronvolt}$ pions. For the $\SI{60}{\giga\electronvolt}$ case, very good agreement is observed between data and interpolation for most of the kinematic variables, with the exception of the hit energy and total energy distributions. The interpolated hit energy distribution does not fully match the low-energy region observed in data, but agrees well with data at higher hit energies. Conversely, the maximum of the interpolated total energy distribution lies at the expected position, but the curve peaks less sharply and is smaller in comparison to data. For $\SI{120}{\giga\electronvolt}$ pions, very good agreement is also the case of the majority of all distributions shown in Figures \ref{fig: kinematic shower variables for equidistant interpolation 120 GeV 1} and \ref{fig: kinematic shower variables for equidistant interpolation 120 GeV 2}, except for the hit energy and total energy distributions. Here, the interpolation faces the same problems as for $\SI{60}{\giga\electronvolt}$ pions. Furthermore, the maximum of the central fraction PDF is slightly shifted towards smaller values with respect to its expectations.
\par
\begin{figure}[ht]
    \centering
    \subfigure[]{\includegraphics[width = 0.49\textwidth]{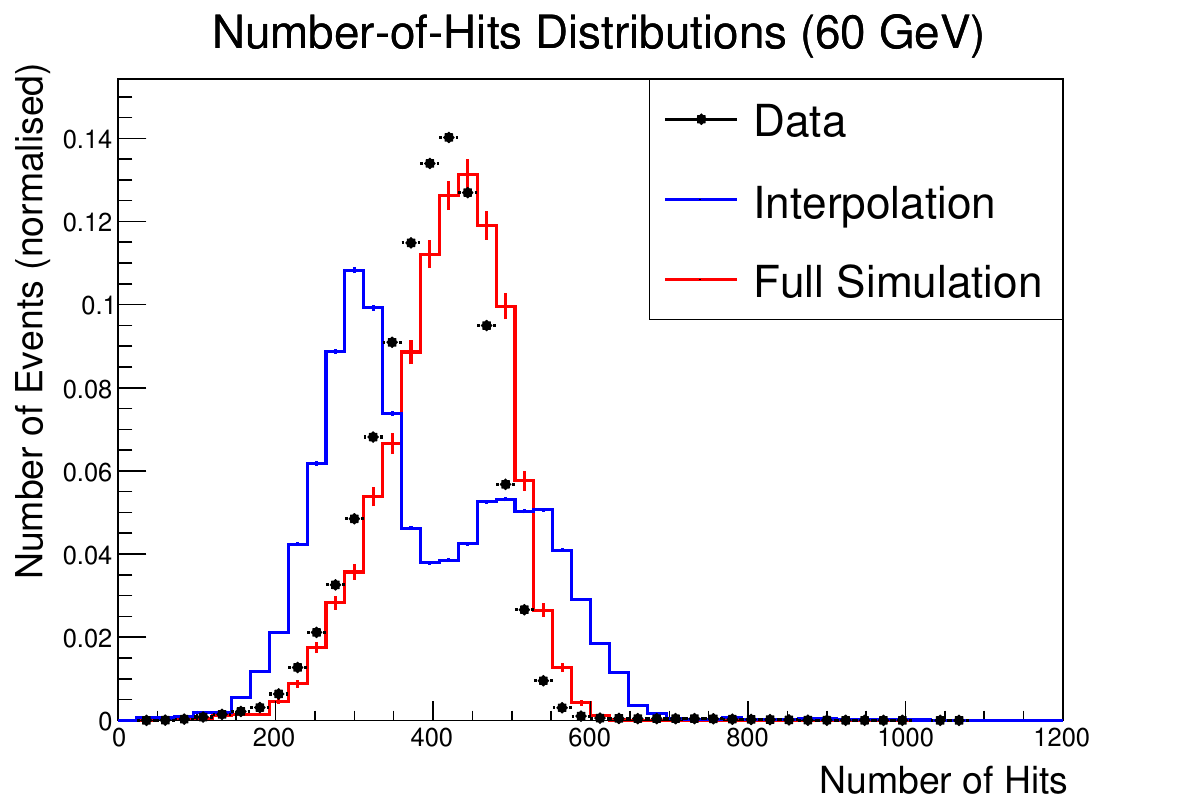}}
    \subfigure[]{\includegraphics[width = 0.49\textwidth]{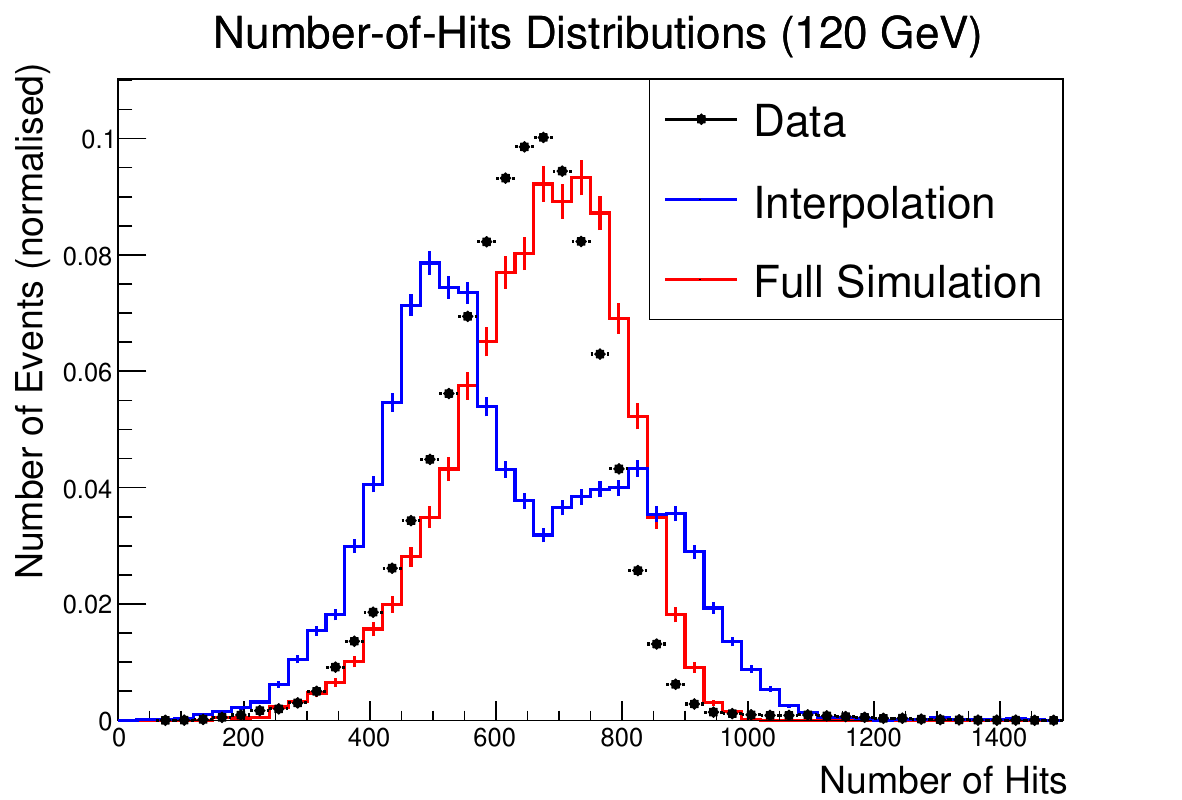}}
    \caption{Distributions for the number of hits per event for $\SI{60}{\giga\electronvolt}$ (left) and $\SI{120}{\giga\electronvolt}$ (right) pions. Black points represent the complete datasets, dark blue the ones obtained from interpolation, and the red curves depict the full simulation. There are significant deviations visible between the interpolation and the data.}
    \label{fig: kinematic shower variables for equidistant interpolation 3}
\end{figure}

There is, however, significant disagreement between the data and the interpolation curves of the number of hits in Figure \ref{fig: kinematic shower variables for equidistant interpolation 3}. While the data curves each display a single maximum, the interpolation curves exhibit two distinct peaks. Moreover, the interpolation peaks are broader and their amplitudes smaller than expected. The left peak originates from the interpolation of events at $E^{\text{small}}$, whereas the right one comes from the interpolation of those at $E^{\text{large}}$. These results suggest that the interpolation fails to reproduce the correct number of hits, which is most likely due to the implementation of the interpolation algorithm.
\par
A similar performance of the interpolation algorithm was also found for non-equidistant approaches, i.e. when one of the reference energies lies closer to the target energy than the other. In these cases, the degree of agreement between fast simulation and data is on a comparable level to the results presented above, demonstrating that the distance between the reference and target energies has no impact on the final results. It was, however, also found that it is more advantageous for the interpolation to operate on reference energies that are as close to the target energy as possible.

\subsection{Interpolated Correlations between Kinematic Shower Variables and Computational Requirements}
\label{subsec: interpolated correlations between kinematic shower variables}

\begin{figure}[hp]
    \centering
    \subfigure[]{\includegraphics[width = 0.49\textwidth, page = 1]{figures/hitlevel/60GeV/CorrelationPlots/KinematicVariables/TotalEnergyCorrelations_Data.pdf}}
    \subfigure[]{\includegraphics[width = 0.49\textwidth, page = 1]{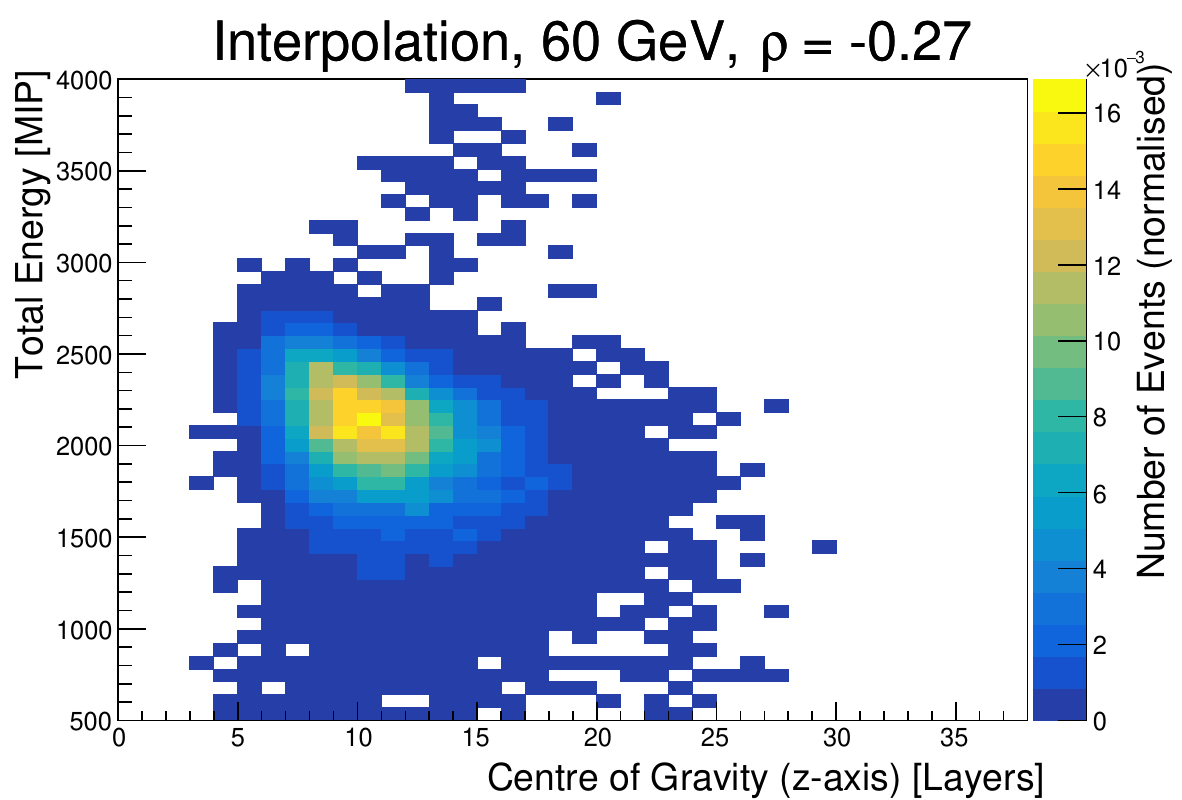}}
    \subfigure[]{\includegraphics[width = 0.49\textwidth, page = 2]{figures/hitlevel/60GeV/CorrelationPlots/KinematicVariables/TotalEnergyCorrelations_Data.pdf}}
    \subfigure[]{\includegraphics[width = 0.49\textwidth, page = 2]{figures/interpolation/40GeV_80GeV/CorrelationPlots/KinematicVariables/TotalEnergyCorrelations_Interpolation.pdf}}
    \subfigure[]{\includegraphics[width = 0.49\textwidth, page = 3]{figures/hitlevel/60GeV/CorrelationPlots/KinematicVariables/TotalEnergyCorrelations_Data.pdf}}
    \subfigure[]{\includegraphics[width = 0.49\textwidth, page = 3]{figures/interpolation/40GeV_80GeV/CorrelationPlots/KinematicVariables/TotalEnergyCorrelations_Interpolation.pdf}}
    \caption{Two-dimensional correlation plots for $\SI{60}{\giga\electronvolt}$ pion showers between the total energy and either the CoG$_{z}$ (top row), the shower radius (middle row), or the central fraction (bottom row). All histograms are shown for data (left column) and interpolation (right column). For each row, very good agreement is visible between data and interpolation.}
    \label{fig: correlations total energy vs. kinematic variables interpolation 60 GeV}
\end{figure}
\begin{figure}[hp]
    \centering
    \subfigure[]{\includegraphics[width = 0.49\textwidth, page = 1]{figures/hitlevel/120GeV/CorrelationPlots/KinematicVariables/TotalEnergyCorrelations_Data.pdf}}
    \subfigure[]{\includegraphics[width = 0.49\textwidth, page = 1]{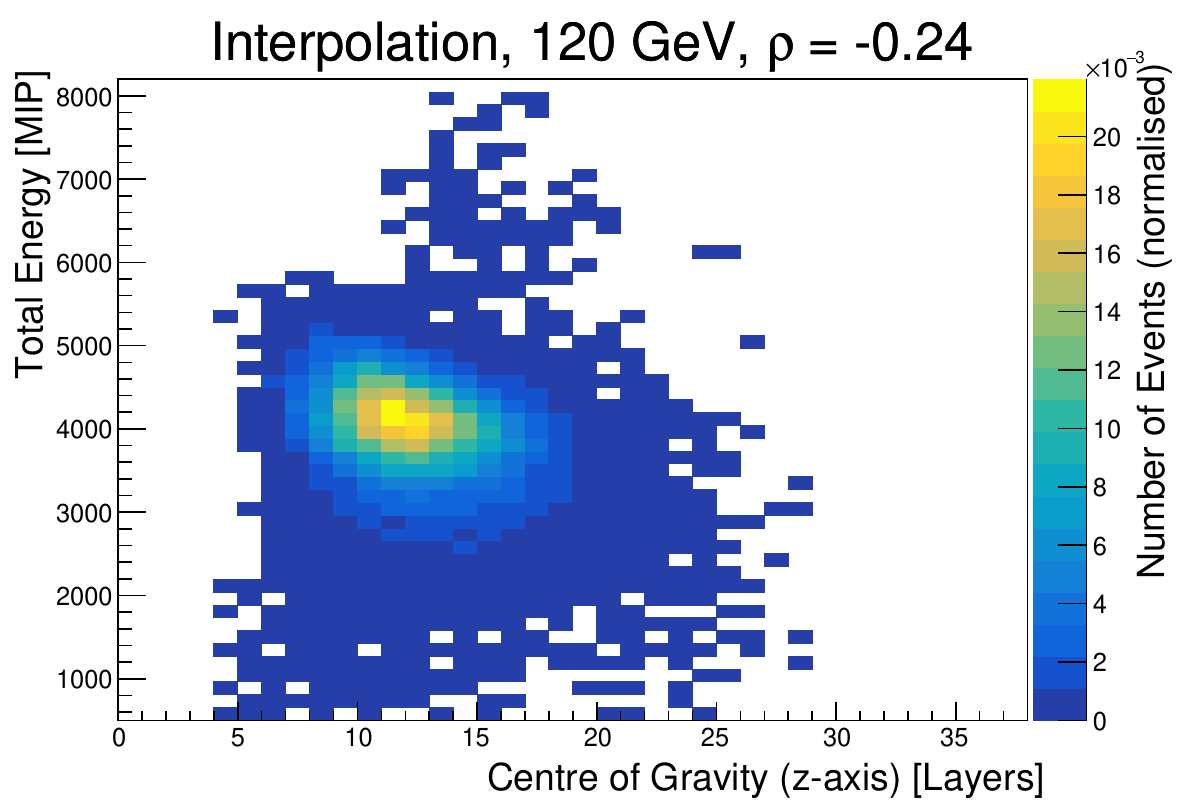}}
    \subfigure[]{\includegraphics[width = 0.49\textwidth, page = 2]{figures/hitlevel/120GeV/CorrelationPlots/KinematicVariables/TotalEnergyCorrelations_Data.pdf}}
    \subfigure[]{\includegraphics[width = 0.49\textwidth, page = 2]{figures/interpolation/120GeV/CorrelationPlots/KinematicVariables/TotalEnergyCorrelations_Interpolation.pdf}}
    \subfigure[]{\includegraphics[width = 0.49\textwidth, page = 3]{figures/hitlevel/120GeV/CorrelationPlots/KinematicVariables/TotalEnergyCorrelations_Data.pdf}}
    \subfigure[]{\includegraphics[width = 0.49\textwidth, page = 3]{figures/interpolation/120GeV/CorrelationPlots/KinematicVariables/TotalEnergyCorrelations_Interpolation.pdf}}
    \caption{Two-dimensional correlation plots for $\SI{120}{\giga\electronvolt}$ pion showers between the total energy and either the CoG$_{z}$ (top row), the shower radius (middle row), or the central fraction (bottom row). All histograms are shown for data (left column) and interpolation (right column). For each row, very good agreement is visible between data and interpolation.}
    \label{fig: correlations total energy vs. kinematic variables interpolation 120 GeV}
\end{figure}
\begin{figure}[ht]
    \centering
    \includegraphics[width = 1\textwidth]{figures/hitlevel/60GeV/CorrelationPlots/KinematicVariables/CorrelationMatrixKinematicVariables_Data_withValues.pdf}
    \includegraphics[width = 1\textwidth]{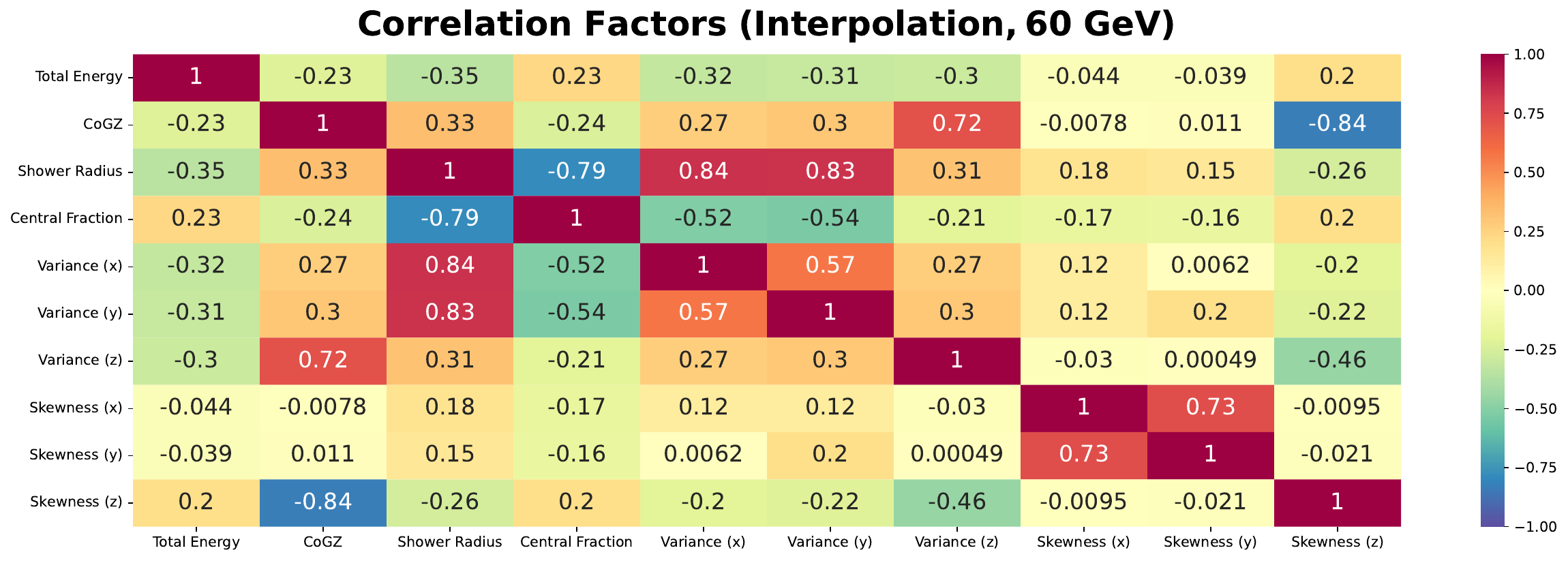}
    \caption{Correlation matrices for kinematic shower variables obtained from data (upper plot) and interpolation (lower plot) for $\SI{60}{\giga\electronvolt}$ pions. Dark red colouring represents strong correlation between two variables, whereas dark blue indicates strong anticorrelation. The matrices are symmetric about their diagonals and agree very well with each other.}
    \label{fig: correlation matrices data and interpolation 60 GeV}
\end{figure}
\begin{figure}[ht]
    \centering
    \includegraphics[width = 1\textwidth]{figures/hitlevel/120GeV/CorrelationPlots/KinematicVariables/CorrelationMatrixKinematicVariables_Data_withValues.pdf}
    \includegraphics[width = 1\textwidth]{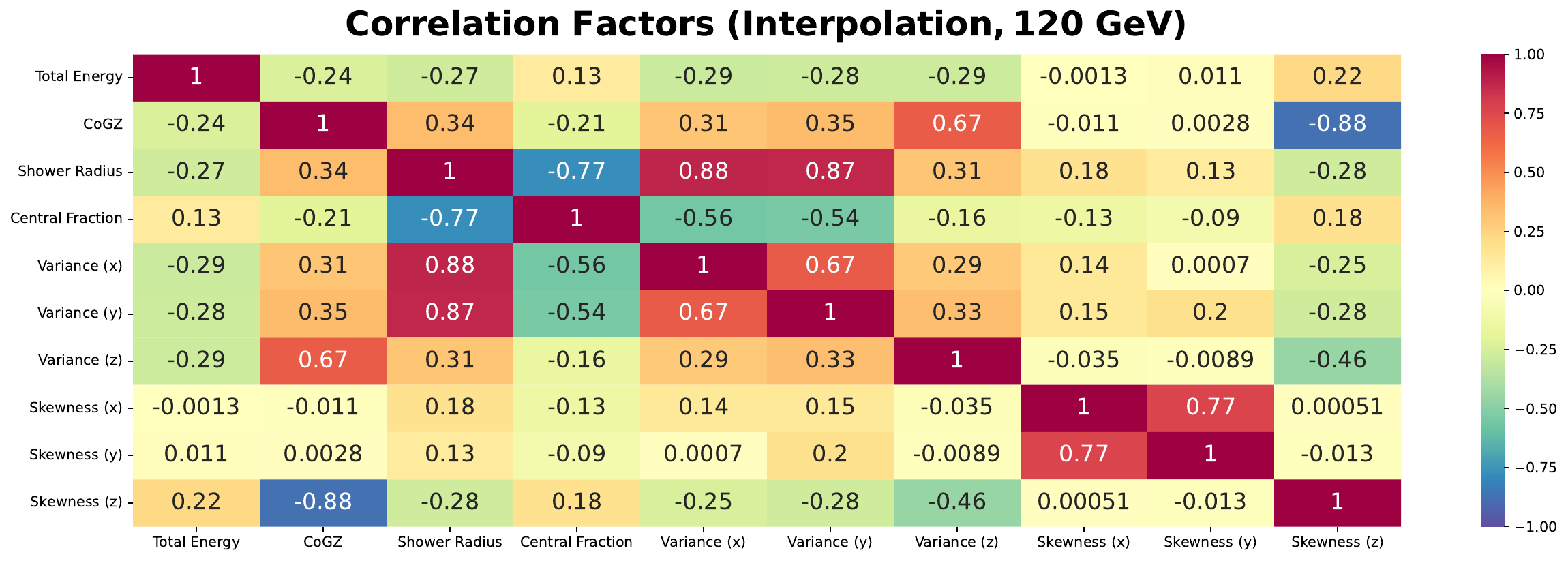}
    \caption{Correlation matrices for kinematic shower variables obtained from data (upper plot) and interpolation (lower plot) for $\SI{120}{\giga\electronvolt}$ pions. Dark red colouring represents strong correlation between two variables, whereas dark blue indicates strong anticorrelation. The matrices are symmetric about their diagonals and agree very well with each other.}
    \label{fig: correlation matrices data and interpolation 120 GeV}
\end{figure}

Similar to Section \ref{subsec: simulated correlations between kinematic shower variables}, one must also compare interpolated correlation factors between different kinematic variables with data. For this, two-dimensional histograms are shown in Figures \ref{fig: correlations total energy vs. kinematic variables interpolation 60 GeV} and \ref{fig: correlations total energy vs. kinematic variables interpolation 120 GeV}. Here, correlation factors between the total energy and other kinematic variables are displayed. The interpolation successfully predicts the correct (anti-)correlations for the total energy, with the largest difference being $\SI{0.05}{}$ at $\SI{60}{\giga\electronvolt}$, and the interpolated correlation matrices in Figures \ref{fig: correlation matrices data and interpolation 60 GeV} and \ref{fig: correlation matrices data and interpolation 120 GeV} also exhibit only small correlation differences overall. In summary, the interpolation algorithm achieves excellent agreement between distributions of interpolated, simulated pion showers and data, except for the number of hits per event, and correctly predicts the linear correlation factors between kinematic variables.
\par
The average computing times for the equidistant interpolation of $\SI{60}{\giga\electronvolt}$ pion showers are listed in Table \ref{tab: computing times for hit level interpolation}. Here, the total times are shown in the first and second row for two and $\SI{20000}{}$ events, respectively. These times, however, include generating the integration limits, computing cumulative distributions, as well as the actual interpolation itself. Since the former two steps can be easily outsourced from the interpolation, the times it takes to only interpolate two and $\SI{20000}{}$ events, respectively, are also given in Table \ref{tab: computing times for hit level interpolation} (rows three and four). Similar results are also obtained for the other test beam energies. In addition, the computation times obtained for full simulation are listed for comparison. Similar to the previous section, the interpolation exceeds the full simulation by a factor of $\mathcal{O}(1000)$.
\begin{table}[ht]
    \centering
    \caption{Average computing times for the interpolation of simulated $\SI{60}{\giga\electronvolt}$ pion showers. For interpolating either two or $\SI{20000}{}$ events, the first two rows show the total algorithm runtime and rows three and four only show the required time for the interpolation itself, excluding the integration limit generation and computation of cumulative PDFs. For comparison, the full simulation runtime is also listed.}
    \begin{tabular}{|c|c|}
        \hline
        Task & Average Computing Time\\
        \hline
        \hline
        \makecell{Generating integration limits, computing cumulative PDFs,\\and interpolating two events} & $\SI{39.2}{\milli\second}$\\
        \makecell{Generating integration limits, computing cumulative PDFs,\\and interpolating $\SI{20000}{}$ events} & $\SI{784}{\second}$\\
        \hline
        Interpolating two events & $\SI{5.2}{\milli\second}$\\
        Interpolating $\SI{20000}{}$ events & $\SI{104}{\second}$\\
        \hline
        Generating one event with full simulation & $\SI{41.56}{\second}$\\
        Generating $\SI{10000}{}$ events with full simulation & $\SI{4.81}{\day}$\\
        \hline
    \end{tabular}
    \label{tab: computing times for hit level interpolation}
\end{table}

%% file: chapters/conclusion.tex
\section{Conclusion}
\label{sec: conclusion}

A data-driven fast simulation approach of pion showers has been presented in this work. Unlike previous studies which only investigated the simulation of hadron showers based on generative models with simulated data samples, this approach has been implemented based on a pion shower dataset that was recorded under realistic, experimental conditions in 2018 at \cern with the \ahcal Technological Prototype. It comprises pion shower data of nine different initial energies ranging between $\SI{10}{\giga\electronvolt}$ and $\SI{200}{\giga\electronvolt}$. From this dataset, energy distributions of individual calorimeter tiles have been calculated.
\par
KDEs have been exploited in order to estimate hit energy distributions. In contrast to neural-network-based approaches, KDEs offer a model-independent description of shower PDFs without relying on the optimisation of network architectures or hyperparameters, as well as the complex training procedures. From these estimated PDFs, simulated events have been generated and compared with data. In particular, kinematic shower variables have been computed and compared between data, fast simulation, and full simulation. These variables include the total shower energy, the (energy-weighted) mean shower radius, various shower moments, and many more. Furthermore, correlation factors between pairs of kinematic variables have also been calculated, both for data as well as fast simulation. The results show excellent agreement between data and the fast simulation approach. For certain variables, the fast simulation performs even better than the full simulation, aided by its data-driven nature.
\par
Based on these results, an interpolation algorithm has been developed, aiming at estimating hit energy distributions of pion energies that are not included in the pion shower dataset. For this, events of a target energy, $E^{\text{int}}$, have been interpolated from two different initial energies, $E^{\text{small}}$ and $E^{\text{large}}$, one of which is smaller and the other larger than the target energy. Depending on their distances to the target energy, $E^{\text{small}}$ and $E^{\text{large}}$ have been weighted differently during the interpolation, such that closer energies have a stronger influence on the final distributions, and vice versa. The results of this procedure also agree very well with data, except for the number-of-hits distribution, which means that the interpolation is capable of predicting shower distributions of any energy that lies within the boundaries of the recorded dataset.
\par
There are various ways of improving the fast simulation even further. One possibility is to develop an extrapolation algorithm, in this case for energies below $\SI{10}{\giga\electronvolt}$ or above $\SI{200}{\giga\electronvolt}$. However, such an algorithm would need to be even more sophisticated than the interpolation already is. This is because extrapolated distributions would only be bounded from below or above (depending on the target energy) by one pion energy, not by two as it is the case for the interpolation. Any biases stemming from $E^{\text{small}}$ or $E^{\text{large}}$, as described in Section \ref{subsec: mathematical background of interpolations}, cannot be as easily corrected as in the interpolation algorithm. Consequently, an extrapolation algorithm will possess only a limited range of applicability; if the distance of the dataset to the target energy is too great, the results will likely be incomplete or even incorrect.
\par
Improvement of the fast simulation algorithm might also be achieved through means of data reduction. That is, by applying some transformation to the original input dataset such that the number of dimensions is reduced, but the associated information loss remains minimal, one may enhance the performance of the fast simulation even more. Such data compression could allow one to still make use of KDEs, but would also reduce the need for computational resources (perhaps even significantly). A dimensionality reduction investigation has already been conducted by applying the Discrete Cosine Transform to test beam data. The results of this investigation will be published in an upcoming PhD thesis.
\par
Hit timing information can also be included into the fast simulation, which would allow to not only simulate the overall structure of a pion shower, but also its temporal development within the \ahcal. In addition to that, full simulation can be used to generate samples of pion energies not included in the given dataset, of different incident angles, or even of different initial particles (or a combination of all of them). More generally, the fast simulation could be utilised in order to simulate any operating condition of the \ahcal. However, this would likely require corrections to the fast simulation algorithm, which might not be trivial, depending on the (pseudo-)dataset the fast simulation is based upon. Furthermore, simulating particles other than pions would likely also require additional test beam data. Still, if such investigations were successful, then this would complete the construction of a first data-driven fast simulation prototype that is able to simulate any kind of particle shower under various test beam conditions in highly granular calorimeters such as the \ahcal.